\pdfoutput=1
\documentclass[cernpreprint, atlasdraft=true, texlive=2016, coverpage=false, UKenglish, , texmf]{atlasdoc}
 
\usepackage{atlaspackage}

\usepackage{atlasbiblatex}
 
\usepackage{atlasphysics}

\graphicspath{{logos/}{figures/}}
 
\usepackage{ANA-STDM-2017-08-PAPER-defs}
\usepackage{comment}
\usepackage{multirow}
\usepackage{longtable}

\AtlasTitle{Two-particle Bose--Einstein correlations in \({\mathit pp}\) collisions at
\(\mathbf{\sqrt{s} = 13}\)~TeV  measured with the ATLAS detector at the LHC
}
 
\AtlasVersion{2.0}
 
\AtlasAbstract{
This paper presents studies
of Bose--Einstein correlations (BEC) in  proton--proton  collisions at  a
centre-of-mass energy of  \(13\)~TeV, using data from the ATLAS detector
at the CERN Large Hadron Collider.
Data were collected in a special low-luminosity configuration with
a minimum-bias trigger and a high-multiplicity track trigger,
accumulating integrated luminosities of \(151\)~\(\mu\)b\(^{-1}\)
and \(8.4\)~nb\(^{-1}\),  respectively.
The BEC are measured for pairs of like-sign charged particles, each with \(|\eta| < 2.5\),
for two kinematic ranges:
the first with particle    \(p_{\mathrm{T}} > 100\)~MeV and
the second with particle \(p_{\mathrm{T}} > 500\)~MeV.
The BEC parameters, characterizing  the  source radius and particle correlation strength,
are investigated as functions of charged-particle
multiplicity (up to 300) and average transverse momentum of the pair
(up to  \(1.5\)~GeV).
The double-differential  dependence on charged-particle multiplicity
and average transverse momentum of the pair is also studied.
The BEC radius is found to be independent of the charged-particle multiplicity
for high charged-particle multiplicity (above 100),
confirming a previous observation at lower energy.
This saturation occurs independent
of the transverse momentum of the pair.
}
 
\author{The ATLAS Collaboration}
 
\AtlasRefCode{STDM-2017-08}
 
\PreprintIdNumber{CERN-EP-2021-172}
 
\AtlasDate{\today}

\AtlasJournalRef{Eur. Phys. J. C 82 (2022) 608}
\AtlasDOI{10.1140/epjc/s10052-022-10472-0}

\hypersetup{pdftitle={ATLAS document},pdfauthor={The ATLAS Collaboration}}
 
\begin{document}
 
\maketitle
 
{
}
 
\section{Introduction}

Particle correlations play an important role in the understanding of multiparticle production
in hadron--hadron, hadron--nucleus and nucleus--nucleus collisions.
Correlations between two identical bosons, called Bose--Einstein correlations,
(BEC),
are a well known phenomenon in high-energy and nuclear physics  (for reviews see
Refs.~\cite{becrev-zajc,qo-jap,becrev-weiner-1,becrev-kit,becrev-tamas,becrev-alex,Kittel:2005fu,Lisa:2005dd}).
The BEC are often  considered to be the  analogue of the Hanbury Brown and Twiss effect
\cite{hbt,hbt1,hbt2} in astronomy,  describing the interference of incoherently emitted identical bosons
\cite{hbt2bec,hbt2bec1,hbt2bec-1,hbt2bec-2}.
The BEC constitute a sensitive probe of the space-time geometry of the hadronization region, and
allow the determination of the size and shape of the source from which particles are emitted.
High-multiplicity data in proton--proton  interactions can serve as a reference for studies of nucleus--nucleus  collisions
\cite{Shuryak:2014zxa}.
The effect of BEC is reproduced in both the hydrodynamical/hydrokinetic
\cite{sinyukov,sinyukov-2,Shapoval:2013jca}
and the Pomeron-based \cite{schegelsky,schegelsky-1} approaches for modelling
of
high-multiplicity hadronic interactions.
The BEC between a pair of particles depend on the particle multiplicity
in the event and on the average transverse momentum of the pair.
The dependence on these parameters is particularly sensitive to the space-time features of the hadronization region, as
discussed by the authors of various models
\cite{kT-th,kT-th-1,bec-mult-2_0,diffppaamods-3,diffppaamods-4,diffppaamods-6}.
 
The BEC can be studied for two, three, or more identical bosons, assuming a hadronization
region parametrized by one, two, or three size parameters.
At the Large Hadron Collider (LHC), the BEC with one size parameter, the source radius, have been studied
by the ATLAS Collaboration
in \(pp\) collisions  at the centre-of-mass energies \(\sqrt{s} = 0.9\) and \(7\)~\TeV\
\cite{Aad:2015sja}, and
in \(p\)+Pb collisions  at \(\sqrt{s_\mathrm{NN}} = 5.02\)~\TeV\ per nucleon--nucleon pair
\cite{Aaboud:2017xpw}.
The BEC source radius parameter, \(R\), was observed by ATLAS to saturate, i.e.\ to reach a plateau level, at high
charged-particle multiplicity, in \(\sqrt{s} = 7\)~\TeV\ \(pp\) collisions \cite{Aad:2015sja}.
The  BEC in \(pp\) collisions have also been studied  in the
ALICE
\cite{alice-bec900,alice-bec3d,Abelev:2014pja,Adam:2015pbc,Adam:2029802},
CMS
\cite{cms-bec1st,cms-JHEP,Sirunyan:2017ies,Sirunyan:2019umv}
and
LHCb \cite{Aaij:2017oqu} experiments.
The CMS results at \(13\)~\TeV\ also show a plateau in \(R\) at high multiplicity \cite{Sirunyan:2019umv}.
 
Pomeron-based models
predict
a plateauing of \(R\) at intermediate multiplicities
as a consequence of the overlap of the colliding protons, followed by the
return to a lower value of \(R\) at high multiplicities \cite{schegelsky,schegelsky-1}.
The IP-Glasma model \cite{Bzdak:2013zma}
predicts
some tendency to
\(R\) saturation at high multiplicities for \(pp\)  interactions.
 
In a different approach, two-particle correlations have been analysed at
\(\sqrt{s} = 7\)~\TeV\ by ATLAS in the
framework of a phenomenological model based on ordered hadron chains. This approach
provides an alternative interpretation
of the correlations commonly
attributed to Bose--Einstein interference \cite{Aaboud:2017plr}.
 
In this paper, studies of the
BEC
in \(pp\) collisions at \(\sqrt{s} = 13\)~\TeV\  using the ATLAS detector  \cite{atlasdet} are presented.
The same methodology is used in this analysis as was used in the previous ATLAS BEC studies \cite{Aad:2015sja},
and the same minimum-bias  (MB) trigger is
used  as was used in an earlier study of  charged-particle distributions
at  \(13\)~\TeV\ \cite{Aad:2016mok,Aaboud:2016itf}.
In addition, the study is extended to the region of high-multiplicity collisions
by using the high-multiplicity track (HMT) trigger previously used for studies
of long-range elliptic azimuthal anisotropies \cite{Aad:2015gqa}.
 
This paper is organized as follows.
The analysis method is described in Section~\ref{analysis}.
The  ATLAS detector is described in  Section~\ref{atlasdetector}.
The  Monte Carlo (MC) and data samples are presented in Section~\ref{datasets}.
The
track selection
criteria and
data correction
procedure are presented in Section~\ref{correction}
and  the reference sample is discussed in Section~\ref{subsect:ref_sample}.
The systematic uncertainties are presented in Section~\ref{systematic}.
The BEC source radius and correlation strength dependence on, separately,
charged-particle multiplicity and average transverse momentum of the pair
is
shown in Section~\ref{results}.
This section also presents the double-differential dependence of the BEC parameters
on multiplicity and average transverse momentum.
A comparison with previous, lower-energy ATLAS results is presented,
as well as a comparison with CMS results at \(13\)~\TeV.
The summary and conclusions are given in Section~\ref{summary}.

\section{Analysis method}
\label{analysis}
The BEC are observed at small values of the square of the four-momentum difference, \(Q^2\),
between two identical bosons,
where  \(Q^2 = - (p_1 - p_2)^2  = {M_{12}^2} - {4 m^2}\),
\(p_1\) and \(p_2\) are the four-momenta of the particles,
\(M_{12}\) is the invariant mass of the  particle pair and \(m\) is the particle
mass.
The BEC parameters are measured in terms of a two-particle correlation function,
\begin{equation}
\label{eq_c2}
C_{2} (Q) = {\rho (Q)  }/{  \rho_0 (Q)  },
\end{equation}
where \(\rho (Q)\) is the distribution of \(Q\)
formed from all like-sign charged-particle pairs,
assuming in this analysis that all charged particles are pions, and \(\rho_0 (Q)\) is a reference distribution
specially constructed to exclude the effect of BEC.
In this analysis, \(\rho_0 (Q)\) is constructed from unlike-sign charged-particle pairs,
as discussed in Section~\ref{subsect:ref_sample}.
The distributions \(\rho (Q)\) and \(\rho_0 (Q)\)  are normalized to unity,
i.e.\ they are  probability density functions.
 
To account for the effects of resonances
(for example  \(\eta\), \(\eta^{\prime}\),  \(\omega\), \(\rho\),    \(K^{*}\), \(\phi\)),
the \(C_2 (Q)\)  correlation  function
is corrected by dividing by the corresponding quantity constructed
from MC simulations that contain resonances but not the BEC:
\begin{equation}
\label{eq_4}
R_{2}(Q) =
\frac{
C_{2}^{\mathrm{data}}(Q)
}{
C_{2}^{\mathrm{MC}}(Q)
}
=
\frac{ \rho^{\mathrm{data}}{
(Q)
} }{\rho_0^{\mathrm{data}} (Q)
}
\Bigg/ \frac{ \rho^{\mathrm{MC}}
{
(Q)
} }{  \rho_0^{\mathrm{MC}}  (Q)
}\, ,
\end{equation}
where \(C_{2}^{\mathrm{data}}(Q)\)
and \(C_{2}^{\mathrm{MC}}(Q)\) are the \(C_{2} (Q)\)
functions reconstructed from data and MC samples, respectively.
 
The BEC effect is usually described  by  a function, \(\Omega\), with two parameters:
the effective \(R\) parameter, and  the  strength parameter, \(\lambda\),
where the  latter is also called the  incoherence or chaoticity parameter  \cite{lambdaintro}.
If measurement can be made down to a very small value of \(Q\)  (\( \lesssim 10\)~\MeV),
the parameter \(\lambda\) may also be sensitive to a ``halo'' contribution due to long-lived
resonances
\cite{becrev-tamas},
but this is not possible in this analysis.
A typical functional form is:
\begin{equation}
\label{eq_3}
R_{2}(Q)
= c_{0} [1 +  \Omega(\lambda, Q R)] (1 + \varepsilon Q) \, .
\end{equation}
In a simplified scheme for fully coherent emission of identical bosons,
\(\lambda =0\), while  for incoherent (chaotic) emission, \(\lambda =1\).
Various  parameterizations of the  \(\Omega (\lambda, Q R)\) function  in Eq.~(\ref{eq_3})
can be found in the  literature (see for example Refs.~\cite{becrev-weiner-1,Kozlov:2007dv}),
each assuming a different shape for the particle-emitting source.
The parameter \(c_{0}\) is a constant depending on the normalizations of
\( \rho (Q) \) distributions and is chosen such that \(R_2 (Q)\) is unity at large \(Q\).
 
A L\'{e}vy-type parameterization for \( \Omega \) introduces a further free parameter, \( \alpha \) :
\begin{equation}
\nonumber
\Omega (\lambda, Q R, \alpha) = \lambda\, \mathrm{e}^{{- (R Q)}^\alpha},
\end{equation}
with \( 0 < \alpha \le 2\).
The L\'{e}vy distribution is a simple exponential when \( \alpha = 1\)   and a Gaussian when \( \alpha = 2\).
Fits to \( R_2 (Q)\)  with \( \alpha \) free usually return a value between 1 and 2,
and the fit quality,
in terms of \( \chi^2 / ndf \), is generally better than the fit using an exponential or Gaussian.
However, the interpretation of the R parameter in terms of the distribution of space-time points is then less clear.
Also, in studies for the 7~\TeV\ analysis it was found that \( \alpha \) varied when \( R_2 (Q)\)
was studied more differentially in terms of charged multiplicity or pair transverse momentum.
So, in common with the other LHC experiments, this analysis constrains \( \alpha \)  to be 1 or 2.
Investigation of the Gaussian parameterization is found to give a significantly poorer
description of the data in the low-\(Q\) region.
(See Section \ref{twoparticle}).
So,
in the study presented here, the BEC parameters are extracted using  an exponential parameterization,
\begin{equation}
\label{eq_exp}
\Omega (\lambda, Q R) = \lambda\, \mathrm{e}^{- R Q},
\end{equation}
of a static source assuming a radial Lorentzian distribution of the source.
The fitted parameter \( \varepsilon \) in Eq.~(\ref{eq_3}) takes into account possible
long-distance correlations that have not been removed fully by \( \rho (Q) \).

\section{The ATLAS detector}
\label{atlasdetector}
 
ATLAS  is a multipurpose particle physics experiment
operating at one of the beam interaction points at the LHC.
The ATLAS detector\footnote{ATLAS uses a right-handed coordinate system
with its origin at the nominal interaction point (IP)
in the centre of the detector and the \(z\)-axis along the beam pipe.
The \(x\)-axis points from the IP to the centre of the LHC ring,
and the \(y\)-axis points upward.
Cylindrical coordinates \((r, \phi)\) are used in the transverse plane, \(\phi\) being
the azimuthal angle around the beam pipe.
The pseudorapidity, \(\eta\), is defined in terms of the polar angle
\(\theta\) with \(\eta=-\ln\tan(\theta/2)\).
}
covers almost the whole solid angle around
the collision point with layers of tracking detectors, calorimeters and muon chambers.
It is designed to study a wide range of physics topics at LHC energies.
For the measurements presented in this paper, the tracking devices and the
trigger system are of particular importance.
 
The   innermost part of the   ATLAS  detector is the inner tracking
detector (ID), which  has full coverage in \(\phi\) and
covers the pseudorapidity range \(|\eta|<2.5\).
The ID is immersed in the \(\mathrm{2\, T}\)  axial magnetic field of a superconducting solenoid
and measures the trajectories of charged particles.
It consists of a silicon pixel detector (Pixel),
a silicon microstrip detector (SCT) and a straw-tube transition radiation tracker (TRT),
each
split into a barrel and two endcap components.
The Pixel, SCT and TRT
are located around the interaction point spanning radial distances of
\(33\)--\(150\)~mm, \(299\)--\(560\)~mm and \(563\)--\(1066\)~mm,
respectively.
The barrel (each endcap)
consists of four (three) pixel layers, four (nine) double layers
of silicon microstrips  and 73 (160) layers of TRT straws.
The Pixel, SCT and TRT have \((r, \phi)\)-position resolutions of
\(10\)~\(\mu\)m,  \(17\)~\(\mu\)m,  and  \(130\)~\(\mu\)m,
respectively.
 
For Run 2 a new innermost pixel layer,
the insertable B-layer  (IBL)  \cite{ibl_1}, was installed  around a new, narrower
and thinner beam pipe.
The IBL is composed of \(14\) lightweight staves arranged in a cylindrical geometry,
each made of \(12\) silicon planar sensors in its central region and \(2\times 4\)
three-dimensional sensors at the ends.
The IBL pixel dimensions are  \(50\)~\(\mu\)m in the \(\phi\)-direction and  \(250\)~\(\mu\)m
in the \(z\)-direction (compared with \(50\)~\(\mu\)m by \(400\)~\(\mu\)m for
the other pixel layers).
The intrinsic spatial resolution of the IBL readout is \(10\)~\(\mu\)m in \((r, \phi)\)-position
and \(75\)~\(\mu\)m in \(z\)-position \cite{ATLAS:2012rar}.
The smaller radius and the reduced pixel size result in improvements
in both the transverse and longitudinal impact parameter resolutions
\cite{Aad:2016mok,Aaboud:2016itf}.
The services for the existing pixel detector were upgraded,
significantly reducing the amount of material in
the region \(|\eta| > 1.5\), in particular at the boundaries of the active tracking volume.
 
A track from a charged particle traversing the barrel detector typically has \(12\) silicon
measurement points (hits), of which \(4\) are Pixel and \(8\) are SCT,
and more than \(30\) TRT  straw hits.
Requirements on an IBL hit and on impact parameters
strongly suppress the number of tracks from secondary particles.
 
The ATLAS detector has a two-level trigger system:  the first-level (L1) trigger and the
high-level trigger (HLT)  \cite{Aaboud:2016leb}.
Events used in this analysis were required to satisfy L1 triggers using the
minimum-bias trigger scintillators (MBTS).
These are mounted at each end of the detector in front of the liquid-argon
endcap-calorimeter cryostats  at \(z = \pm3.56\)~m,
and are segmented into  two  rings in pseudorapidity
(\(2.07 < |\eta| < 2.76\) and \(2.76 < |\eta| < 3.86\))
with \(8\) azimuthal sectors in the inner ring and \(4\) in the outer.
The MB events were selected on the basis of the MBTS alone,
while the  HMT events satisfied additional HLT criteria.
Further details are given in Section~\ref{dataset}.
 
An extensive software suite~\cite{ATL-SOFT-PUB-2021-001} is used in the reconstruction and analysis of
real and simulated data, in detector operations and in the trigger and data acquisition systems of the experiment.
 
\section{Monte Carlo and data samples}
\label{datasets}
\subsection{Data samples and event selection}
\label{dataset}
 
This study uses the \(pp\)-collision datasets  at \(\sqrt{s} = 13\)~\TeV\
that were used in previous ATLAS studies of
MB
events  \cite{Aad:2016mok,Aaboud:2016itf,Aaboud:2017fwp}
and
HMT
events  \cite{Aaboud:2016yar}.
The MB and HMT datasets were taken in a special configuration
of the LHC in June 2015,
with low beam currents and reduced beam focusing, producing a
low mean number of interactions per  bunch-crossing, \(\langle\mu\rangle\):
in the range \(0.003\)--\(0.007\) for MB events
and \(0.01\)--\(0.04\) for  HMT events.
This configuration and the event selection used
are documented in detail in the ATLAS
\(\sqrt{s} = 13\)~\TeV\  MB analysis
\cite{Aad:2016mok,Aaboud:2016itf}.
For the MB data sample,  events were selected from colliding proton bunches using the
MBTS trigger.
This trigger selection required one counter above threshold from
either side of the detector, and is referred to as a single-arm trigger.
The HMT trigger is a combination of an L1 requirement on the number
of hits in the MBTS
and an HLT requirement on the multiplicity of HLT-reconstructed
charged-particle tracks.
This trigger required more than 900 SCT space-points and
more than 60 reconstructed good-quality
charged tracks with
\(p_{\mathrm{T}}> 400\)~\MeV\
associated with the reconstructed vertex with the highest multiplicity in the event.
These selections correspond to integrated luminosities of
\(151\)~\(\mu\)b\(^{-1}\) and \(8.4\)~nb\(^{-1}\)
for the MB and  HMT triggers, respectively.
 
In the offline selection,
track candidates are reconstructed in the Pixel and SCT detectors and extended to
include measurements in the TRT
\cite{Aaboud:2017all}.
The selected tracks  satisfy the following criteria:
\(p_{\mathrm{T}} > 100\)~\MeV\ and \(|\eta|< 2.5\);
at least one Pixel hit and an IBL  hit if expected;\footnote{A
hit is expected if the extrapolated track crosses a known active
region of a pixel module.
If an innermost pixel layer hit is not expected, a next-to-innermost pixel layer
hit is required if expected.
}
at least two, four or six SCT hits for \(p_{\mathrm{T}} < 300\)~\MeV,
\(p_{\mathrm{T}} < 400\)~\MeV\ or  \(p_{\mathrm{T}} > 400\)~\MeV,
respectively, in order to account for the dependence of track length on
\(p_{\mathrm{T}}\).
All events are required to have at least one vertex \cite{Aaboud:2016rmg},
formed from a minimum of two tracks.
The vertex position is required to be consistent with the beam-spot
position within the ATLAS detector
\cite{Aaboud:2017all}.
If more than one vertex is reconstructed, the primary vertex (PV)
is defined as the vertex
with the largest number of associated reconstructed tracks,
weighted by the square of their transverse momenta
\cite{Aaboud:2017all}.
Selected tracks are then also required to satisfy
\(|d_0^{\mathrm{BL}}| < 1.5\)~mm, where the transverse impact parameter
\(d_0^{\mathrm{BL}}\) is calculated with respect to the measured beam line (BL)
\cite{ATLAS:2010gaa};
and \(|z_0^{\mathrm{BL}}\times \sin{\theta}| < 1.5\)~mm,
where \(z_0^{\mathrm{BL}}\) is the difference between the
longitudinal coordinate of the track along the beam line at the point where
\(d_0^{\mathrm{BL}}\) is measured and the longitudinal coordinate
of the PV, and \(\theta\) is the polar angle of the track.
High-momentum tracks with mismeasured \(p_{\mathrm{T}}\)
are removed by requiring the track-fit \(\chi^2\) probability to be larger than
$0.01$ for tracks with
\(p_{\mathrm{T}} > 10\)~\GeV.
A special configuration of the track
reconstruction algorithms is
used in this analysis to reconstruct low-momentum tracks \cite{Aaboud:2017all}.
The track reconstruction efficiency  at  \(p_{\mathrm{T}}\approx 100\)~\MeV\
is  about \(28\)\% and increases rapidly
to almost \(75\)\% at  \(p_{\mathrm{T}}\approx 300\)~\MeV\
\cite{Aaboud:2016itf}.
For the  MB trigger, the low-\(p_{\mathrm{T}}\) selected dataset consists of \(9.3\times 10^{6}\) events with a total of
\(2.4 \times 10^{8}\)  tracks  with \(p_{\mathrm{T}} > 100\)~\MeV.
The high-\(p_{\mathrm{T}}\) selection requires in addition at least two tracks
with \(p_{\mathrm{T}} > 500\)~\MeV. This yields \(8.9 \times 10^{6}\) events with
\(1.1 \times 10^{8}\) tracks with \(p_{\mathrm{T}} > 500\)~\MeV.
For the  HMT trigger, which requires at least 60 tracks
with \(p_{\mathrm{T}} > 400\)~\MeV\ at the trigger level,
the  low-\(p_{\mathrm{T}}\) selection requires at least 60 tracks
with \(p_{\mathrm{T}} > 100\)~\MeV\ and results in \(9.1 \times 10^{6}\) events with
\(9.8 \times 10^{8}\)  tracks  with \(p_{\mathrm{T}} > 100\)~\MeV.
For \(p_{\mathrm{T}} > 500\)~\MeV\ the requirement is at least
40 tracks with \(p_{\mathrm{T}} > 500\)~\MeV;
the sample consists of \(9.1 \times 10^{6}\) events
with \(5.3 \times 10^{8}\)  tracks with \(p_{\mathrm{T}} > 500\)~\MeV.
 
To reduce contamination from multiple proton--proton interactions (pile-up) in the colliding bunches,
events with a second vertex containing four or more tracks are removed.
Such events represent less than \(0.3\)\% of the sample.
Secondary vertices can be reconstructed with high efficiency provided the distance, \(\Delta z\),
between the  \(z\)-coordinates of the primary and pile-up vertices
satisfies \(|\Delta z|\ge 4\)~mm.
To estimate  any  possible residual influence of multiple  \(pp\) interactions in
MB- and  HMT-triggered data, a dedicated study
of the \(\Delta z\) distribution was performed,
using a sample with no pile-up rejection.
This distribution was extrapolated to estimate the level of pile-up
that will survive the maximum track-multiplicity limit on a second vertex.
For the MB events, the fraction of  events with pile-up vertices in the region
\(|\Delta z|\le 4\)~mm around the PV
is  about \(0.006\)\%.
For the HMT events, it is about \(0.05\)\%.\footnote{Since
\(\langle\mu\rangle\) is much smaller than 1 for both
the MB and HMT samples,
the ratio of the number of events with more than one vertex to the number of events
with at least one vertex is
\( \approx \langle\mu\rangle /2 \).
The requirement \(|\Delta z| \le 4\)~mm reduces this fraction by a further factor
of \( \approx 100 \).
}
The average track multiplicities per pile-up vertex
for the MB and HMT samples are \(9.4\) and \(23\), respectively.
For the primary vertices these multiplicities are \(26\) and \(108\).
The fractions of surviving pile-up tracks in the MB and HMT samples are approximately
\(0.002\)\% and \(0.01\)\% per event, respectively,
and these levels have a negligible influence on the BEC studies.
 
The contributions from beam--gas collisions and from non-collision background
(cosmic rays  and detector noise) were investigated in Ref.~\cite{atlas927} and found to be negligible.
 
\subsection{Monte Carlo samples}
 
The  \textsc{Pythia\,8}  MC generator
\cite{pythia8}
was used to  calculate the  \(C_2^{\mathrm{MC}}(Q)\) correlation functions
(Eq.~(\ref{eq_4})).
The generator was used with parameter values set to the \textsc{A2} tune \cite{ATLAS:2011krm} and
with the \textsc{MSTW2008lo  PDF}   set
\cite{Martin:2009iq}.
\textsc{Pythia\,8} models the effects of colour coherence,
which is important in dense parton environments and effectively reduces
the number of particles produced in multiple parton--parton interactions.
In \textsc{Pythia\,8}, the simulation is split into non-diffractive and diffractive processes.
The non-diffractive process is  dominated by  \(t\)-channel  gluon exchange,
comprising  approximately  80\% of the selected events.
The  diffractive processes,
which are further split into single- and double-diffractive processes, are
described by a Pomeron-based approach
\cite{Corke:2010yf}.
The contributions from non-diffractive, single-diffractive and
double-diffractive processes
were included in proportion to the cross sections predicted by
\textsc{Pythia\,8} with the  \textsc{A2} tune.
Large MC samples of MB and HMT events were generated
and passed through the ATLAS simulation
program \cite{Aad:2010ah}, which is based on \textsc{Geant4} \cite{geant4},
and the reconstruction chain, which is exactly that used for collision
dataset.
In addition, all selection criteria applied to the data were also applied as
appropriate to the MC samples.

In order to estimate the systematic uncertainty due to the choice of MC model,
large MC samples,
comparable to the size of the collision dataset, were also generated with
three
other programs.
These were the \textsc{Pythia\,8} Monash tune
\cite{Skands:2014pea},
the \textsc{EPOS}  MC generator
\cite{EPOS}
and
\textsc{Herwig++} \cite{Bahr:2008pv}
using the
\textsc{UE-EE-5} tune \cite{Gieseke:2012ft}.
The  Monash underlying-event tune
is based on  the   \textsc{NNPDF2.3lo PDF}
\cite{Ball:2012cx}
and incorporates updated fragmentation parameters.
\textsc{EPOS} implements a parton-based Gribov--Regge \cite{Gribov:1968fc}
theory, which is an effective field theory describing both hard and soft scattering.
The \textsc{EPOS} generator was used with  the \textsc{LHC}  tune \cite{Pierog:2013ria}.
The \textsc{Herwig++ UE-EE-5}  tune
is based on  the  \textsc{CTEQ6L1 PDF}.
All sets of generated events were passed through
the ATLAS simulation program and reconstruction chain.
 
The purity of the analysis sample of
particle pairs is defined as  the fraction of identical particle pairs
(\(\pi^\pm \pi^\pm\),  \(K^\pm K^\pm\), \(pp\) and \(\bar{p}\bar{p}\))
in the sample relative to all like-charge particle (LCP) pairs.
The purity of the analysis sample in terms of identical boson pairs is estimated from
\textsc{Pythia\,8} with the \textsc{A2} tune
to be about 74\% (where about  73\% are \(\pi^\pm \pi^\pm\)
and about  1\% are \(K^\pm K^\pm\)), using the selection criteria described above
for MB \(p_{\mathrm{T}}>100\)~\MeV.
In studies for the \(7\)~\TeV\ analysis \cite{Aad:2015sja} it was found that the impurity
reduces the value of \(\lambda\)  by a factor that is
approximately equal to the impurity
of the sample,
where a \(100\)\% purity has no effect, but the parameter \(R\) is unaffected.
 
\section{Data correction procedure}
\label{correction}
 
To correct for the inefficiencies in the trigger and
in the reconstruction of tracks and vertices, event-by-event and track-by-track
weighting procedures are applied as appropriate and
as used in the previous ATLAS measurements \cite{Aad:2016mok,Aaboud:2016itf}.
 
The efficiency of the MB trigger was studied with an independent control trigger
\cite{Aad:2016mok,Aaboud:2016itf}.
The control trigger selects events randomly at L1,
which are then filtered in the HLT
by requiring at least one reconstructed track with
\(p_{\mathrm{T}} > 200\)~\MeV.
The MB trigger efficiency increases from  \(97\)\%  for events with
low track multiplicity, \(n_{\mathrm{sel}} =2\),
to \(100\)\%  for events with  higher multiplicity, \(n_{\mathrm{sel}} \ge 7\),
where \(n_{\mathrm{sel}}\) is the number of tracks reconstructed offline and
satisfying the selection criteria given in Section~\ref{dataset}.
The efficiency of the HMT trigger is studied using MB-triggered events as the control sample.
The efficiency is found to be
\(15\)\%
for events
with \(n_{\mathrm{sel}} \simeq 85\)
and reaches a plateau at \(100\)\%
for \(n_{\mathrm{sel}} \geqslant  110\).
The measured MB and HMT trigger inefficiencies
as a function of \(n_{\mathrm{sel}}\)
are used  to correct each event.
These corrections are found
to have a negligible impact on the extraction of the BEC parameters discussed
in Section~\ref{results}.
The effect of the vertex-reconstruction efficiency on the BEC parameters is also negligible.
 
In addition, each track is assigned a weight that corrects for the
track reconstruction efficiency, for the fraction of secondary particles, for the
effect of detector resolution smearing primary
particles\footnote{In the  MC simulations,
primary charged particles are defined as  charged particles with a mean
lifetime \(\tau > 0.3 \times 10^{-10}\)~s either directly produced in \(pp\)
collisions or from the subsequent decay of particles with a shorter lifetime.
}
into and out of the fiducial region
(\( p_{\mathrm{T}} > 100 \)~\MeV, \( |\eta| < 2.5 \))
and for the fraction of
fake
tracks
formed by a random combination of
hits
\cite{Aad:2016mok,Aaboud:2016itf}.
The effect of the loss of  reconstructed pairs
at small opening  angles is negligible
in comparison with the track reconstruction corrections.
 
The \(Q\)
resolution is
found to be similar for all MC generators.
For the BEC region \( Q < 0.2 \)~\GeV, where the
\(Q\)
distribution rises steeply,
the \(Q\) resolution is better than 5~\MeV\ and much smaller than the bin size.
For \( Q > 0.2 \)~\GeV, where the \(Q\) distribution is essentially flat,
the resolution increases linearly with \(Q\), rising to
\({\simeq}40\)~\MeV\
for
\( Q \simeq 2 \)~\GeV.
Overall, no unfolding of \(Q\) is required.
 
The track
reconstruction inefficiency is such
that \(n_{\mathrm{sel}}\) differs significantly from the
true charged-particle multiplicity, \(n_{\mathrm{ch}}\).
In order to present and study \( R_2 (Q) \) (Eq.~(\ref{eq_4}))
as a function of  the true charged-particle multiplicity, \(n_{\mathrm{ch}}\),
the correlation functions \( R_2 (Q, n_{\mathrm{sel}}) \) must be unfolded.
In the earlier ATLAS MB study at \(13\)~\TeV\ \cite{Aaboud:2016itf,Aad:2016mok}
an
unfolding  matrix that reflects the probability
of reconstructing \(n_{\mathrm{sel}}\) charged tracks
in an event with a generated charged-particle  multiplicity
\(n_{\mathrm{ch}}\) was constructed, using
MC simulation with the \textsc{Pythia\,8} ATLAS  \textsc{A2} tune.
For the present analysis this unfolding matrix was applied
to both the MB and HMT data.
It is found that, on average, an event with
\(n_{\mathrm{sel}} = 100\)
reconstructed
tracks corresponds to
\(n_{\mathrm{ch}} = 127.6 \pm 0.1\)
charged particles at particle level.
 
As a check of the entire analysis chain, the distribution of \(C_2(Q)\)
(Eq.~(\ref{eq_c2}))
using the generated particles, \(C_2^\mathrm{MC, gen}(Q)\), was compared with the \(C_2^\mathrm{MC, rec} (Q)\)
distribution using the simulated detector response to reconstruct the tracks and vertices,
which were then corrected for inefficiencies as described above, and unfolded to the
particle level.
This study was carried out for several \(n_{\mathrm{ch}}\)
and \(k_{\mathrm{T}}\) intervals, where \(k_{\mathrm{T}}\) is the average transverse momentum
of the pair:
\begin{equation}\nonumber
k_{\mathrm{T}} = |{\mathbf p}_{{\mathrm{T}}, 1} + {\mathbf p}_{{\mathrm{T}},  2}|/2.
\end{equation}
In the small-\(Q\) region the difference
between the unfolded and
particle-level
distributions
varies,
depending on \(n_{\mathrm{ch}}\) and \(k_{\mathrm{T}}\), but is a few percent on
average.
The effect is significant for the first
two
\(Q\)-bins
used,
\(0.02 \le Q < 0.04\)~\GeV\
and
\(0.04 \le Q < 0.06\)~\GeV,
and for these bins it is
accounted for by adding it in quadrature to the statistical uncertainty
in
the \(R_2 (Q)\) distribution.
It is therefore not included in the systematic uncertainties discussed
in Section~\ref{systematic}   and listed in Table~\ref{tab:systr2}.
A large part of the difference in the first \(Q\)-bin, \(0.02 \le Q < 0.04\)~\GeV, is due to
photon conversions in the detector material,
which are included in the simulated detector response but are not part of the set of generated particles.
Since \(C_2\) is a ratio of like-sign pairs to unlike-sign pairs,
the denominator in \(C_2^\mathrm{MC,rec} (Q)\) contains the contribution of pairs from photon conversions,
which show up at small \(Q\), whereas \(C_2^\mathrm{MC,gen} (Q)\) does not.
For MB events the difference for the first bin
is 4\%, of which 3.5\% is the contribution of photon conversions.
For HMT events, the corresponding numbers are 3.25\% and 2\%.
Conversion photons are less significant for the HMT events, because
the number of track combinations increases
quadratically with \(n_{\mathrm{ch}}\) while the number of
conversion photons increases only linearly.
The effect of photon conversions in the region
\(Q > 0.04\)~\GeV\  is negligible.
 
The  long-range Coulomb force causes a momentum shift between particles
in the LCP and unlike-charge particle (UCP) pairs.
The density distributions are corrected for this effect by applying the Gamow penetration
factor per track pair with a weight \(1/G(Q)\) such that
\(\rho_{\mathrm{corr}} (Q)   = {\rho (Q) / G(Q)}\)
\cite{coulomb,coulomb-0,coulomb-1}  (for a review see Ref.~\cite{HI-revs-wied}).
The Gamow factor is given by
\begin{equation}\nonumber
G(Q) = 2 \pi \zeta / [\exp{(2 \pi \zeta)} - 1],
\end{equation}
with the dimensionless  parameter \(\zeta\) defined as  \(\zeta  = \pm \alpha m/ Q\),
where \(\alpha\)  is the electromagnetic fine-structure constant and
\(m\) is the pion mass.
The corrected two-particle  density function,  \(\rho_{\mathrm{corr}} (Q)\),
is normalized to unity.
The sign of \(\zeta\) is positive for LCP pairs and negative for UCP pairs.
The Gamow factor particularly affects pairs with
\(Q < 0.1\)~\GeV,
and the correction is
\(  24\)\% for \(0.02 \le Q < 0.04\)~\GeV,
\( 14\)\% for \(0.04 \le Q < 0.06\)~\GeV\
and
\( 10\)\% for \(0.04 \le Q < 0.06\)~\GeV.
The Coulomb interaction is not included in the generated
MC event samples, and so the correction is  applied only to data events.
\section{Reference sample}
\label{subsect:ref_sample}
 
A good choice of reference sample is important in obtaining an unbiased estimate
of the BEC signal.
Ideally, \(\rho_0  (Q)\) should include all momentum correlations present
in the signal dataset except for those arising from BEC.
Several different methods of constructing an appropriate reference sample
have been investigated by various collaborations and authors.
 
The main methods are:
\begin{itemize}
\item take UCP pairs from the same event;
\item take LCP pairs from the same event,  but invert the momentum
of one of the selected particles:  (\( p_{\mathrm{x}}, p_{\mathrm{y}}, p_{\mathrm{z}}
\rightarrow -p_{\mathrm{x}}, -p_{\mathrm{y}}, -p_{\mathrm{z}} \)).
This is the opposite hemisphere (OHP) technique \cite{Kittel:2005fu};
\item take LCP pairs  from the same event,
but  invert  the transverse momentum  of one of the selected particles:
(\( p_{\mathrm{x}}, p_{\mathrm{y}}, p_{\mathrm{z}}
\rightarrow -p_{\mathrm{x}}, -p_{\mathrm{y}}, p_{\mathrm{z}} \)).
This is the \(z\)-rotation method \cite{Kittel:2005fu};
\item take LCP pairs
from different events,  randomly mixed; this is the `mixed event' (MIX) technique \cite{mixed}.
\end{itemize}
In addition, most of the recently used methods compare the data with a reference sample
having global event parameter values that are the same as, or similar to, those in data --
such as the event multiplicity, the invariant mass of the pair, or the rapidity of the pair.
For the previous ATLAS analysis \cite{Aad:2015sja} the UCP, OHP and MIX procedures
were investigated in detail  using dedicated MC studies, and it was observed
that MIX and OHP behaved similarly.
For this analysis, the UCP and  OHP  were studied.
 
As the MC models do not implement BEC,
the single-ratio correlation functions
\(C_2^{\mathrm{MC}} (Q)\)
are expected to be close to 1, and to be a rather flat function of \(Q\),
particularly in the low-\(Q\), BEC-sensitive region.
This is indeed the behaviour observed in the earlier analysis \cite{Aad:2015sja} over much of the
parameter space in \(n_{\mathrm{ch}}\) and \(k_{\mathrm{T}}\)
for the MIX or OHP technique, and for the OHP  technique in this analysis.
However, in high-multiplicity and high-\(k_{\mathrm{T}}\) intervals,
which are relevant for this analysis,
\(C_2^{\mathrm{MC}} (Q)\) constructed with the MIX or OHP technique
exhibits an increase in the low-\(Q\), BEC-sensitive region.
 
\begin{figure*}[t!]
\begin{minipage}[h]{0.50\linewidth}
\center{\includegraphics[width=1\linewidth]{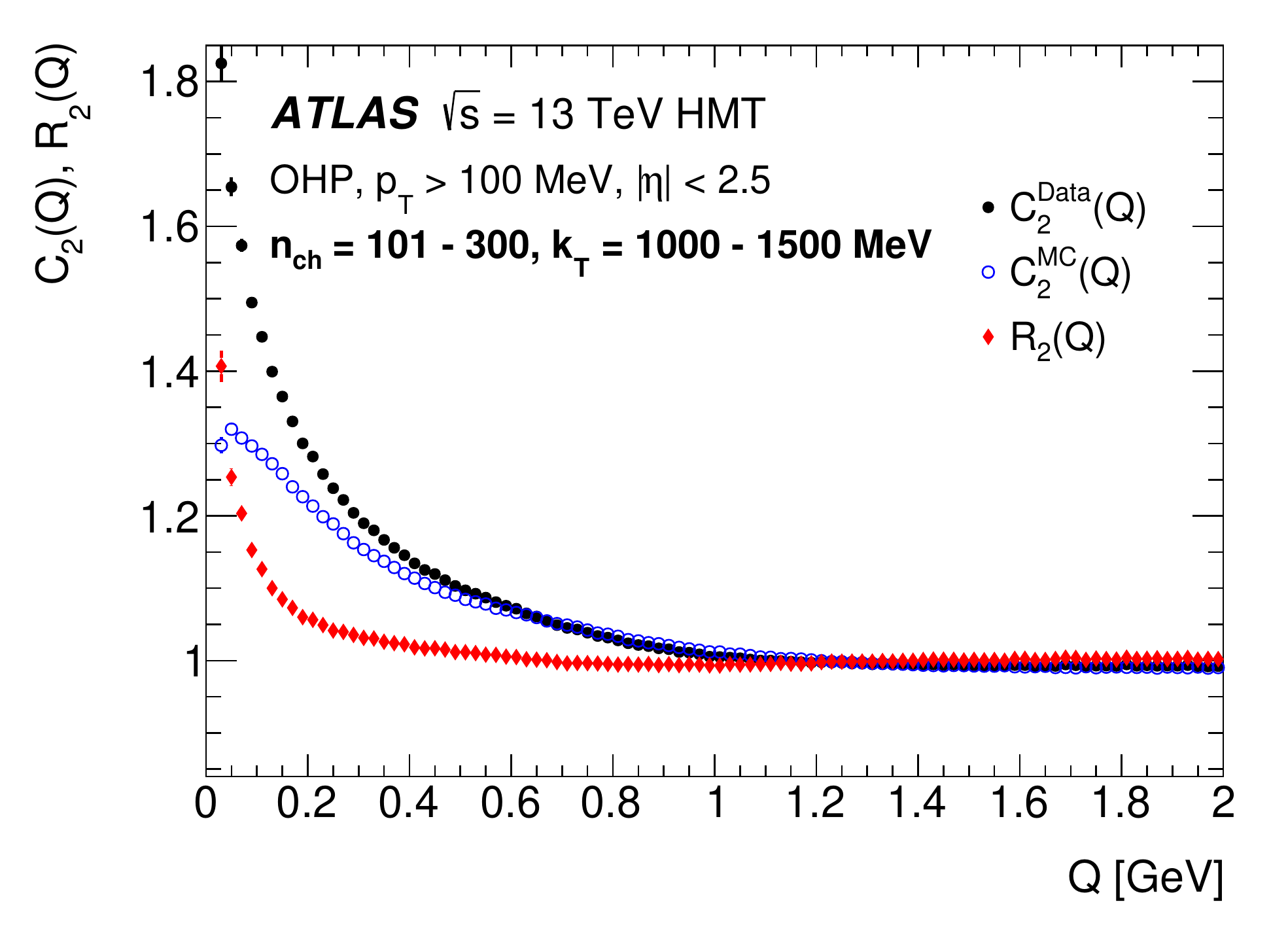}}
(a) \\
\end{minipage}
\hfill
\begin{minipage}[h]{0.50\linewidth}
\center{\includegraphics[width=1\linewidth]{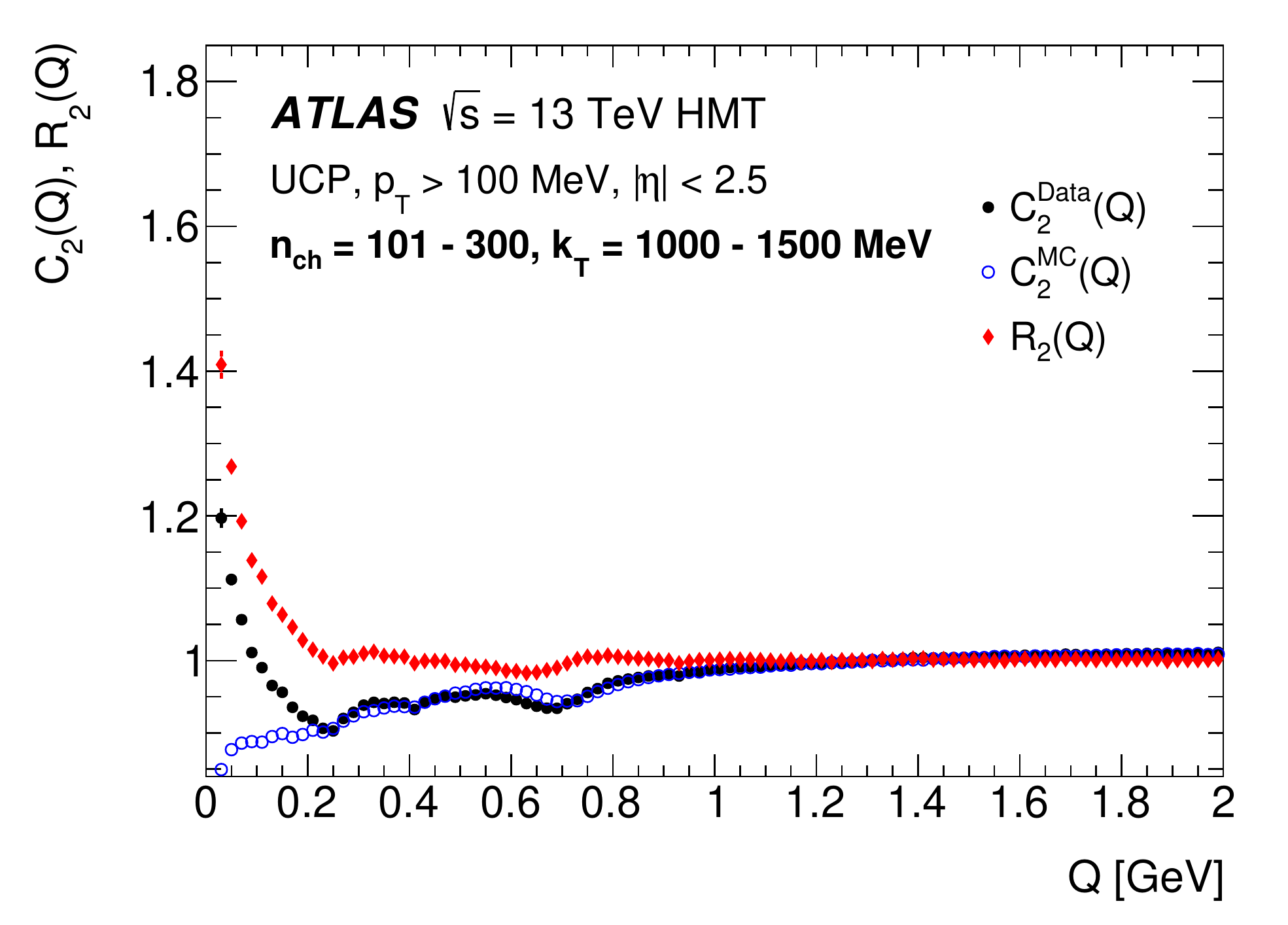}}
(b) \\
\end{minipage}
\caption{
Comparison of  single-ratio two-particle  correlation functions,
\(C_2^{\mathrm{data}}(Q)\)
and
\(C_2^{\mathrm{MC}}(Q)\),
with  the
two-particle double-ratio  correlation function,
\(R_2(Q)\),
for  the high-multiplicity track (HMT) events
using
(a) the opposite hemisphere  (OHP)  like-charge particles pairs reference sample
for
\(k_{\mathrm{T}}\)-interval
\(1000 < k_{\mathrm{T}} \le 1500\)~\MeV\
and
(b) the unlike-charge particle (UCP) pairs reference sample
for  \(k_{\mathrm{T}}\)-interval
\(1000 < k_{\mathrm{T}} \le 1500\)~\MeV.
}
\label{fig:Fig_1}
\end{figure*}
 
As an example, Figure~\ref{fig:Fig_1}(a)
shows the correlation functions using the OHP
technique for the reference sample.
The correlation functions shown are the single ratios
\(C_2^{\mathrm{data}}(Q)\) and \(C_2^{\mathrm{MC}}(Q)\),
and the double ratio  \(R_2(Q)\), all for HMT events, and for
the average transverse momentum of the pair
\(1000 < k_{\mathrm{T}} \le 1500\)~\MeV.
The \(C_2^{\mathrm{MC}} (Q)\) exhibits a rise similar to that seen
in \(C_2^{\mathrm{data}} (Q)\) in the BEC-sensitive region, although no such
effect should be present in  \(C_2^{\mathrm{MC}}(Q)\).
Such a rise would have a marked effect on the BEC parameters extracted from \(R_2(Q)\).
Figure~\ref{fig:Fig_1}(a) also shows that  \(R_2 (Q)\) is less than \(1\) in the region \(0.7<Q<1.2\)~\GeV.
 
The use of  UCP pairs for the reference sample was also investigated in detail.
As in the other techniques, \(C_2^{\mathrm{MC}} (Q)\) behaves satisfactorily over much of the
parameter space in \(n_{\mathrm{ch}}\) and \(k_{\mathrm{T}}\), although resonances distort the
region \(Q \approx 0.5\)~\GeV, particularly at low multiplicity or at high \(k_{\mathrm{T}}\).
An example is shown in Figure~\ref{fig:Fig_1}(b), using the UCP reference sample,
and for the same
\(n_{\mathrm{ch}}\) and \(k_{\mathrm{T}}\) intervals as in Figure~\ref{fig:Fig_1}(a).
Although there is some distortion in the resonance region,
where  \(C_2^{\mathrm{MC}} (Q)\) and \(C_2^{\mathrm{data}} (Q)\) do not match perfectly,
the resultant \( R_2 (Q) \) matches well the BEC region of \(C_2^{\mathrm{data}} (Q)\).
Accordingly, the UCP method is used in this analysis, as it was for the previous
ATLAS analysis  \cite{Aad:2015sja}, and the other methods are not suitable.
 
The  UCP  reference sample used in this analysis is taken from the same
events used to form pairs of like-charge particles,
i.e.\  \(\rho_0 (Q)  \equiv  \rho  (p_1^\pm, p_2^\mp)\).
By construction, this UCP sample has the same topology and global parameters
as the  LCP sample \(\rho (p_1^\pm, p_2^\pm)\), but is naturally free of any BEC effect.
However, the UCP sample contains oppositely charged hadron pairs from the decay of resonances
such as  \(\eta\), \(\eta^{\prime}\), \(\omega\), \(\rho\), \(K^{*}\), and \(\phi\),
which are not present in the  LCP combinations.
Details concerning resonances are given in Appendix~\ref{app:resonances}.
These  resonances  contribute to the low-\(Q\) region, and can give a spurious BEC
signature with a large effective \(R\)
\cite{resonances,Gyulassy:1988yr,Padula:1988uf,resonances-1,resonances-2,resonances-3,resonances-4}.
The effect of the resonant states in the \(Q\) variable
was checked in MC studies, and the most affected regions
are excluded from the analysis of the BEC effect, as described in Section~\ref{twoparticle}.
 
\section{Systematic uncertainties}
\label{systematic}
In this section the systematic uncertainties in values of the BEC parameters
\(R\) and  \(\lambda\) are summarized.
These parameter values are obtained from a fit to the exponential model
(\ref{eq_exp}).
This fit is applied to the `inclusive' distributions,
corresponding to \(2 \le n_{\mathrm{ch}} \le 250\) for MB
and \(100 < n_{\mathrm{ch}} \le 300\) for HMT, and all \(k_{\mathrm{T}}\).
The fit is also applied to selected ranges of \(n_{\mathrm{ch}}\) or \(k_{\mathrm{T}}\),
and to the double-differential (\(n_{\mathrm{ch}}\), \(k_{\mathrm{T}}\))-distributions.
The discussion in
this section is mainly in terms of the `inclusive' fits, but
the systematic uncertainties have been evaluated for all selections applied to the data.
 
The systematic uncertainties resulting from the track reconstruction efficiency,
which are parameterized as a function of \(p_{\mathrm{T}}\) and \(\eta\), were determined
in earlier analyses \cite{Aad:2016mok,Aaboud:2016itf}
and these are implemented as uncertainties in
the track weights  for particle pairs  in the \(Q\) distributions
that enter the correlation functions.
The effects of track splitting and merging
have been studied in detail.
They are largest for \(Q\) values below \(0.04\)~\GeV,
but even for the first used bin (\( 0.02 \le Q < 0.04\)~\GeV)
the effect is only 0.4\% and is negligible in comparison with other uncertainties.
 
The effects of uncertainties in the unfolding matrix
on the BEC parameters \(\lambda\) and \(R\)
are found to be negligible for \(n_{\mathrm{ch}} < 200\).
For the highest-multiplicity  MB (\( 201 \le n_{\mathrm{ch}} < 250\))
and HMT (\( 231 \le n_{\mathrm{ch}} < 300\)) events
this systematic uncertainty is very small and not more than
\(0.01\)\% for \(\lambda\) and \(0.03\)\% for \(R\).


\begin{figure*}[t!]
\begin{minipage}[h]{0.50\linewidth}
\center{\includegraphics[width=1\linewidth]{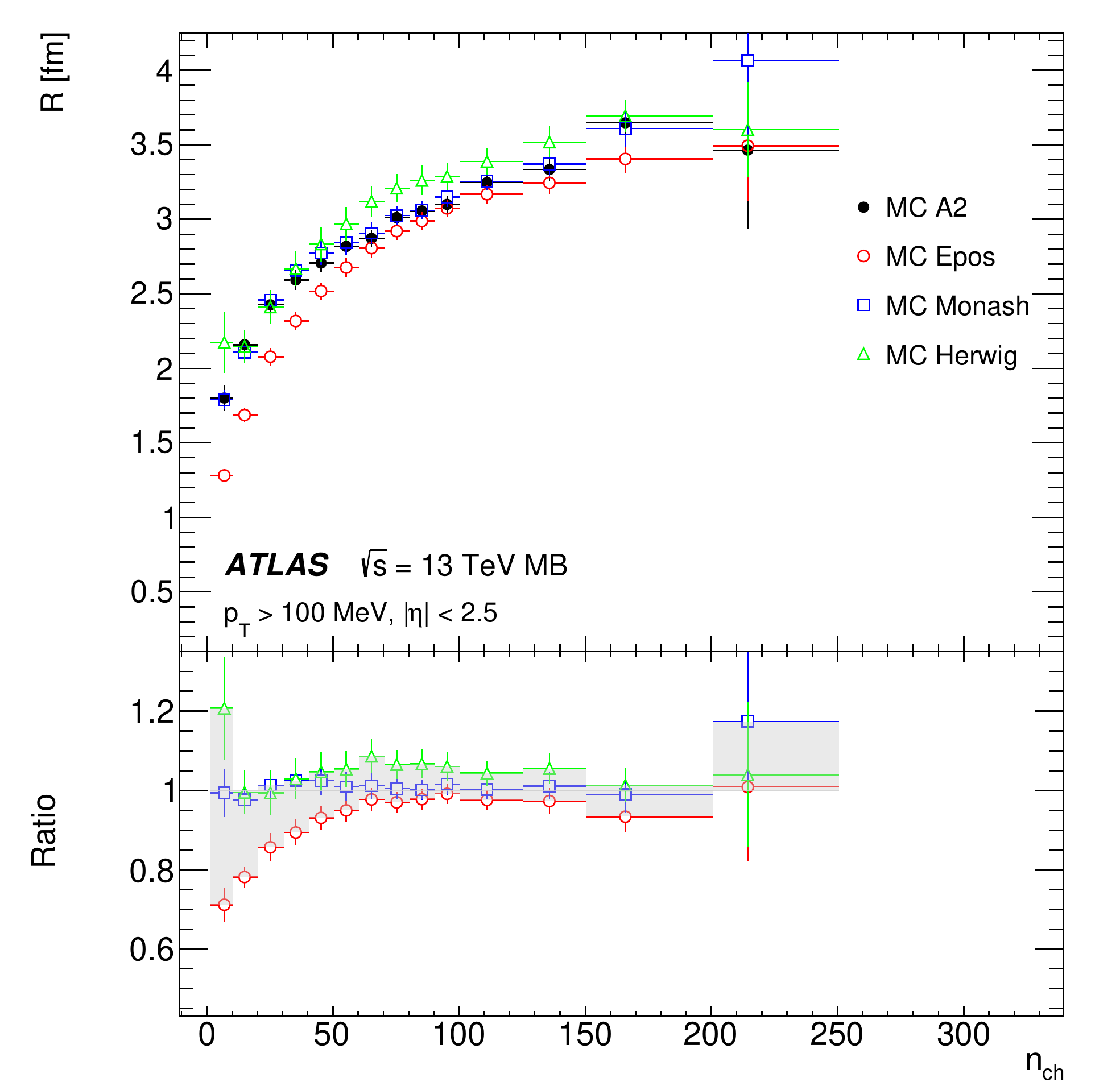}}
(a)
\\
\end{minipage}
\hfill
\begin{minipage}[h]{0.50\linewidth}
\center{\includegraphics[width=1\linewidth]{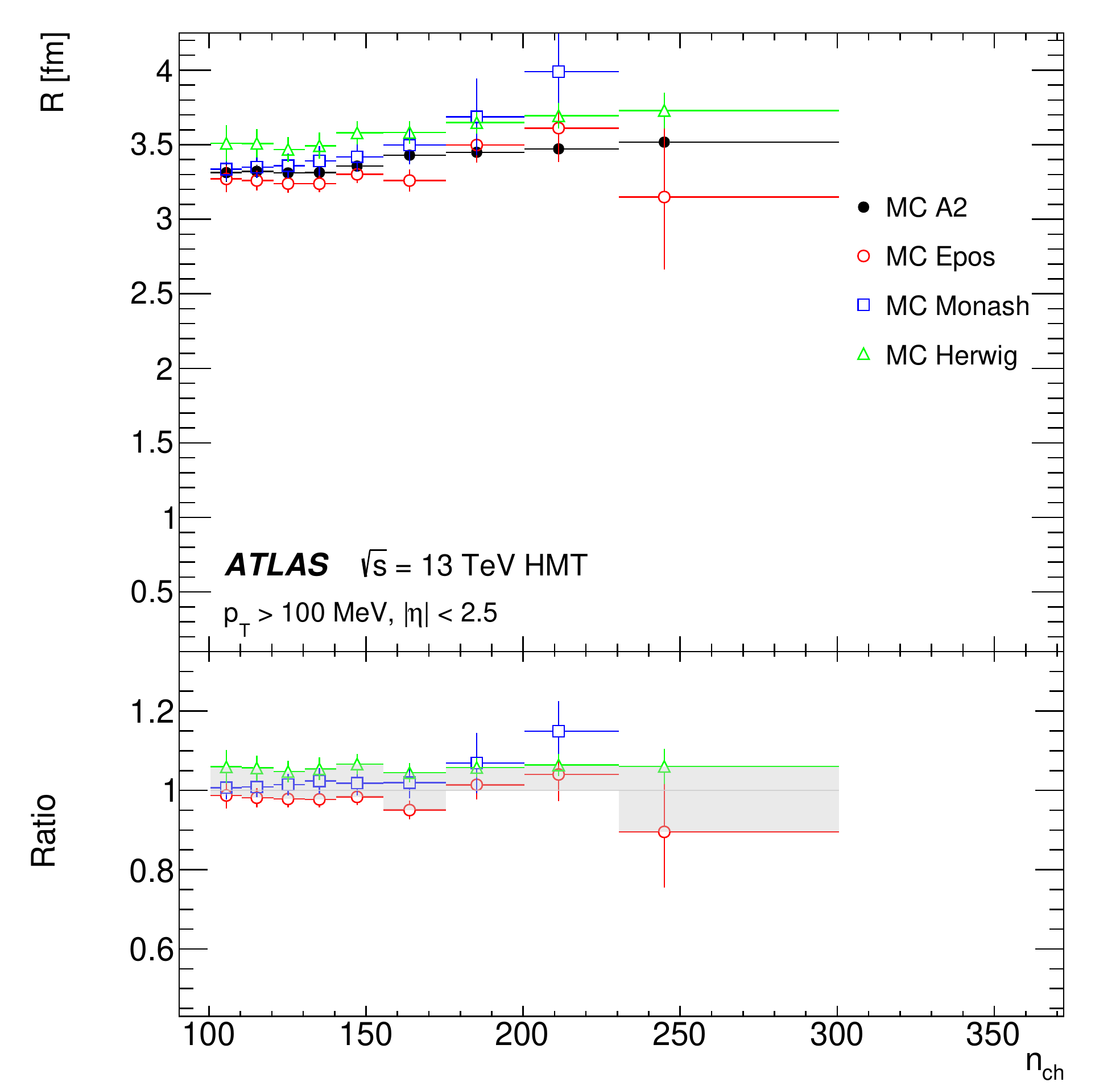}}
(b)
\\
\end{minipage}
\vfill
\begin{minipage}[h]{0.50\linewidth}
\center{\includegraphics[width=1\linewidth]{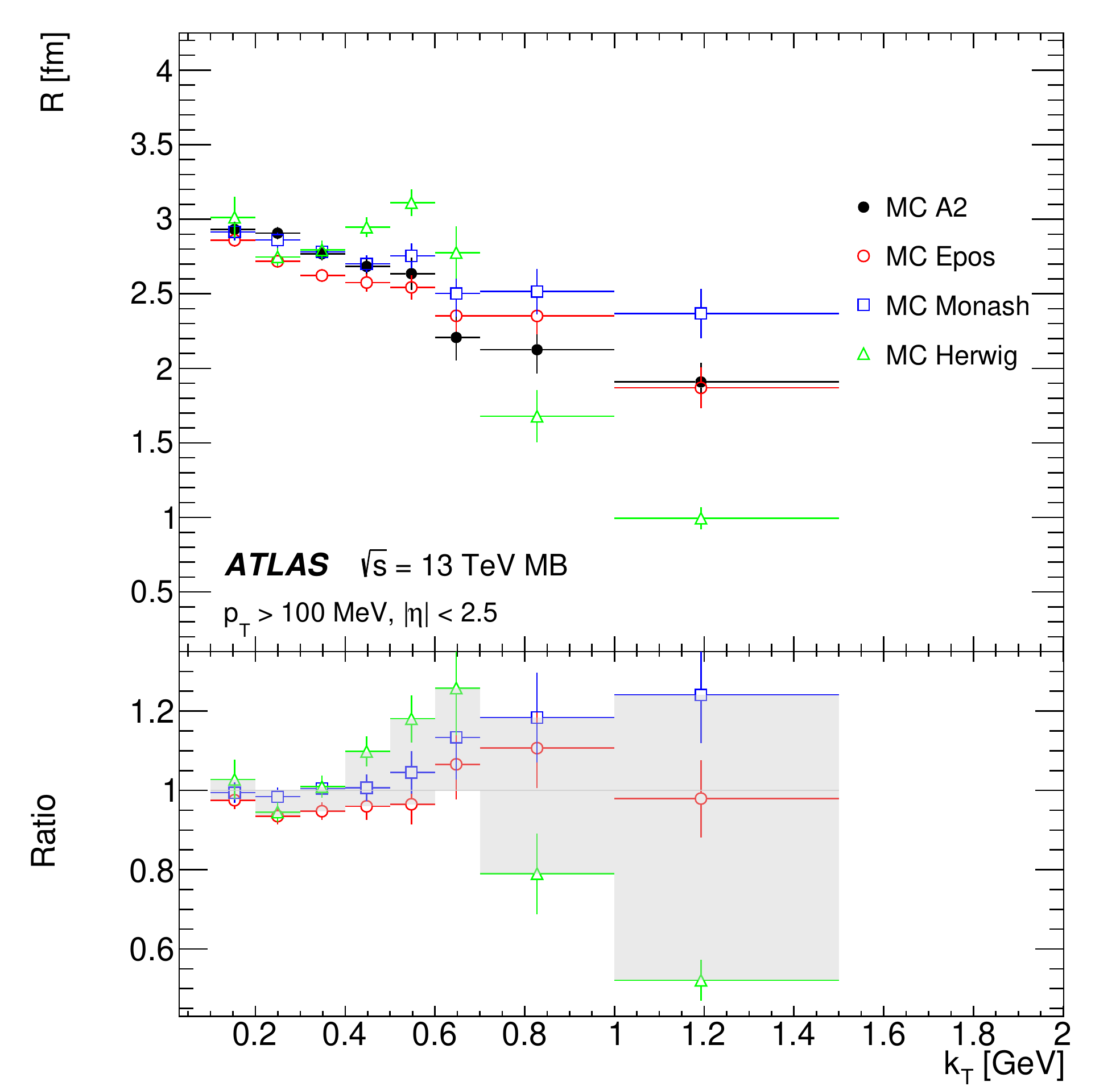}}
(c)
\\
\end{minipage}
\hfill
\begin{minipage}[h]{0.50\linewidth}
\center{\includegraphics[width=1\linewidth]{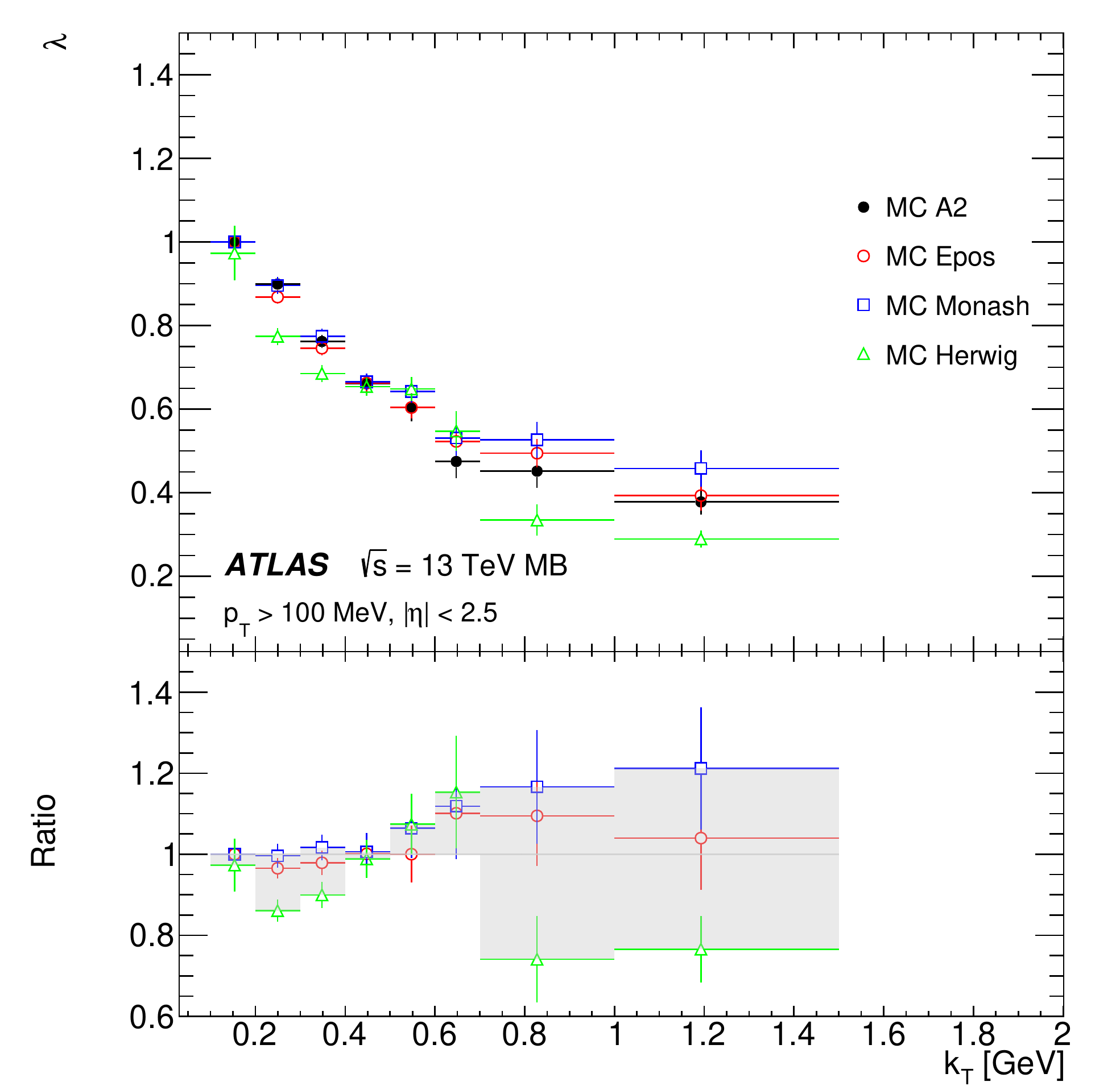}}
(d)
\\
\end{minipage}
\caption{
The  Bose--Einstein correlation (BEC) parameters
\(\lambda\) and \(R\)
as a function of \(n_{\mathrm{ch}}\) and \(k_{\mathrm{T}}\)
using different MC generators in the calculation of \(R_2 (Q)\).
(a) \(R\) versus \(n_{\mathrm{ch}}\) for MB events,
(b) \(R\) versus \(n_{\mathrm{ch}}\) for HMT events,
(c) \(R\) versus \(k_{\mathrm{T}}\) for MB events
and
(d) \( \lambda \) versus \(k_{\mathrm{T}}\) for MB events.
The uncertainties shown are statistical.
The lower panel of each plot shows the ratio of the BEC parameters obtained using
\textsc{EPOS LHC} (red circles),
\textsc{Pythia\,8} Monash (blue squares)
and
\textsc{Herwig++ UE-EE-5} (green triangles)
compared with the parameters obtained using
\textsc{Pythia\,8 A2}.
The grey band in the lower panels
is the MC systematic uncertainty, obtained as explained in the text.
}
\label{fig:Fig_MCcomp_01}
\end{figure*}
 
The leading source of uncertainty in \(\lambda\) and \(R\) is the choice of MC generator
used in the calculation of \(R_2(Q)\).
Figure~\ref{fig:Fig_MCcomp_01}
shows a selection of plots of the BEC parameters as a function of
\(n_{\mathrm{ch}}\) or \(k_{\mathrm{T}}\),
for different choices of MC generator.
The values of the parameters \(\lambda\) and \(R\) obtained using
\textsc{EPOS LHC},
\textsc{Pythia\,8} Monash
or
\textsc{Herwig++ UE-EE-5}
are compared with the parameter values obtained using
\textsc{Pythia\,8 A2}
in terms of the ratios shown in the lower panel of each plot
in Figure~\ref{fig:Fig_MCcomp_01}.
 
As can be seen in this figure, there are some regions of \(n_{\mathrm{ch}}\) and \(k_{\mathrm{T}}\)
where there are pronounced and significant differences between the results obtained using the different generators.
In order to assign the systematic uncertainty due to the MC modelling,
a conservative approach is adopted:
for each generator the weighted deviation is evaluated with respect to \textsc{Pythia\,8 A2},
where the weight is inversely proportional to the statistical uncertainty for that generator.
The systematic uncertainty is taken as the largest (in magnitude) of these weighted deviations,
treating positive and negative deviations separately, which leads to
asymmetric uncertainties. This procedure is followed for MB and HMT events separately.
 

An important source of systematic uncertainty for the
determination of the  parameter \(\lambda\) arises from the Coulomb correction.
This  uncertainty  is estimated by varying the Gamow factor \(G(Q)\)  by
\(\pm 0.15\),
which is the magnitude of
the effect on UCP or LCP pairs at \(Q = 0.02\)~\GeV.
This is the standard variation for a conservative estimate of the
systematic uncertainty in the Coulomb correction \cite{Aad:2015sja,cms-bec1st},
and accounts also for effects that include the extended size of the emission source,
as discussed in   Ref.~\cite{becrev-weiner}.
The resulting systematic  uncertainties for the inclusive fit
of the parameter \(\lambda\) are the same for both triggers
and are equal to
\(1.8\)\%.
For the parameter \(R\) the resulting systematic uncertainties
are small compared to the cumulative systematic uncertainties,
and equal to \(0.1\)\% and  \(0.3\)\% for MB and HMT events, respectively.
 
The effect of the upper limit of the fit range in \(Q\) is estimated by changing
it from the nominal \(2\)~\GeV\  by  \(\pm 0.1\)~\GeV,
which corresponds to a total variation of
about five times
the nominal  \(Q\)-resolution for \(Q \approx 2\)~\GeV.
This gives an estimate of the residual uncertainty due to
long-range correlations  accounted for by \(\varepsilon\), the parameter in
the linear term of Eq.~(\ref{eq_3}).
Other effects contributing to the systematic uncertainties
are the lowest value of \(Q\) for the fit,
the bin size, and  the exclusion of  \(Q\) regions
where resonances are not modelled perfectly, as discussed in
Section~\ref{twoparticle}.
These uncertainties are estimated by varying the lowest \(Q\) value in the fit by
\(\pm 0.01\)~\GeV, by changing the bin size by \(\pm 0.01\)~\GeV,
and by broadening the excluded intervals by
\(0.02\)~\GeV\  on both sides.

\begin{table*}[t!]
\centering
\caption{
Systematic uncertainties  (in percent)   in  the correlation strength, \(\lambda\),
and source radius, \(R\), for the exponential fit of the two-particle   double-ratio correlation functions, \(R_2 (Q)\),
for \(p_{\mathrm{T}} > 100\)~\MeV\ at \(\sqrt{s}= 13\)~\TeV\  for the MB and HMT events.
The choice of MC generator gives rise to asymmetric uncertainties, denoted by $\uparrow$ and $\downarrow$.
This asymmetry propagates through to the cumulative uncertainty.
The columns under ‘Uncertainty range’ show the range of systematic uncertainty from the fits
in the various \(n_{\mathrm{ch}}\) intervals.
}
\label{tab:systr2}
\medskip
\begin{tabular}{lr@{}c@{}lr@{}c@{}lcrrr@{}c@{}lr@{}c@{}lrr}
\hline
\hline
 
&
\multicolumn{8}{c}{
Uncertainties  for MB events
}
&
&
\multicolumn{8}{c}{
Uncertainties for HMT events
}
\\

&
\multicolumn{8}{c}{
[\%]
}
&
&
\multicolumn{8}{c}{
[\%]
}
\\
 
\cline{2-18}
\multicolumn{1}{c}{Sources of systematic}
&
\multicolumn{3}{c}{
\(\lambda\)
}
&
\multicolumn{3}{c}{
\(R\)
}
&
\(\lambda\)
&
\(R\)
&
&
\multicolumn{3}{c}{
\(\lambda\)
}
&
\multicolumn{3}{c}{
\(R\)
}
&
\(\lambda\)
&
\(R\)
\\
 
\cline{2-18}
\multicolumn{1}{c}{uncertainty}
&
\multicolumn{6}{c}{
Uncertainty range
}
&
\multicolumn{2}{c}{
Inclusive
}
&
&
\multicolumn{6}{c}{
Uncertainty range
}
&
\multicolumn{2}{c}{
Inclusive
}
\\
\hline
 
Track splitting and merging
&
\multicolumn{8}{c}{negligible}
&
&
\multicolumn{8}{c}{negligible}
\\
Track reconstr.\ efficiency
&0		&--&0.4		&0.1	&--&0.4		&0.3	&0.1	&&0.1	&--&0.2		&0.01	&--&0.1		&0.2	&0.01 \\
Unfolding matrix	 			&0		&--&0.3		&0		&--&0.1		&0.01	&0.0	&&0	&--&0.4		&0		&--&0.1		&0.01	&0.03 	\\
MC generators ($\uparrow$)	&0		&--&28		&0		&--&21		&2.7	&4.9	&&0	&--&7.1		&4.4	&--&6.6		&5.2	&5.6  \\
MC generators ($\downarrow$)	&0		&--&12		&0		&--&29		&5.4	&5.8	&&1.4	&--&9.9		&0		&--&11		&1.6	&2.3  \\
Coulomb  correction 			&1.3	&--&2.0		&0.01	&--&0.6		&1.8	&0.1	&&1.7	&--&1.9		&0.2	&--&0.4		&1.8	&0.3 \\
Fitted range  of \(Q\) 			&0		&--&0.5		&0.02	&--&0.9		&0.2	&0.3	&&0	&--&0.2		&0		&--&0.2		&0.02	&0.03 \\
Starting value  of \(Q\)			&0		&--&1.9		&0.01	&--&1.1		&0.3	&0.2	&&0.5	&--&1.4		&0.3	&--&0.7		&0.7	&0.4\\
Bin size         					&0		&--&2.4		&0.1	&--&1.5		&0.8	&0.4	&&1.0	&--&1.7		&0.4	&--&0.9		&1.3	&0.6 \\
Exclusion intervals 				&0		&--&1.1		&0.3	&--&0.8		&0.1	&0.3	&&0.4	&--&0.6		&0.3	&--&0.6		&0.5	&0.5 \\
\hline
\hline
Cumulative uncertainty
($\uparrow$)					&1.3	&--&28		&1.2	&--&21		&3.9	&5.0	&&2.9	&--&7.6		&4.6&--&6.7	&5.6	&5.7\\
\hline
Cumulative uncertainty
($\downarrow$)				&2.4	&--&12		&1.2	&--&29		&5.9	&5.8	&&2.8	&--&10		&1.1&--&11	&2.6	&2.5\\
\hline
\hline
\end{tabular}
\end{table*}
 
 
The systematic uncertainties for the inclusive study  are summarized in  Table~\ref{tab:systr2}.
The systematic uncertainties are combined by adding them in quadrature, and
the resulting values are given in the bottom row of the table.
The cumulative systematic  uncertainties  are
\(\mbox{}^{+ 3.9}_{- 5.9}\)\%
and
\(\mbox{}^{+ 5.6}_{- 2.6}\)\%
for  the inclusive determination of \(\lambda\),
and
\(\mbox{}^{+ 5.0}_{- 5.8}\)\%
and
\(\mbox{}^{+ 5.7}_{- 2.5}\)\%
for  \(R\), for the MB and HMT events,  respectively.
 
The same sources  of uncertainty are considered for the single- and double-differential
measurements in  \(n_{\mathrm{ch}}\) and \(k_{\mathrm{T}}\).
The results for the \(n_{\mathrm{ch}}\) distribution are also shown in Table~\ref{tab:systr2}.
The pattern of uncertainties is similar to the inclusive case, but shows a wider spread.
Away from the BEC region, the MC description of \(C_2^{data}\) improves as the multiplicity increases,
and this is reflected in Table~\ref{tab:systr2} by smaller systematic uncertainties for the HMT results.
This is also true for the \(k_{\mathrm{T}}\) distribution and the double-differential distribution.

\section{Results}
\label{results}
The BEC parameters, \( \lambda \) and \( R \), are extracted from
a fit to the \(R_2 (Q)\) distribution in intervals
of \( n_{\mathrm{ch}}\) and \(k_{\mathrm{T}}\).
In addition, there are two possible selections on the \(p_{\mathrm{T}}\) threshold of
the final-state particles.
In order to give a reasonably complete picture of all of this information,
the parameters describing the dependencies of \( \lambda \) and \( R \) on variables of interest
are themselves fitted to simple functions (usually exponentials),
and the parameters of these fits are presented.
 
In Section~\ref{twoparticle} the procedure used to fit the \(R_2 (Q)\) distribution is described.
The dependence on multiplicity is presented in
Section~\ref{sect:multiplicity}.
The comparison with previous ATLAS results at \(7\)~\TeV\ is discussed in
Section~\ref{sect:comparison}.
The
behaviour with \(k_{\mathrm{T}}\) is presented in  Section~\ref{sect:transverse}.
The `double-differential' dependence of the BEC parameters
on multiplicity and \(k_{\mathrm{T}}\) at \(13\)~\TeV\ is explored in
Section~\ref{sect:2D_mch_kT},
by looking at the dependence on one variable for different intervals  of the other.
Finally, a comparison with CMS results at \(13\)~\TeV\ is shown in Section~\ref{sect:CMScomparison}.

\subsection{Procedure for BEC parameter extraction from \(R_2 (Q)\)}
\label{twoparticle}
 
\begin{figure*}[t!]
\begin{minipage}[h]{0.50\linewidth}
\center{\includegraphics[width=1\linewidth]{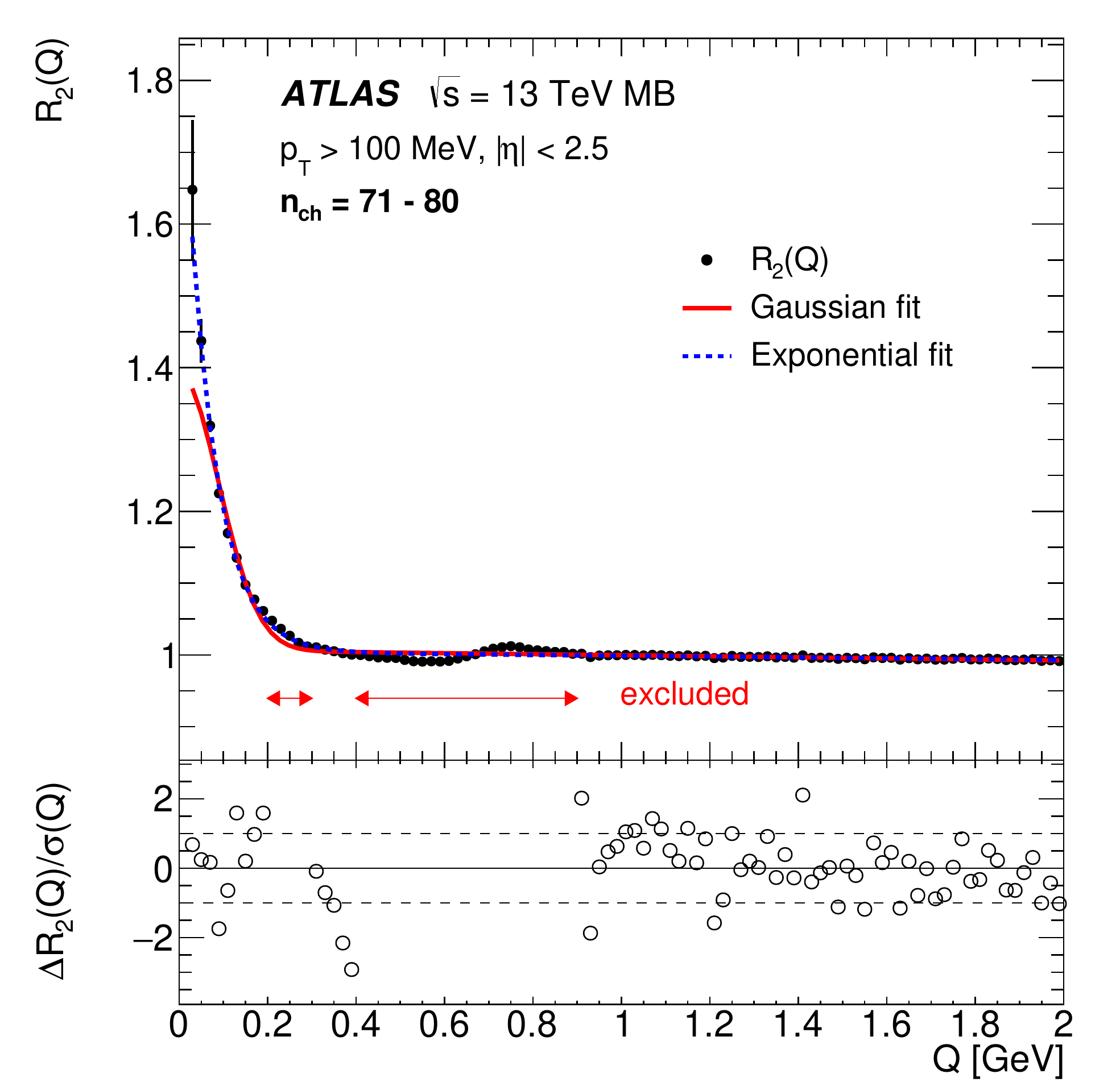}}
(a) \\
\end{minipage}
\hfill
\begin{minipage}[h]{0.50\linewidth}
\center{\includegraphics[width=1\linewidth]{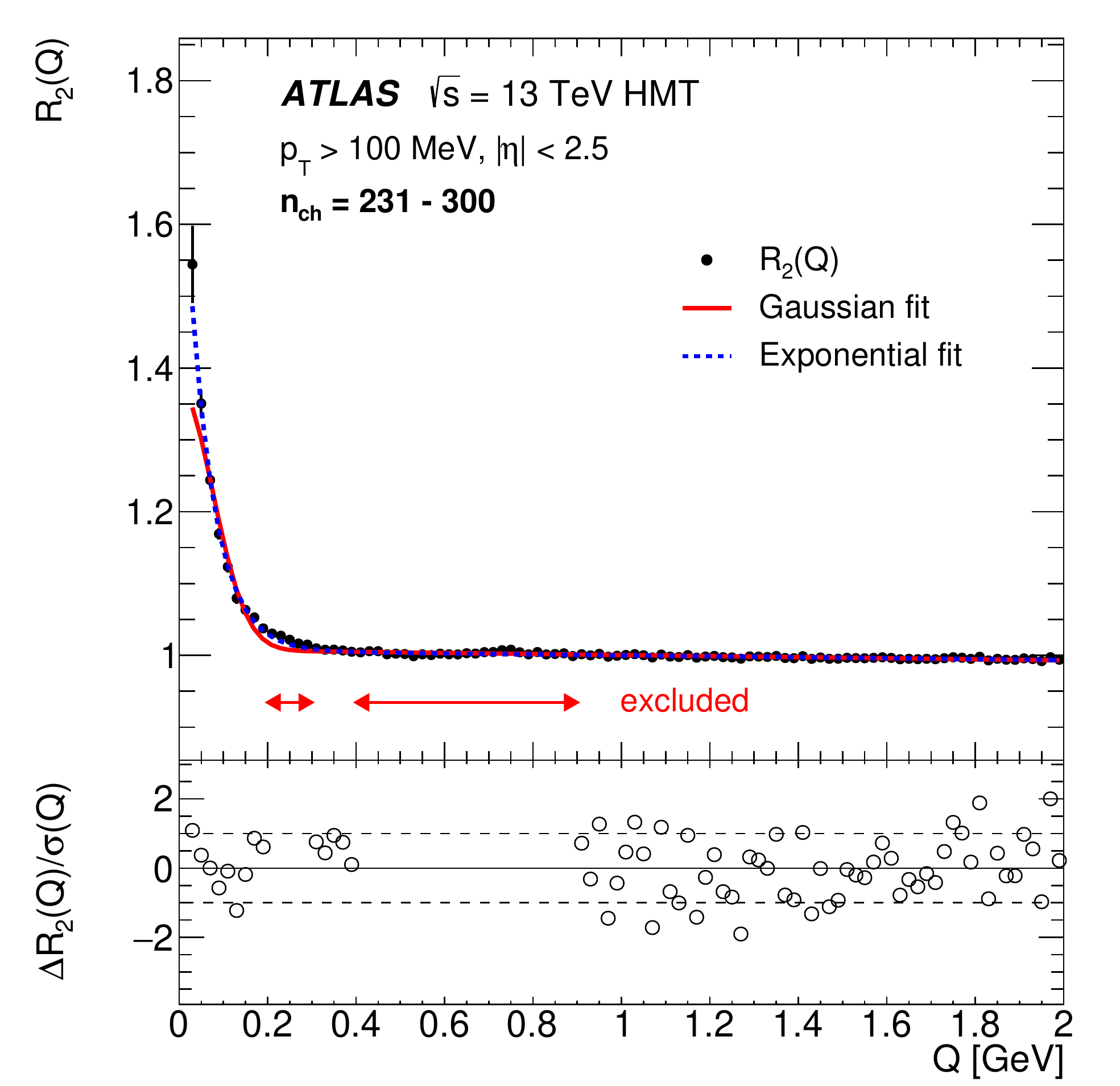}}
(b) \\
\end{minipage}
\caption{
Top panel:
The two-particle double-ratio correlation function,  \(R_2 (Q)\),
for   \(pp\) collisions  for track \(p_{\mathrm{T}} >100\)~\MeV\
at \(\sqrt{s}=13\)~\TeV\
in  the multiplicity intervals
(a)  \(71 \le n_{\mathrm{ch}} < 80\)  for the minimum-bias (MB) events,
and  (b)  \(231 \le n_{\mathrm{ch}} < 300\)   for the  high-multiplicity track (HMT) events.
The blue dashed
and red solid lines
show the results of the exponential and Gaussian fits, respectively.
The region excluded from the fits is shown.
The statistical uncertainty and the systematic uncertainty
for imperfections in  the data reconstruction procedure are added in quadrature.
Bottom panel:
The difference between the experimental  \( R_2 (Q) \)
function  and the
result of the exponential fit  normalized to the experimental uncertainty,
\(\Delta R_2 (Q)/\sigma(Q)\), is presented.
}
\label{Fig_2_1b}
\end{figure*}

Figure~\ref{Fig_2_1b} shows examples of
the \(R_2(Q)\) correlation functions
measured at  \(\sqrt{s} = 13\)~\TeV\
for  two different multiplicity intervals,
and the fits to the Gaussian and exponential forms.
In  the lower panels of Figure~\ref{Fig_2_1b}
the differences between the measured \(R_2 (Q)\)
and  the fit to the exponential form,
normalized to the experimental uncertainty \(\sigma (Q)\),
\(\Delta R_2 (Q)/\sigma(Q)=
[R_2^{\mathrm{data}} (Q) - R_2^{\mathrm{fit}} (Q)]/\sigma(Q)\),
are shown  for both multiplicity intervals.
It is evident from  Figure~\ref{Fig_2_1b} that the Gaussian function does not give a good
description of  the low-\(Q\), BEC-sensitive region, \(Q \le 0.1\)~\GeV,
while the exponential function matches the data well in this region.
Therefore, the quantity \(\Delta R_2 (Q)/\sigma(Q)\)
is not shown for the Gaussian case,
and the Gaussian fit is not considered further.
 
Figure~\ref{Fig_2_1b} shows structure in \(R_2 (Q)\), compared with the
smooth fitting functions, in the regions
\(Q\approx 0.25\)~\GeV\ and \(0.4 \le Q \le 0.9\)~\GeV.
If the MC modelling
of resonance production, particularly the \(\eta\), \(\rho\) and
\(\omega\), does not agree very well with data, then this \(Q\) region will be affected,
as appears to be the case.
Accordingly, the regions \(0.2 \le Q \le 0.3\)~\GeV\ and
\(0.4 \le Q \le 0.9\)~\GeV\  are excluded from the fits  to the \(R_2(Q)\)
correlation functions for the MB and HMT events.
For the MB events, similar arguments lead to the additional
exclusion of the region \(1.0 \le Q \le 1.16\)~\GeV, as it appears to be influenced
by the reflection of the \(f_0(980)\) and \(f_2(1270)\) mesons  in the case of
\(100 \le k_{\mathrm{T}} \le 200\)~\MeV\  and in the multiplicity region
\(n_{\mathrm{ch}} \le 40\).
These excluded regions lie outside the BEC-sensitive region and the main effect
of the exclusions is to markedly reduce the \( \chi^2 \) without significantly changing
the fitted BEC parameters.
 
The results for the BEC parameters obtained using  the exponential  fit with the  UCP reference sample
for the MB and  HMT samples,  summed over all multiplicities  for \(p_{\mathrm{T}} > 100\)~\MeV,
are
\begin{equation}\nonumber
\lambda = 1.03^{+0.04}_{-0.06}\ , \  \
R = 3.32^{+0.17}_{-0.19}~{\mathrm{[fm]}} \  \
{\mathrm{for}} \  \
2 \le n_{\mathrm{ch}}(p_{\mathrm{T}}>100~\MeV) \le 250 \ \
{\mathrm{(MB)}}
\end{equation}
and
\begin{equation}\nonumber
\lambda = 1.00^{+0.09}_{-0.07}\ , \ \
R = 3.76^{+0.22}_{-0.14}~{\mathrm{[fm]}} \  \
{\mathrm{for}} \  \
100 < n_{\mathrm{ch}}(p_{\mathrm{T}}>100~\MeV) \le 300 \ \
{\mathrm{(HMT)}} .
\end{equation}
The results for the BEC parameters for tracks with \(p_{\mathrm{T}} > 500\)~\MeV\ are
\begin{equation}\nonumber
\lambda = 0.61^{+0.05}_{-0.06}\ , \  \
R = 2.38^{+0.20}_{-0.25}~{\mathrm{[fm]}}  \ \
{\mathrm{for}} \  \
2 \le n_{\mathrm{ch}}(p_{\mathrm{T}}>500~\MeV) \le 115 \ \
{\mathrm{(MB)}}
\end{equation}
and
\begin{equation}\nonumber
\lambda = 0.66^{+0.08}_{-0.08}\ , \ \
R = 2.82^{+0.39}_{-0.40}~{\mathrm{[fm]}} \ \
{\mathrm{for}} \  \
46 < n_{\mathrm{ch}}(p_{\mathrm{T}}>500~\MeV) \le 136 \ \
{\mathrm{(HMT)}} .
\end{equation}
In both cases the uncertainties are the statistical and systematic  uncertainties
summed in quadrature.
The overall quality of the fits is not good, as the high number of events in
the inclusive distributions exposes the imperfections in the models used.
As noted in Section \ref{analysis},
the fits would be better using a L\'{e}vy distribution with an additional free parameter.
For all fits throughout Section~\ref{results},  the statistical uncertainty is multiplied by the factor
\(\sqrt{\chi^2/ndf}\) if this factor is  significantly  greater than 1 \cite{Patrignani:2016xqp}.
This applies in relatively few cases,
see  e.g.\ 4 of the 8 entries in  Table~\ref{tab:parameters_kT}.

\subsection{Multiplicity dependence}
\label{sect:multiplicity}
 
The dependence of \(R\) and \(\lambda\) on \( n_{\mathrm{ch}}\) is studied
in multiplicity intervals that are chosen to be well populated, and
comparable to those used by other LHC experiments
\cite{cms-bec1st,cms-JHEP,Sirunyan:2019umv,alice-bec900,alice-bec3d}.
 
To compare the BEC parameters from different \(p_{\mathrm{T}}\)-thresholds,
a scaled charged-particle multiplicity is introduced as follows.
At \(13\)~\TeV,
the average charged-particle multiplicity
per unit of pseudorapidity
in  the region \(|\eta|< 0.2\),
\( \langle \left. \mathrm{d}n_{\mathrm{ch}} / \mathrm{d}\eta \right|_{|\eta|<0.2} \rangle \),
is \(6.5\) for particles with \(p_{\mathrm{T}} > 100\)~\MeV\
\cite{Aaboud:2016itf}
and \(3.24\) for particles with \(p_{\mathrm{T}} > 500\)~\MeV\
\cite{Aad:2016mok}.
A scaled charged-particle multiplicity for \(|\eta|<2.5\),  i.~e.\ \(5\)
units of pseudorapidity is then defined as:
\begin{equation}\nonumber
m_{\mathrm{ch}} \equiv n_{\mathrm{ch}}/
( 5 \cdot \langle \left. \mathrm{d}n_{\mathrm{ch}} / \mathrm{d}\eta \right|_{|\eta|<0.2} \rangle ) .
\end{equation}

\begin{figure*}[t!]
\begin{minipage}[h]{0.50\linewidth}
\center{\includegraphics[width=1\linewidth]{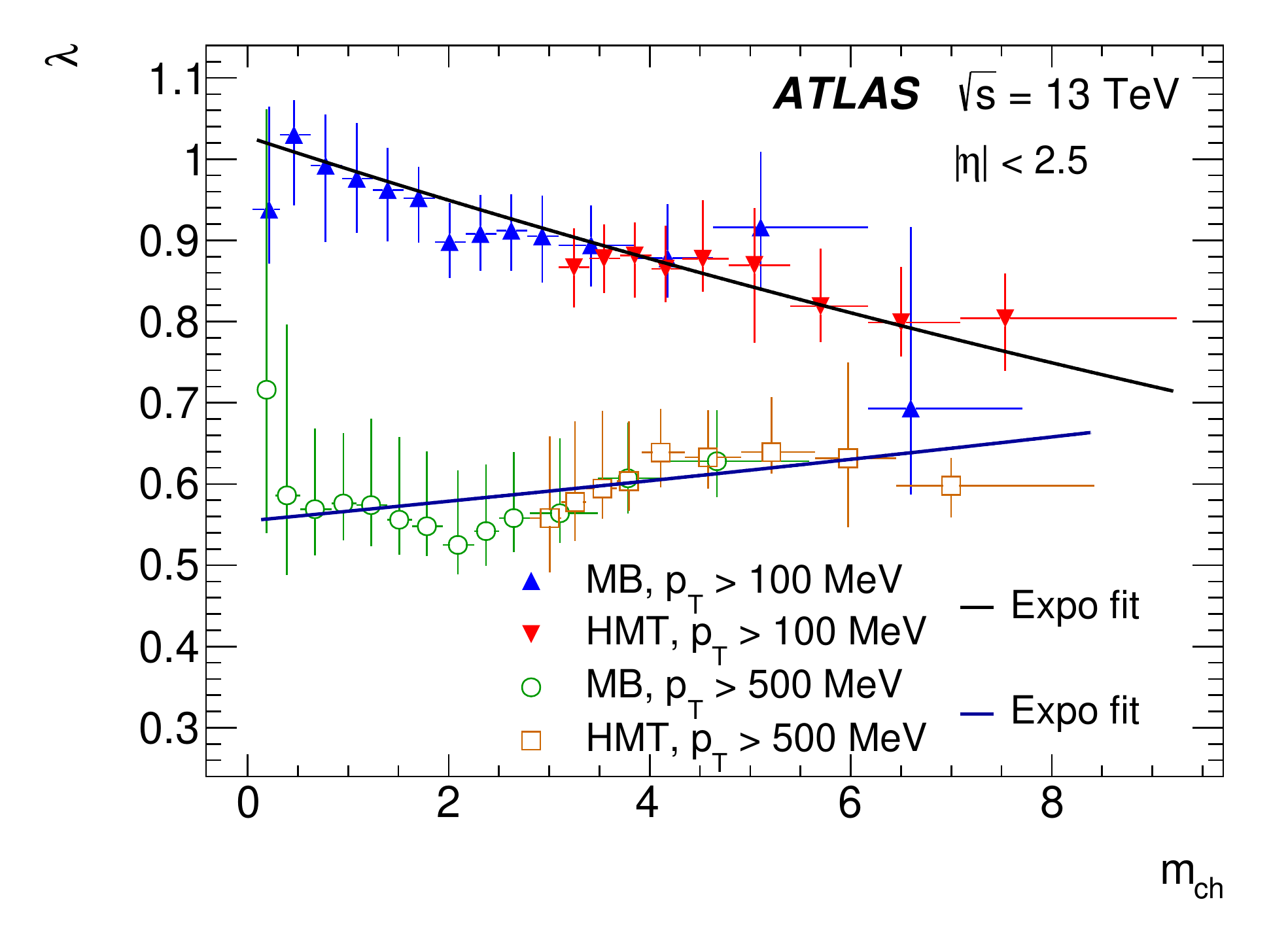}}
(a) \\
\end{minipage}
\hfill
\begin{minipage}[h]{0.50\linewidth}
\center{\includegraphics[width=1\linewidth]{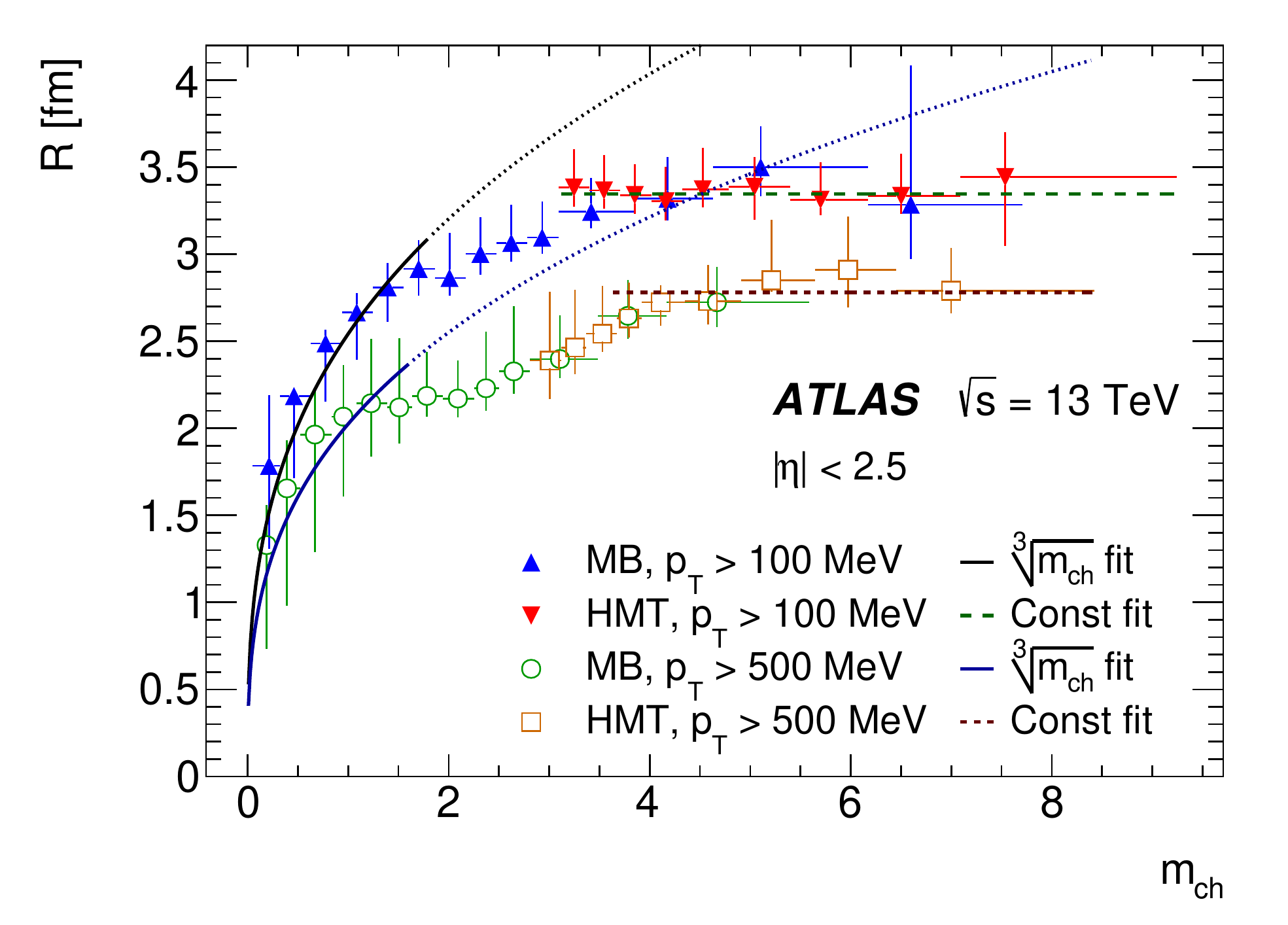}}
(b) \\
\end{minipage}
\vfill
\hfill
\begin{minipage}[h]{0.50\linewidth}
\center{\includegraphics[width=1\linewidth]{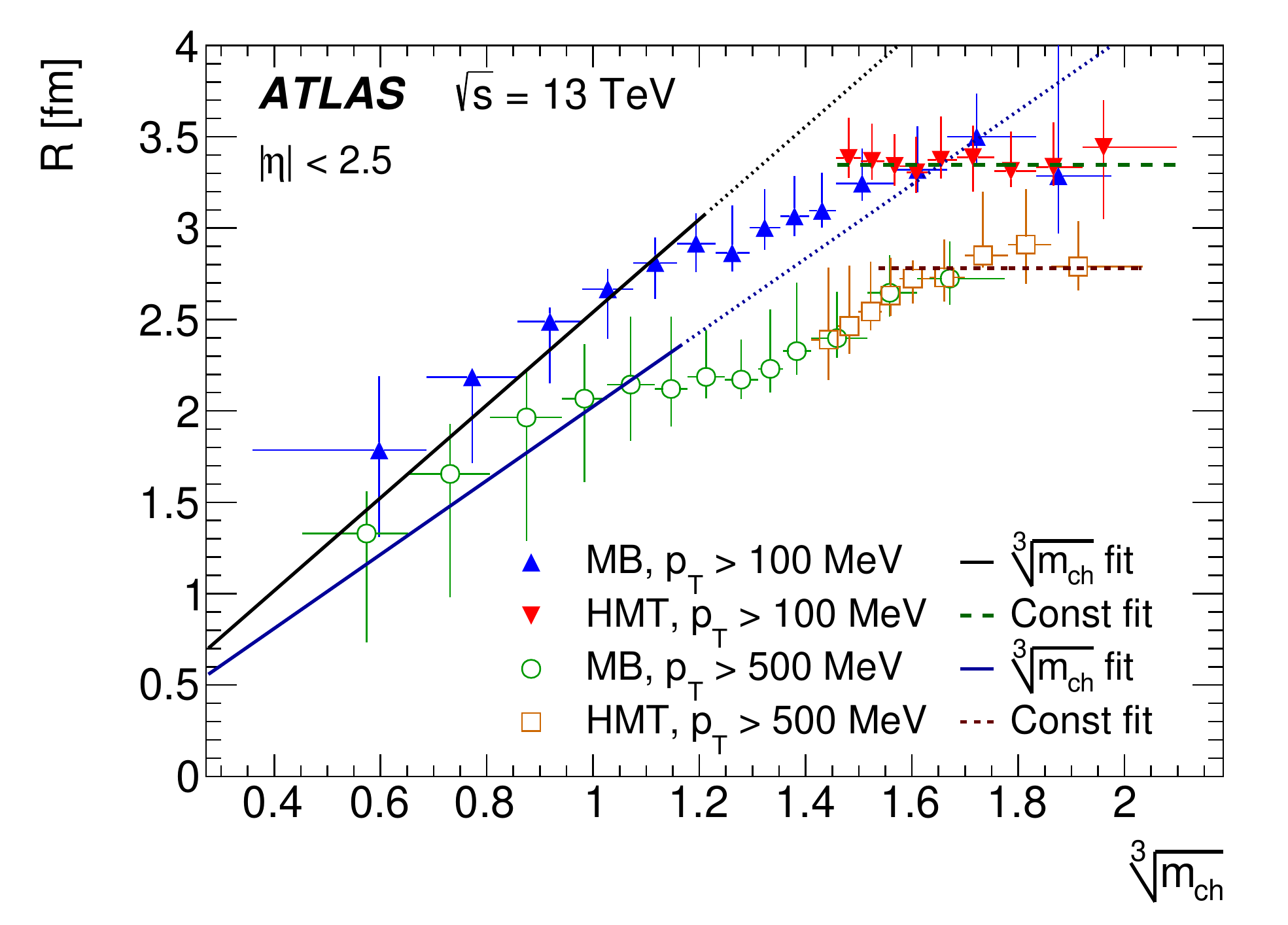}}
(c) \\
\end{minipage}
\caption{
(a) The dependence of the
correlation strength,
\(\lambda  (m_{\mathrm{ch}})\), on
rescaled multiplicity,  \(m_{\mathrm{ch}}\),  obtained from the exponential fit of  the
\(R_2 (Q)\)  correlation functions  for tracks with
\(p_{\mathrm{T}} > 100\)~\MeV\  and \(p_{\mathrm{T}} > 500\)~\MeV\
at \(\sqrt{s} = 13\)~\TeV\
for  the   minimum-bias  (MB) and  high multiplicity track (HMT) data.
(b) The dependence of the   source radius, \(R (m_{\mathrm{ch}})\),  on
\(m_{\mathrm{ch}}\).
(c) The dependence of the   \(R (m_{\mathrm{ch}})\)  on
\(\sqrt[3]{m_{\mathrm{ch}}}\).
The uncertainties  represent the sum in quadrature of the
statistical  and
asymmetric
systematic contributions.
The black and blue solid curves
in
(a)
represent the exponential fit  of
\(\lambda  (m_{\mathrm{ch}})\)
for
\(p_{\mathrm{T}} >100\)~\MeV\
and  \(p_{\mathrm{T}} >500\)~\MeV, respectively.
The black
and blue solid curves
in
(b)
represent  the fit of
\(R  (m_{\mathrm{ch}})\)
for \(\sqrt[3]{m_{\mathrm{ch}}} < 1.2\)
for \(p_{\mathrm{T}} >100\)~\MeV\ and
\(p_{\mathrm{T}} >500\)~\MeV, respectively.
The black and blue dotted curves are
extensions
of
the black and blue solid curves
beyond
\(\sqrt[3]{m_{\mathrm{ch}}} > 1.2\), respectively.
The black and brown dashed curves represent the
saturation value of
\(R  (m_{\mathrm{ch}})\)
for \(\sqrt[3]{m_{\mathrm{ch}}} > 1.45\)
with \(p_{\mathrm{T}} >100\)~\MeV\
and for \(\sqrt[3]{m_{\mathrm{ch}}} > 1.6\)
with \(p_{\mathrm{T}} >500\)~\MeV, respectively.
}
\label{Fig_pTcuts_com_nch}
\end{figure*}

Using the exponential fit, the extracted BEC parameters
\(\lambda (m_{\mathrm{ch}}) \) and \(R  (m_{\mathrm{ch}}) \)
at \(13\)~\TeV\ are shown as a function of scaled multiplicity in
Figure~\ref{Fig_pTcuts_com_nch}
for \(p_{\mathrm{T}} > 100\) and \(500\)~\MeV.
The MB and HMT data agree well where they overlap.
 
For \(p_{\mathrm{T}} > 100 \)~\MeV\  the \(\lambda\) parameter decreases with increasing
\(m_{\mathrm{ch}}\),
and the dependence is well described by the exponential function:
\begin{equation}
\label{eq_Lnch}
\lambda (m_{\mathrm{ch}}) =  \gamma\, {\mathrm{e}}^{- \delta m_{\mathrm{ch}}}.
\end{equation}
The  \(R\) parameter  increases with  multiplicity  up to about
\(m_{\mathrm{ch}}   \simeq 3.08 \).
For
\(m_{\mathrm{ch}}  \le 1.9\),
the
\(m_{\mathrm{ch}}\)
dependence of \(R\)   is well described by:
\begin{equation}
\label{eq_n3}
R (m_{\mathrm{ch}}) = \alpha \sqrt[3]{m_{\mathrm{ch}} }\, ,
\end{equation}
which is also the function used in heavy-ion studies  \cite{alice-bec3d,Lisa:2005dd}.
A linear increase of \(R\) with
\( \sqrt[3]{m_{\mathrm{ch}}} \)
follows naturally if the hadronization volume is proportional to the number of produced particles.
 
For higher multiplicities, the measured \(R\) parameter is observed to be independent
of multiplicity.
This saturation of the BEC \(R\) parameter at high multiplicity
is similar to that observed by ATLAS at \(\sqrt{s} = 7\)~\TeV\ \cite{Aad:2015sja}
and by CMS at \(\sqrt{s} = 13\)~\TeV\ \cite{Sirunyan:2019umv}.
The dependence  of \(R\)
for the region
\(m_{\mathrm{ch}} >  3.08\)
is then fitted  simply by a  constant:
\begin{equation}
\label{eq_Rconst}
R (m_{\mathrm{ch}}) = \beta\, .
\end{equation}
For the combined MB and HMT data at \(\sqrt{s} = 13\)~\TeV\ and
\(p_{\mathrm{T}} > 100 \)~\MeV\
the parameters of the
\(m_{\mathrm{ch}}\)
dependence  of \(R\) and \(\lambda\) are given in the upper part of
Table~\ref{tab:parameters_mean-nch},
with statistical and
asymmetric
systematic uncertainties
summed
in quadrature.
 
\begin{table*}[t!]
\centering
\caption{
The  results of  the  fits to the dependencies of the
correlation strength, \(\lambda\),   and  source radius, \(R\),
on  the  average  rescaled charged-particle multiplicity,    \(m_{\mathrm{ch}} \),
for \(|\eta| < 2.5\) and both \(p_{\mathrm{T}} > 100\)~\MeV\
and \(p_{\mathrm{T}} > 500\)~\MeV\ at \(\sqrt{s} = 13\)~\TeV\
for the minimum-bias (MB) and the  high-multiplicity  track (HMT)  events.
The parameters \(\gamma\) and \(\delta\) resulting from a joint fit to the
MB and HMT data
are presented.
The total uncertainties are shown.
}
\label{tab:parameters_mean-nch}
\medskip
\begin{tabular}{ccccccr}
\hline
\hline
\(p_{\mathrm{T}}\)
& BEC
& Fit
& \( m_{\mathrm{ch}} \)
& MB 
& HMT 
& \(\chi^2/ndf\)
\\
Threshold
&
Parameter
&
Equation
& Region
& Events
& Events
&
\\
 
\hline
& & & & & & \\[-3mm]
 
\(> 100\)~[\MeV]
& \(\lambda( m_{\mathrm{ch}} )\)
&  (\ref{eq_Lnch})
&
&
\multicolumn{2}{c}{\( \mbox{~~~~~~~~~~~~}
\gamma = 1.027^{+0.043}_{-0.078} \) }
&
16/21
\\[1mm]

&
&
&
&
\multicolumn{2}{c}{\( \mbox{~~~~~~~~~~~~}
\delta = 0.039^{+0.050}_{-0.083} \) }
 
&
\\[2mm]

& \(R( m_{\mathrm{ch}} )\)
& (\ref{eq_n3})
& \(\leq 1.9\)
&
\(\alpha = 2.54^{+0.12}_{-0.22}\)~[fm]
 
& ---
&
3.7/5
\\[2mm]
 
& \(R( m_{\mathrm{ch}} )\)
& (\ref{eq_Rconst})
&  \( \geq 3.08\)
&
\multicolumn{2}{c}{\(\mbox{~~~~~~~~~~~~~}
\beta = 3.35^{+0.20}_{-0.09} \)~[fm]
}
&
4.7/12
\\[2mm]

\hline
& & & & & & \\[-3mm]

\(> 500\)~[\MeV]
& \(\lambda( m_{\mathrm{ch}} )\)
& (\ref{eq_Lnch})
&
&
\multicolumn{2}{c}{\(\mbox{~~~~~~~~~~~~}
\gamma = 0.555^{+0.124}_{-0.028} \) }
&
20/20
\\[1mm]
 
&
&
&
&
\multicolumn{2}{c}{\(\mbox{~~~~~~~~~~~~~}
\delta = - 0.021^{+0.022}_{-0.007} \) }
&
\\[1mm]

& \(R( m_{\mathrm{ch}} )\)
&  (\ref{eq_n3})
& \(\leq 1.9\)
&
\(\alpha = 2.02^{+0.29}_{-0.39}\)~[fm]
 
& ---
&
5.8/5
\\[2mm]
 
& \(R( m_{\mathrm{ch}} )\)
& (\ref{eq_Rconst})
&  \( \geq 4.17\)
&
\multicolumn{2}{c}{\(\mbox{~~~~~~~~~~~~}
\beta = 2.78^{+0.23}_{-0.09} \)~[fm]
}
&
8.2/7
\\[2mm]

\hline
\hline
 
\end{tabular}
\end{table*}


For \(p_{\mathrm{T}} > 500\)~\MeV\ the variation of the BEC parameters is fitted to the
same functional forms as used for \(p_{\mathrm{T}} > 100\)~\MeV\
(Eqs.~(\ref{eq_Lnch}), (\ref{eq_n3}), (\ref{eq_Rconst})),
and the results are
shown in the lower part of
Table~\ref{tab:parameters_mean-nch}.
 
From Figure~\ref{Fig_pTcuts_com_nch}(a) it can be seen that
\(\lambda (m_{\mathrm{ch}}) \)
shows only a weak dependence on multiplicity for \(p_{\mathrm{T}}>500\)~\MeV.
It is constant within measurement uncertainties, indicating a relatively high level of incoherence.
Figures~\ref{Fig_pTcuts_com_nch}(b) and \ref{Fig_pTcuts_com_nch}(c) show similar behaviour of
\(R (m_{\mathrm{ch}}) \) for both \(p_{\mathrm{T}}\) thresholds.
There is a linear rise
with \(\sqrt[3]{m_{\mathrm{ch}}}\) up to \(1.2\),
after which there is some
tapering off towards a plateau
for  \(\sqrt[3]{m_{\mathrm{ch}}} > 1.6\):
\begin{equation}\nonumber
R =
2.78^{+0.23}_{-0.09}
(\mathrm{total})~{\mathrm{[fm]}}
\end{equation}
for \(p_{\mathrm{T}}>500\)~\MeV,
and
for  \(\sqrt[3]{m_{\mathrm{ch}}} > 1.45\):
\begin{equation}\nonumber
R =
3.35^{+0.20}_{-0.09}
(\mathrm{total})~{\mathrm{[fm]}}
\end{equation}
for \(p_{\mathrm{T}}>100\)~\MeV.
 
\subsection{Comparison of  \(C_2 (Q)\) functions at 13 and 7 \TeV}
\label{sect:comparison}
 
\begin{figure*}[t!]
\begin{minipage}[h]{0.50\linewidth}
\center{\includegraphics[width=1\linewidth]{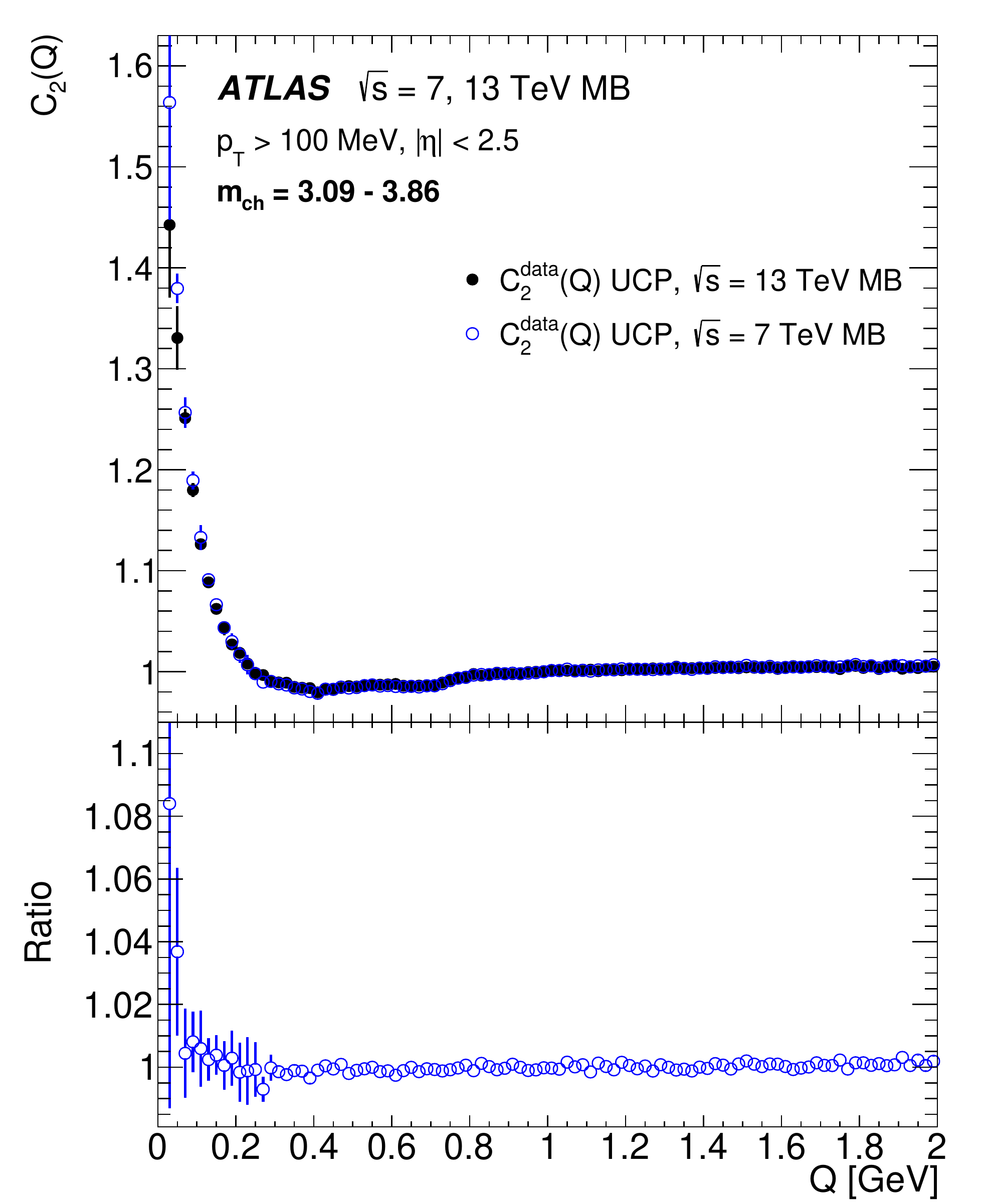}}
(a)
\\
\end{minipage}
\hfill
\begin{minipage}[h]{0.50\linewidth}
\center{\includegraphics[width=1\linewidth]{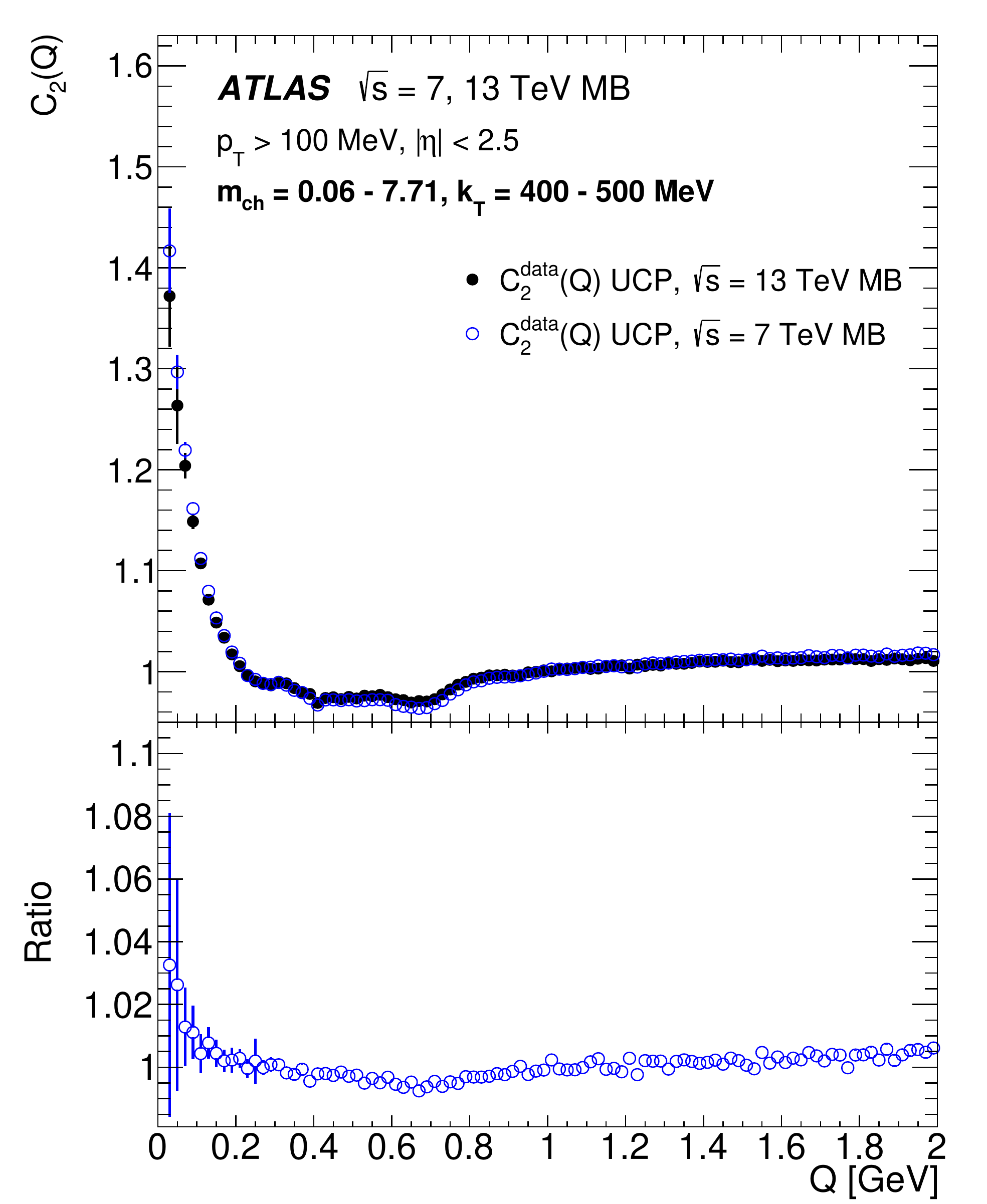}}
(b)
\\
\end{minipage}
\caption{
Comparison of  single-ratio two-particle  correlation functions,
using the unlike-charge particle (UCP) pair reference sample,
for minimum-bias (MB) events, showing
\(C_2^{\mathrm{data}} (Q)\)  (top panel)
at 13~\TeV\ (black circles)
and
7~\TeV\ (open blue circles),
and the ratio of
\(C_{\mathrm{2}}^{\mathrm{7~\TeV}} (Q)\)
to
\(C_{\mathrm{2}}^{\mathrm{13~\TeV}} (Q)\)
(bottom panel).
Comparison of   \(C_2^{\mathrm{data}} (Q)\) (a) for  representative
multiplicity region
\(3.09  <    m_{\mathrm{ch}}    \le 3.86 \)
and
(b)
for  representative
\(k_{\mathrm{T}}\)  region
\(400  <    k_{\mathrm{T}}    \le 500 \)~\MeV.
The statistical and systematic uncertainties, combined in quadrature, are presented.
The systematic uncertainties include
track efficiency,
Coulomb correction,
non-closure
and
multiplicity-unfolding uncertainties.
}
\label{fig:Fig_C2comp_01}
\end{figure*}
 
The BEC parameters \(\lambda\) and \(R\) were measured previously
by ATLAS at 0.9 and 7~\TeV\ \cite{Aad:2015sja}.
The data sample at 7~\TeV\ was large enough to explore high-multiplicity events, and a clear plateau in \(R\)
was observed for \(n_{\mathrm{ch}} \ge 50 \).
This plateau at high multiplicity is confirmed by the present analysis at 13~\TeV, as shown in
Figures~\ref{Fig_pTcuts_com_nch}(b) and \ref{Fig_pTcuts_com_nch}(c).
However, a direct comparison of the BEC parameters obtained at the two  centre-of-mass energies
is complicated by small but  significant differences between the Monte Carlo generators used to construct
\(C_2^{\mathrm{MC}} (Q)\), and hence \(R_2 (Q)\), in the two analyses (see Eqs.~(\ref{eq_c2}) and (\ref{eq_4})).
The BEC parameters obtained at \(7\)~\TeV\ are systematically lower than those at \(13\)~\TeV\
by about \(1.5\) standard
deviations of the estimated \(7\)~\TeV\ uncertainties.
For the previous \(7\)~\TeV\ analysis only limited MC modelling was available.
For this paper, several MC models and tunings are used to estimate the modelling uncertainty.
 
This  finding is supported by a comparison between the quantities \(C_2^{\mathrm{data}} (Q)\)
at the two  centre-of-mass energies.
Figures~\ref{fig:Fig_C2comp_01}(a) and \ref{fig:Fig_C2comp_01}(b)
show \(C_2^{\mathrm{data}} (Q)\) for MB events at 7 and 13~\TeV.
Figure~\ref{fig:Fig_C2comp_01}(a) shows a
representative
plot for a scaled-multiplicity interval
(\( 3.09 <  m_{\mathrm{ch}} \le 3.86 \)) and
Figure~\ref{fig:Fig_C2comp_01}(b)
shows a  representative
plot for a \( k_{\mathrm{T}} \) interval (\(400 < k_{\mathrm{T}} \le 500 \)~\MeV).
In Figure~\ref{fig:Fig_C2comp_01} the uncertainties shown are the statistical and
systematic uncertainties combined in quadrature.
The systematic uncertainties include the effects of the track efficiency, Coulomb correction, non-closure
and multiplicity unfolding.
The plots shown in Figures~\ref{fig:Fig_C2comp_01}(a) and \ref{fig:Fig_C2comp_01}(b) are representative of all
multiplicity and \(k_{\mathrm{T}} \) intervals.
The data from the two
centre-of-mass energies
are in agreement everywhere within the uncertainties,
and the ratio of 7~\TeV\ to 13~\TeV\ data is 1
to better than 1\% except in the very first one or two \(Q\) bins.


\subsection{Dependence on the pair average transverse momentum
}
\label{sect:transverse}
 
\begin{figure*}[t!]
\begin{minipage}[h]{0.50\linewidth}
\center{\includegraphics[width=1\linewidth]{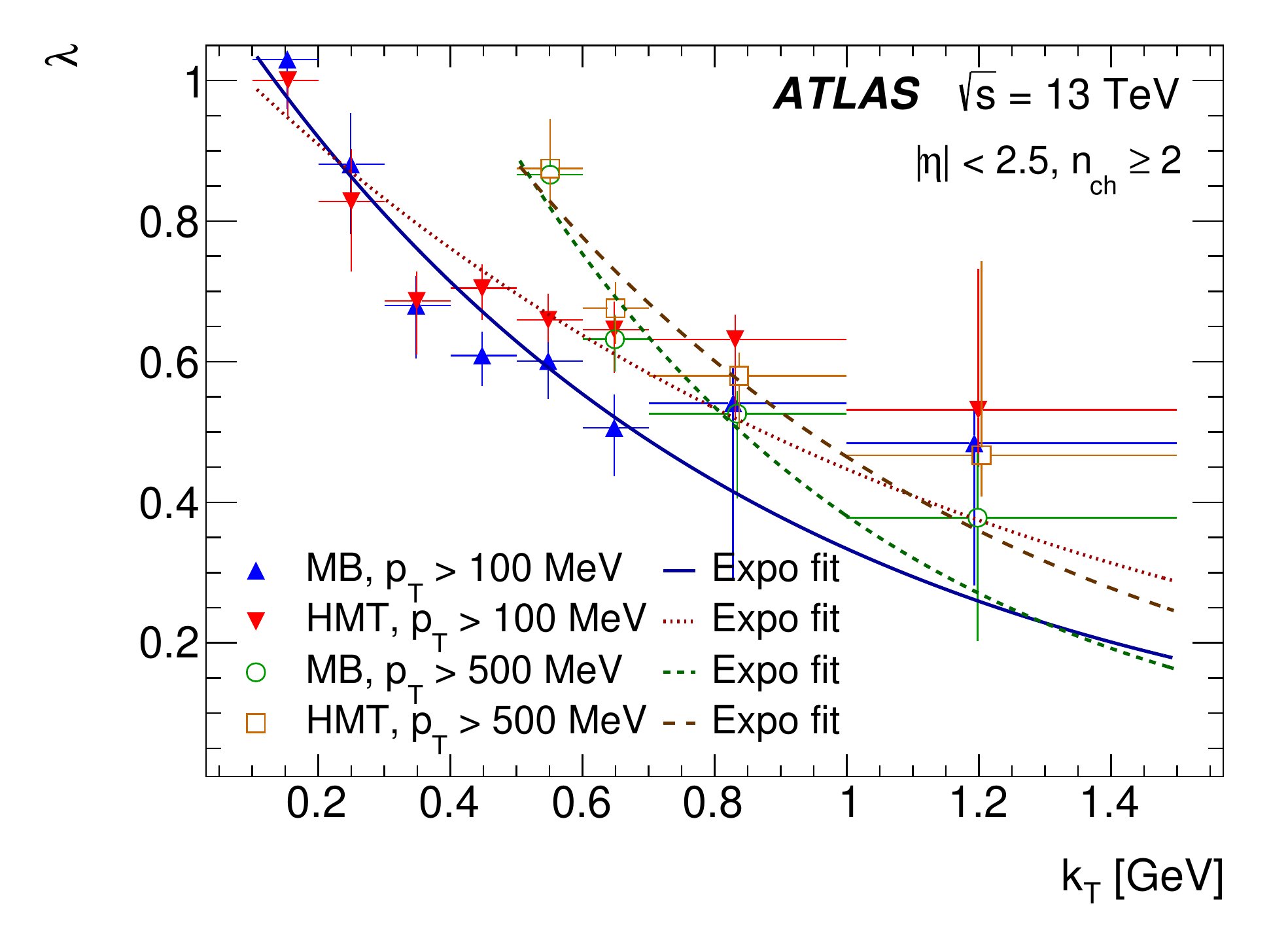}}
(a) \\
\end{minipage}
\hfill
\begin{minipage}[h]{0.50\linewidth}
\center{\includegraphics[width=1\linewidth]{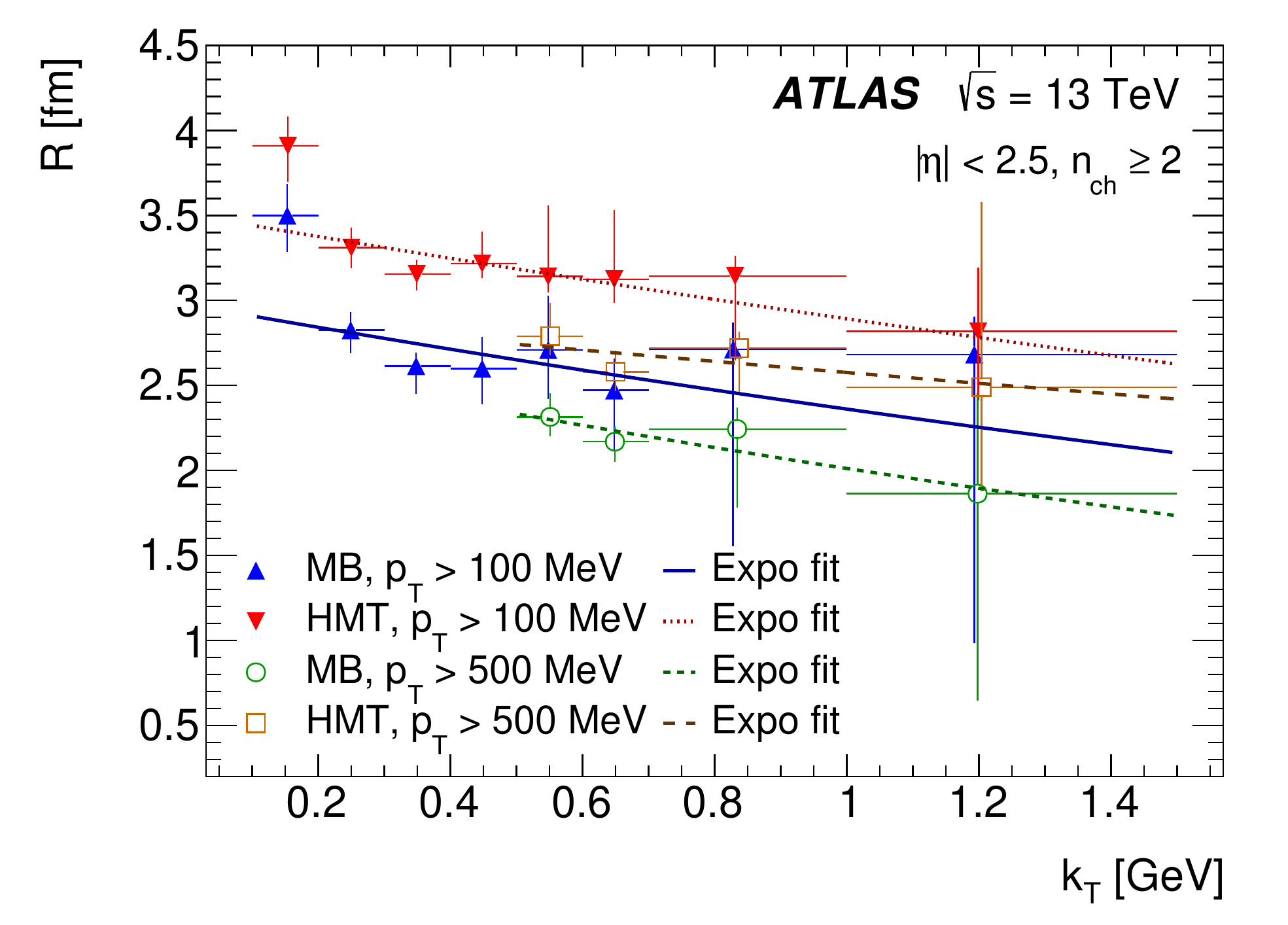}}
(b) \\
\end{minipage}
\caption{
The \(k_{\mathrm{T}}\) dependence of
(a)
the
correlation strength,
\(\lambda (k_{\mathrm{T}})\),
and
(b)
the
source radius, \(R (k_{\mathrm{T}})\),
obtained from the exponential
fit to  the \(R_2 (Q)\)
correlation functions
for events with
multiplicity  \(n_{\mathrm{ch}} \ge 2\)
and transfer momentum of tracks with
\(p_{\mathrm{T}} >100\)~\MeV\  and
\(p_{\mathrm{T}} >500\)~\MeV\
at \(\sqrt{s}=13\)~\TeV\
for the minimum-bias (MB) and
high-multiplicity track (HMT) events.
The uncertainties   represent the sum in quadrature of the  statistical
and systematic contributions.
The curves represent the exponential fits
to
\(\lambda (k_{\mathrm{T}})\) and \(R(k_{\mathrm{T}})\).
}
\label{Fig_pTcuts_com_kT}
\end{figure*}
 
\begin{table*}[t!]
\centering
\caption{
The
results of   the
fits to the dependencies
of the
correlation strength, \(\lambda\),
and
source radius, \(R\),
on
the pair average transverse momentum,  \(k_\mathrm{T}\),
for  various functional forms and  for minimum-bias (MB) and
high-multiplicity track  (HMT) events for
\(p_{\mathrm{T}} > 100\)~\MeV\
and
\(p_{\mathrm{T}} > 500\)~\MeV\
at \(\sqrt{s} = 13\)~\TeV.
The total
uncertainties are shown.
}
\label{tab:parameters_kT}
\medskip
\begin{tabular}{cccllr}
\hline
\hline
BEC
&
Fit
&
\(p_{\mathrm{T}}\)
&
\multicolumn{1}{c}{MB events}
&
\multicolumn{1}{c}{HMT events}
&
\(\chi^2/ndf\)
\\
 
Parameter
&
Equation
&
Threshold
&
\multicolumn{1}{c}{ }
&
\multicolumn{1}{c}{ }
&
\\
\hline
& & & & & \\[-4mm]
 
\(\lambda (k_{\mathrm{T}}) \)
&  (\ref{eq_LkT})
&
\(> 100\)~[\MeV]
&
\(\mu = 1.18^{+0.10}_{-0.11} \)
 
&
\(\mu = 1.09^{+0.08}_{-0.09} \)
&
MB:
11/6
\\[1mm]
 
&
&
&
\(\nu = 1.27^{+0.20}_{-0.30} \)~[\GeV\(^{-1}\)]
 
&
\(\nu = 0.89^{+0.19}_{-0.17} \)~[\GeV\(^{-1}\)]
&
HMT:
14/6
\\[2mm]
 
&
&
\(> 500\)~[\MeV]
&
\(\mu = 2.10^{+0.70}_{-1.29} \)
 
&
\(\mu = 1.68^{+0.50}_{-0.53} \)
&
MB:
2.0/2
\\[1mm]
 
&
&
&
\(\nu = 1.71^{+0.52}_{-0.92} \)~[\GeV\(^{-1}\)]
 
&
\(\nu = 1.29^{+0.47}_{-0.50} \)~[\GeV\(^{-1}\)]
&
HMT:
2.1/2
\\[1mm]
 
\hline
& & & & & \\[-4mm]
 
\(R (k_{\mathrm{T}}) \)
&  (\ref{eq_RkT})
&
\(> 100\)~[\MeV]
&
\(\xi = 2.98^{+0.36}_{-0.78} \)~[fm]
 
&
\(\xi = 3.51^{+0.18}_{-0.21} \)~[fm]
 
&
MB:
21/6
\\[1mm]
 
&
&
&
\(\kappa = 0.23^{+0.29}_{-0.75} \)~[\GeV\(^{-1}\)]
 
&
\(\kappa = 0.19^{+0.16}_{-0.17} \)~[\GeV\(^{-1}\)]
 
&
HMT:
13/6
\\[2mm]
 
&
&
\(> 500\)~[\MeV]
&
\(\xi = 2.71^{+0.56}_{-2.82} \)~[fm]
 
&
\(\xi = 2.92^{+0.33}_{-0.87} \)~[fm]
 
&
MB:
1.5/2
\\[1mm]
 
&
&
&
\(\kappa = 0.30^{+0.38}_{-1.26} \)~[\GeV\(^{-1}\)]
 
&
\(\kappa = 0.13^{+0.20}_{-0.47} \)~[\GeV\(^{-1}\)]
 
&
HMT:
2.6/2
\\[1mm]
 
\hline
\hline
\\
\end{tabular}
\end{table*}

 
The BEC are studied  in \(k_{\mathrm{T}}\) intervals from \(0.1\) to \(1.5\)~\GeV\
that are chosen so that they are populated roughly equally.
The fitted dependences of  \(\lambda (k_\mathrm{T}) \) and \(R (k_\mathrm{T}) \)
for the two
\(p_{\mathrm{T}}\) thresholds
are shown in
Figure~\ref{Fig_pTcuts_com_kT}
for \(p_{\mathrm{T}} > 100\) and \(500\)~\MeV.
In the small-\(Q\),
BEC-sensitive region, the minimum \(k_{\mathrm{T}}\)
is highly correlated with the \(p_{\mathrm{T}}\) threshold,
and so \(k_{\mathrm{T}}\) thresholds of 100 and 500~\MeV\ are applied
for the \(p_{\mathrm{T}}\) thresholds of 100 and 500~\MeV, respectively.
The values of  \(\lambda (k_\mathrm{T}) \) and \(R (k_\mathrm{T}) \)
decrease  with  increasing   \(k_{\mathrm{T}}\).
The decrease of \(\lambda (k_\mathrm{T}) \)  with \(k_{\mathrm{T}}\)  is
described by  the function:
\begin{equation}
\label{eq_LkT}
\lambda (k_{\mathrm{T}}) = \mu\, {\mathrm{e}}^{- \nu k_{\mathrm{T}}}\ .
\end{equation}
The \(R (k_\mathrm{T}) \)  parameter  shows similar behaviour:
\begin{equation}
\label{eq_RkT}
R (k_{\mathrm{T}}) =
\xi\, {\mathrm{e}}^{- \kappa k_{\mathrm{T}}}\ .
\end{equation}
The fitted parameters in Eqs.~(\ref{eq_LkT}) and (\ref{eq_RkT})
are given in  Table~\ref{tab:parameters_kT}.
 
 
\subsection{Double-differential dependence on  \( m_{\mathrm{ch}} \)   and   \( k_{\mathrm{T}} \) }
\label{sect:2D_mch_kT}

There is sufficient data at \(13\)~\TeV\ to study the double-differential dependence:
\( \lambda (m_{\mathrm{ch}}, k_{\mathrm{T}}) \)
and \( R (m_{\mathrm{ch}}, k_{\mathrm{T}}) \).
 
\begin{figure*}[t!]
\begin{minipage}[h]{0.50\linewidth}
\center{\includegraphics[width=1\linewidth]{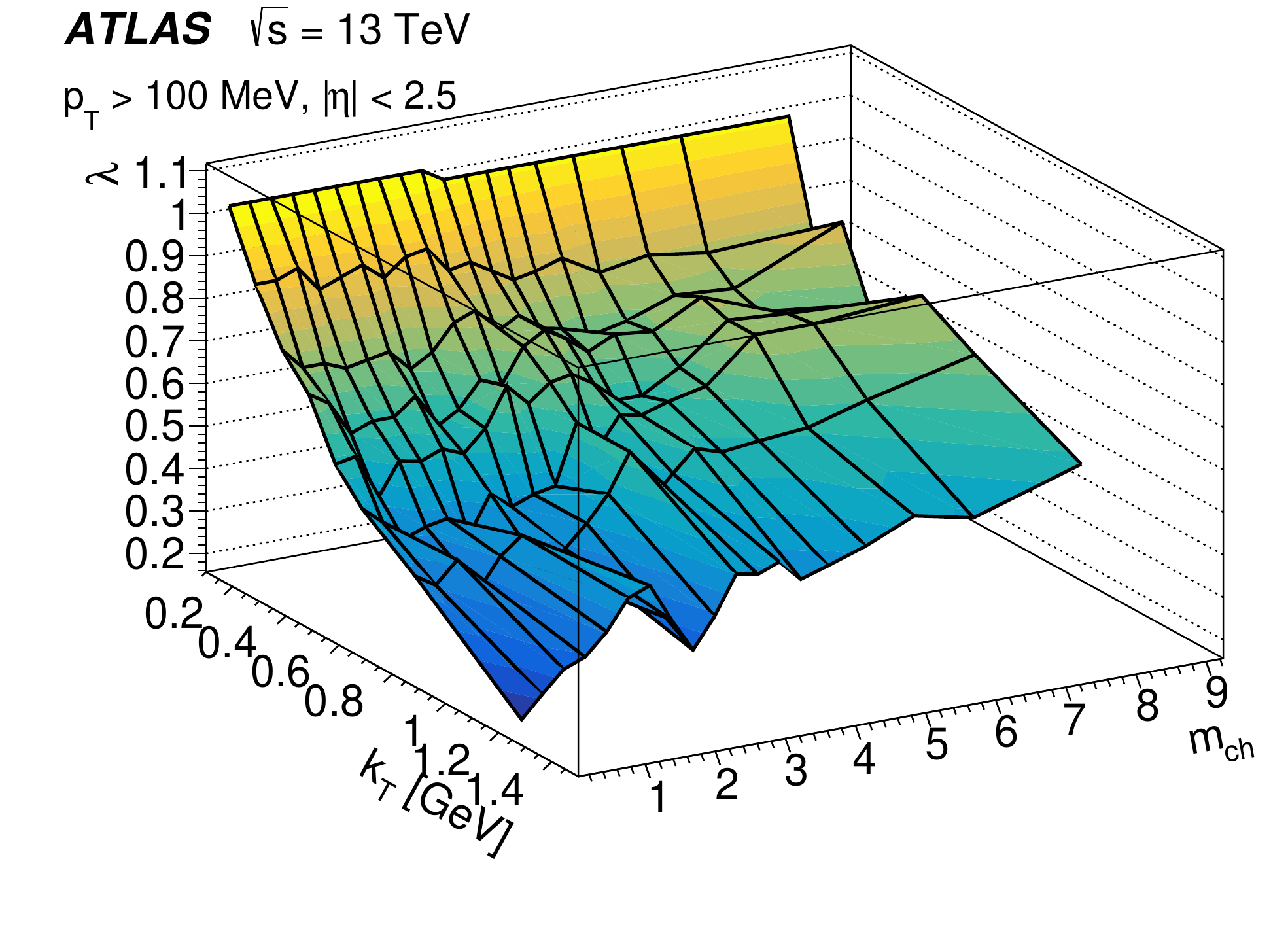}}
(a) \\
\end{minipage}
\hfill
\begin{minipage}[h]{0.50\linewidth}
\center{\includegraphics[width=1\linewidth]{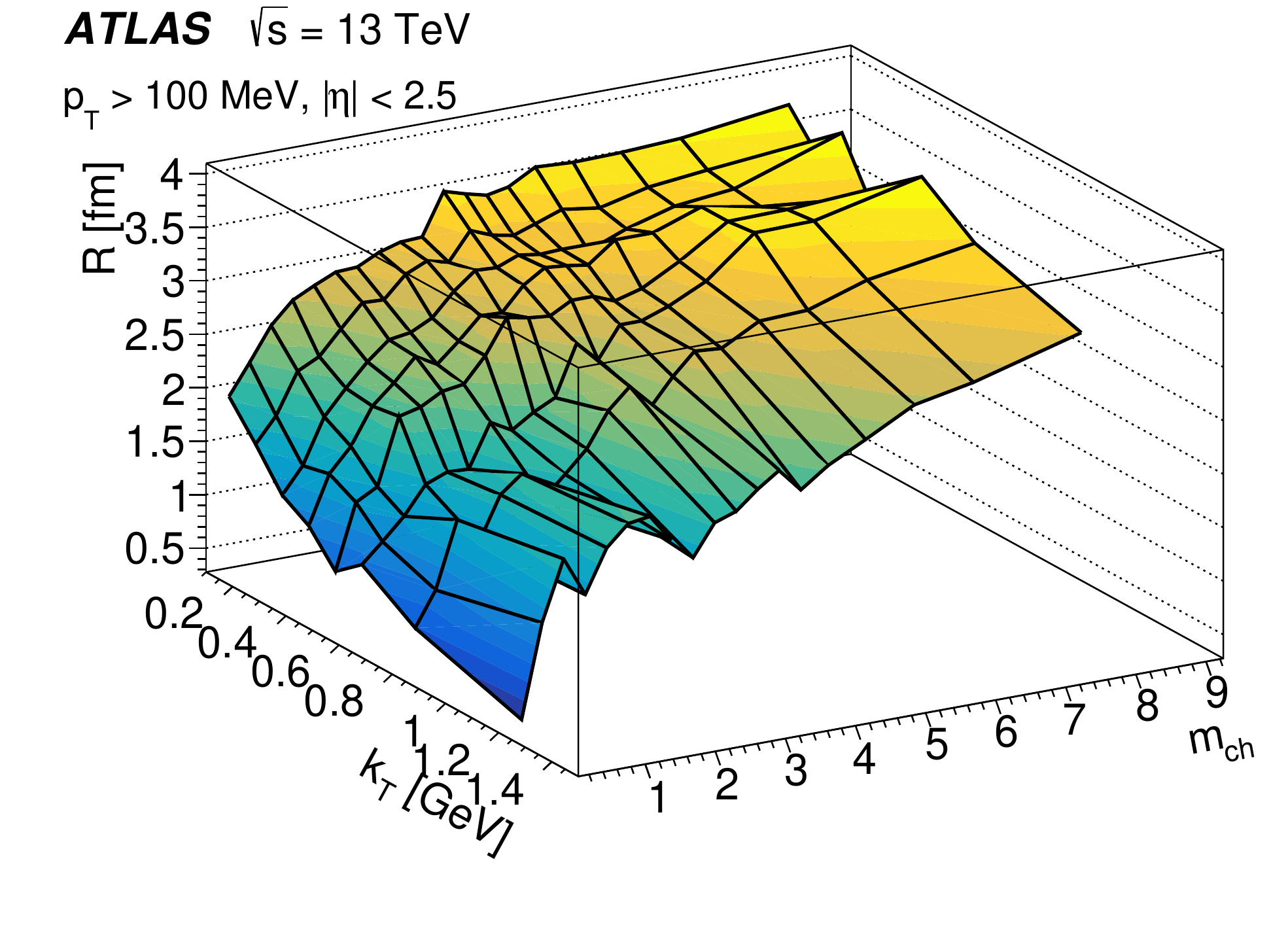}}
(b) \\
\end{minipage}
\caption{
The two-dimensional
dependence on \( m_{\mathrm{ch}}\) and \( k_{\mathrm{T}}\) for
\( p_{\mathrm{T}} > 100 \)~\MeV\
for (a) the correlation strength,
\( \lambda \),
and (b) the source radius,
\( R \),
obtained from the exponential fit to the \( R_2 (Q) \) correlation functions
using the MB sample for \( m_{\mathrm{ch}} \le 3.08 \)
and the HMT sample for \( m_{\mathrm{ch}} > 3.08 \).
}
\label{fig:LambdaR_2Dsurfaces_pt100}
\end{figure*}
 
 
\begin{figure*}[t!]
\begin{minipage}[h]{0.50\linewidth}
\center{\includegraphics[width=1\linewidth]{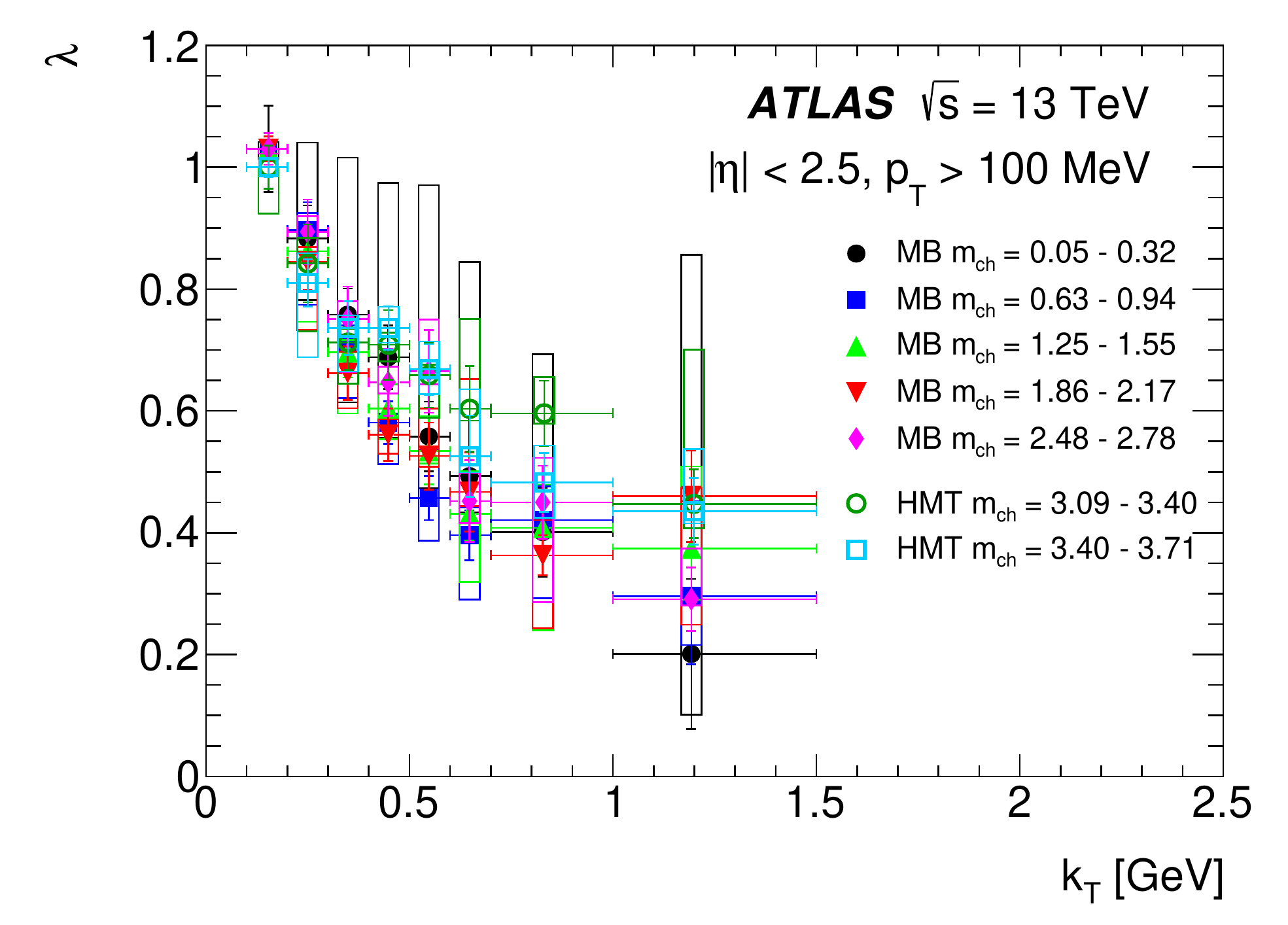}}
(a)
\\
\end{minipage}
\hfill
\begin{minipage}[h]{0.50\linewidth}
\center{\includegraphics[width=1\linewidth]{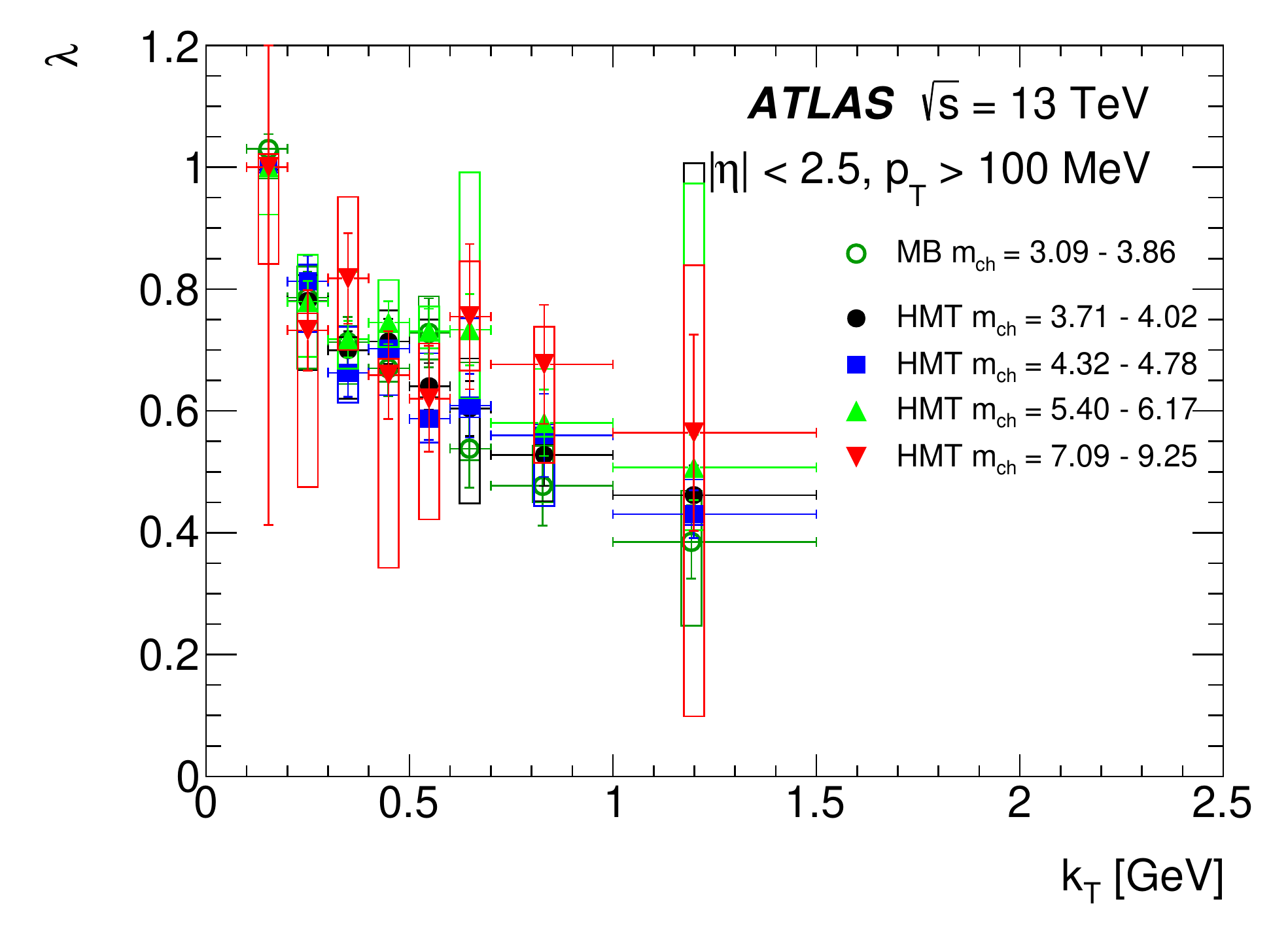}}
(b)
\\
\end{minipage}
\vfill
\begin{minipage}[h]{0.50\linewidth}
\center{\includegraphics[width=1\linewidth]{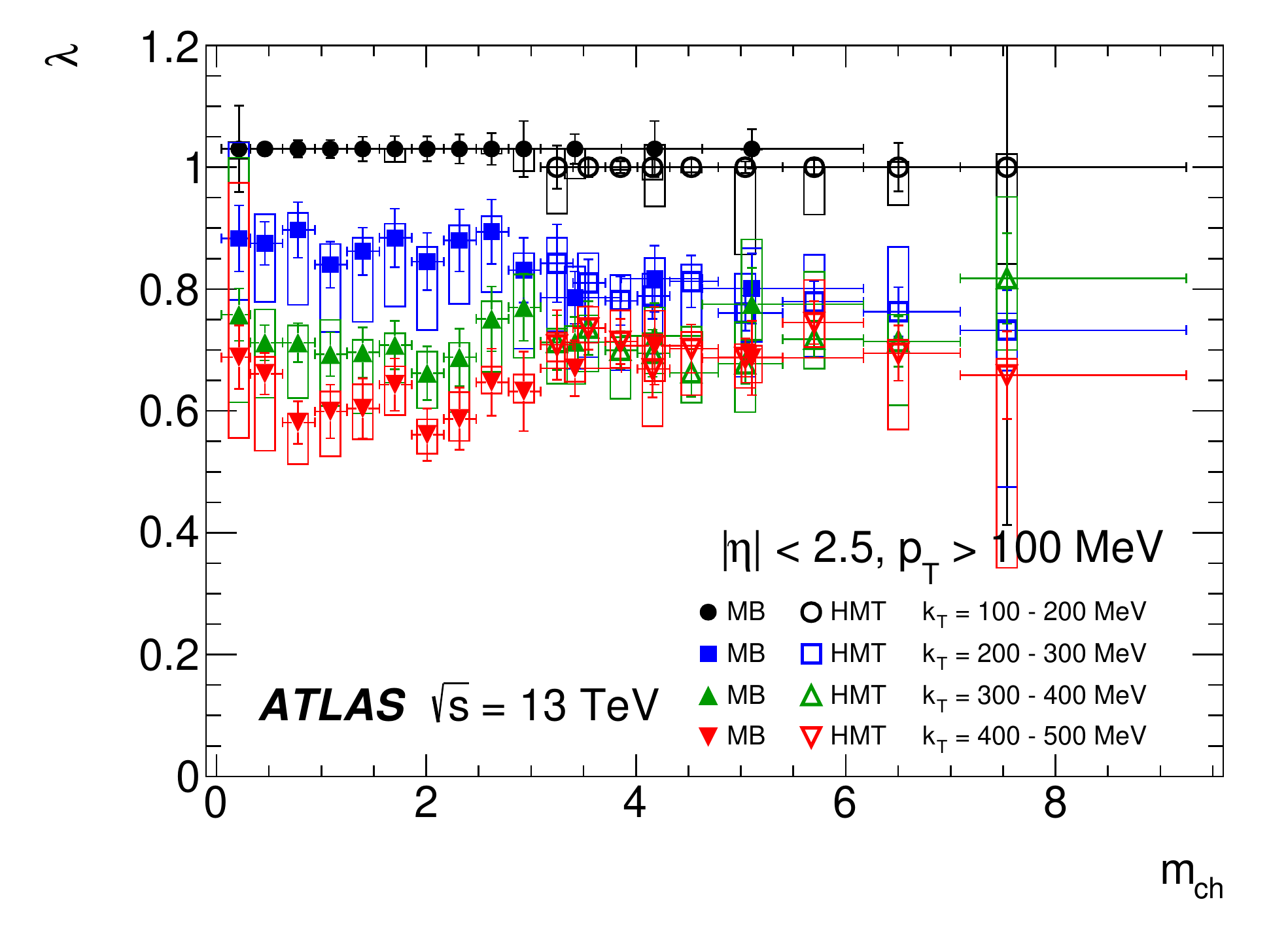}}
(c)
\\
\end{minipage}
\hfill
\begin{minipage}[h]{0.50\linewidth}
\center{\includegraphics[width=1\linewidth]{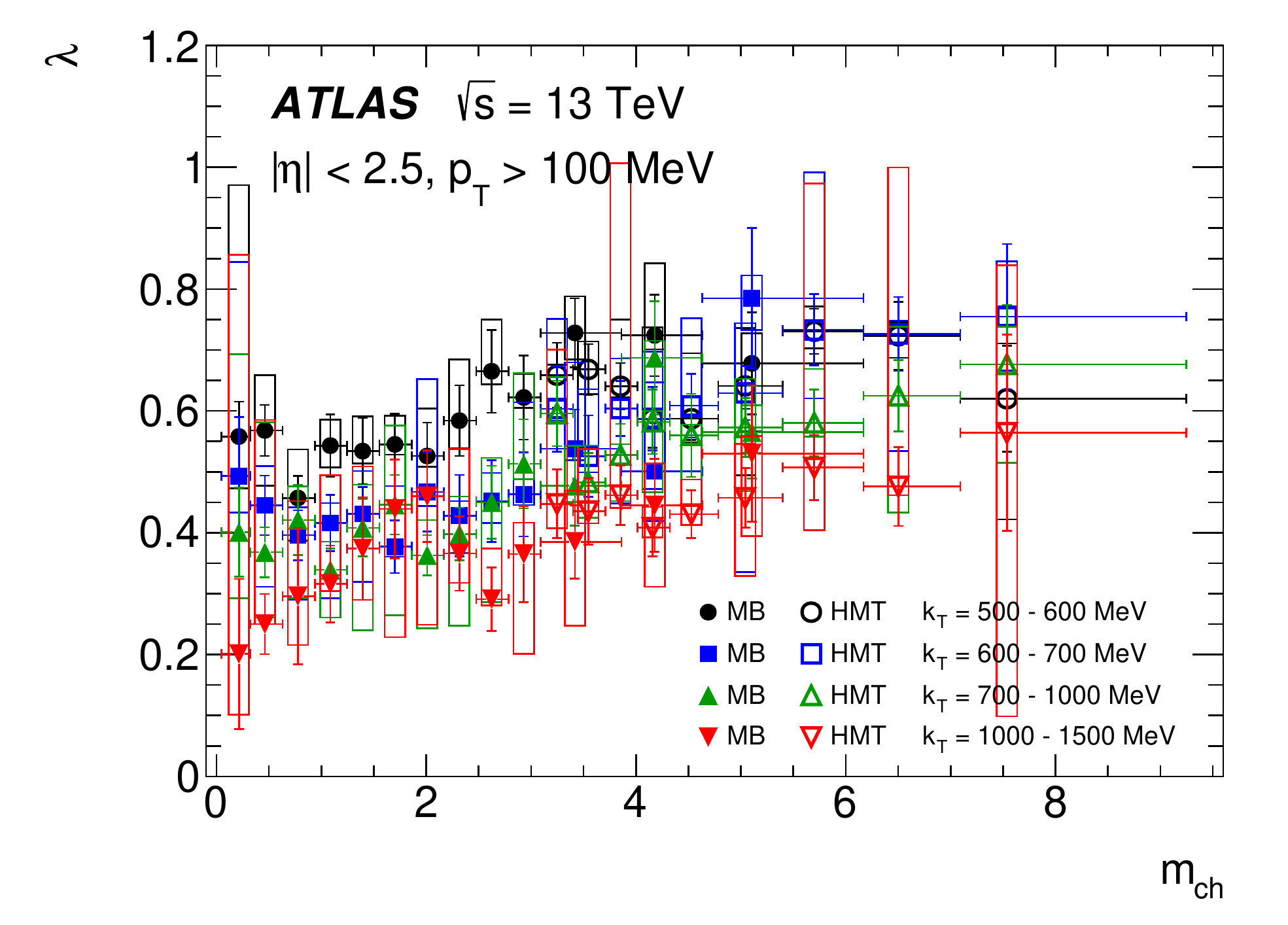}}
(d)
\\
\end{minipage}
\caption{
The parameter \(\lambda\) for \(p_{\mathrm{T}} > 100\)~\MeV:
(a) as a function of \(k_{\mathrm{T}}\) in selected low \(m_{\mathrm{ch}}\) intervals,
(b) as a function of \(k_{\mathrm{T}}\) in selected high \(m_{\mathrm{ch}}\) intervals,
(c) as a function of \(m_{\mathrm{ch}}\) in \(k_{\mathrm{T}}\) intervals between 0.1 and 0.5~\GeV,
and (d) as a function of \(m_{\mathrm{ch}}\) in \(k_{\mathrm{T}}\) intervals between 0.5 and 1.5~\GeV.
The error bars and boxes represent the  statistical  and systematic  contributions, respectively.
}
\label{fig:Lambda_1Dprojections_pt100}
\end{figure*}

 
Figure~\ref{fig:LambdaR_2Dsurfaces_pt100}
shows the  two-dimensional (2D)  dependence of \( \lambda \) and \(R\)   on
\(m_{\mathrm{ch}} \) and \( k_{\mathrm{T}} \), for \(p_{\mathrm{T}} > 100 \)~\MeV.
In order to convey a clear picture of the 2D dependence, the values are presented without uncertainties,
but they are generally less than  10\%.
 
The following features can be seen in Figure~\ref{fig:LambdaR_2Dsurfaces_pt100}:
\begin{itemize}
\item
The parameter \(\lambda\) decreases monotonically with \(k_{\mathrm{T}}\)
in all \(m_{\mathrm{ch}}\) intervals,
similar to the behaviour seen in Figure~\ref{Fig_pTcuts_com_kT}(a).
\item
The variation of \(\lambda\) with \(m_{\mathrm{ch}}\) is rather flat for
\(k_{\mathrm{T}}   \lesssim  500\)~\MeV,
which may be compared with the clear decrease of \(\lambda\) as a function of \(m_{\mathrm{ch}}\)
seen in Figure~\ref{Fig_pTcuts_com_nch}(a) for \(p_{\mathrm{T}} > 100\)~\MeV.
This is discussed below.
\item
The variation of \(\lambda\) with \(m_{\mathrm{ch}}\) for \(k_{\mathrm{T}} > 500\)~\MeV\
indicates a slight rise with \(m_{\mathrm{ch}}\) similar to what is seen in
Figure~\ref{Fig_pTcuts_com_nch}(a) for \(p_{\mathrm{T}} > 500\)~\MeV.
\item
The parameter \(R\) also decreases monotonically with \(k_{\mathrm{T}}\) in all \(m_{\mathrm{ch}}\) intervals,
with the decrease being more pronounced in lower \(m_{\mathrm{ch}}\) intervals.
A similar decrease is seen in Figure~\ref{Fig_pTcuts_com_kT}(b).
\item
\(R\) increases with \(m_{\mathrm{ch}}\) in all \(k_{\mathrm{T}}\) intervals, starting from a lower value
at larger \(k_{\mathrm{T}}\).
\item
This increase of \(R\) flattens off to a plateau level in all \(k_{\mathrm{T}}\) intervals.
This is similar to the plateau seen in Figures~\ref{Fig_pTcuts_com_nch}(b) and \ref{Fig_pTcuts_com_nch}(c).
Within the precision of the present data, the onset of the plateau occurs at
\( m_{\mathrm{ch}} \sim 3 \)
in all \( k_{\mathrm{T}} \) intervals.
\end{itemize}

\begin{figure*}[t!]
\begin{minipage}[h]{0.50\linewidth}
\center{\includegraphics[width=1\linewidth]{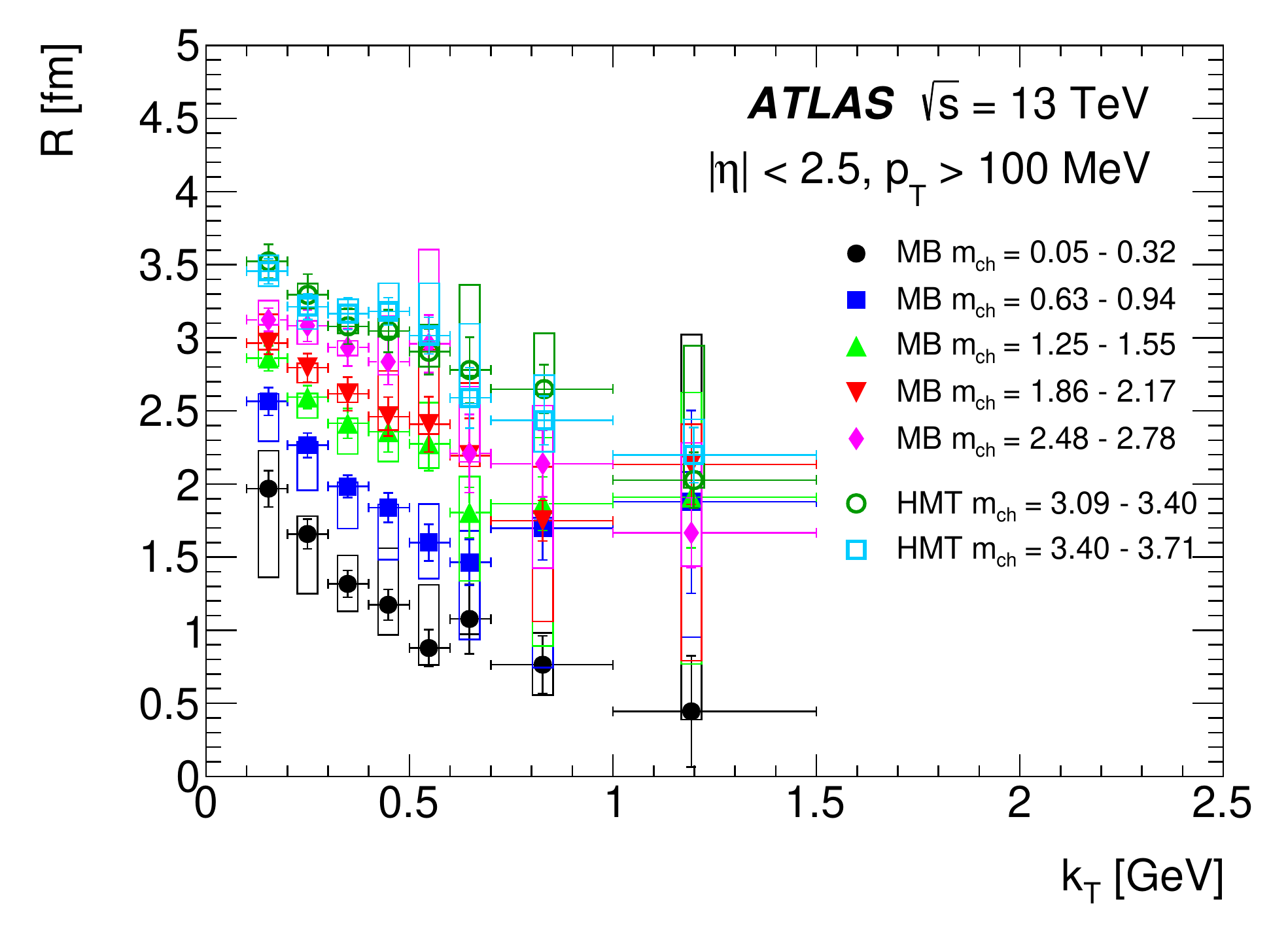}}
(a)
\\
\end{minipage}
\hfill
\begin{minipage}[h]{0.50\linewidth}
\center{\includegraphics[width=1\linewidth]{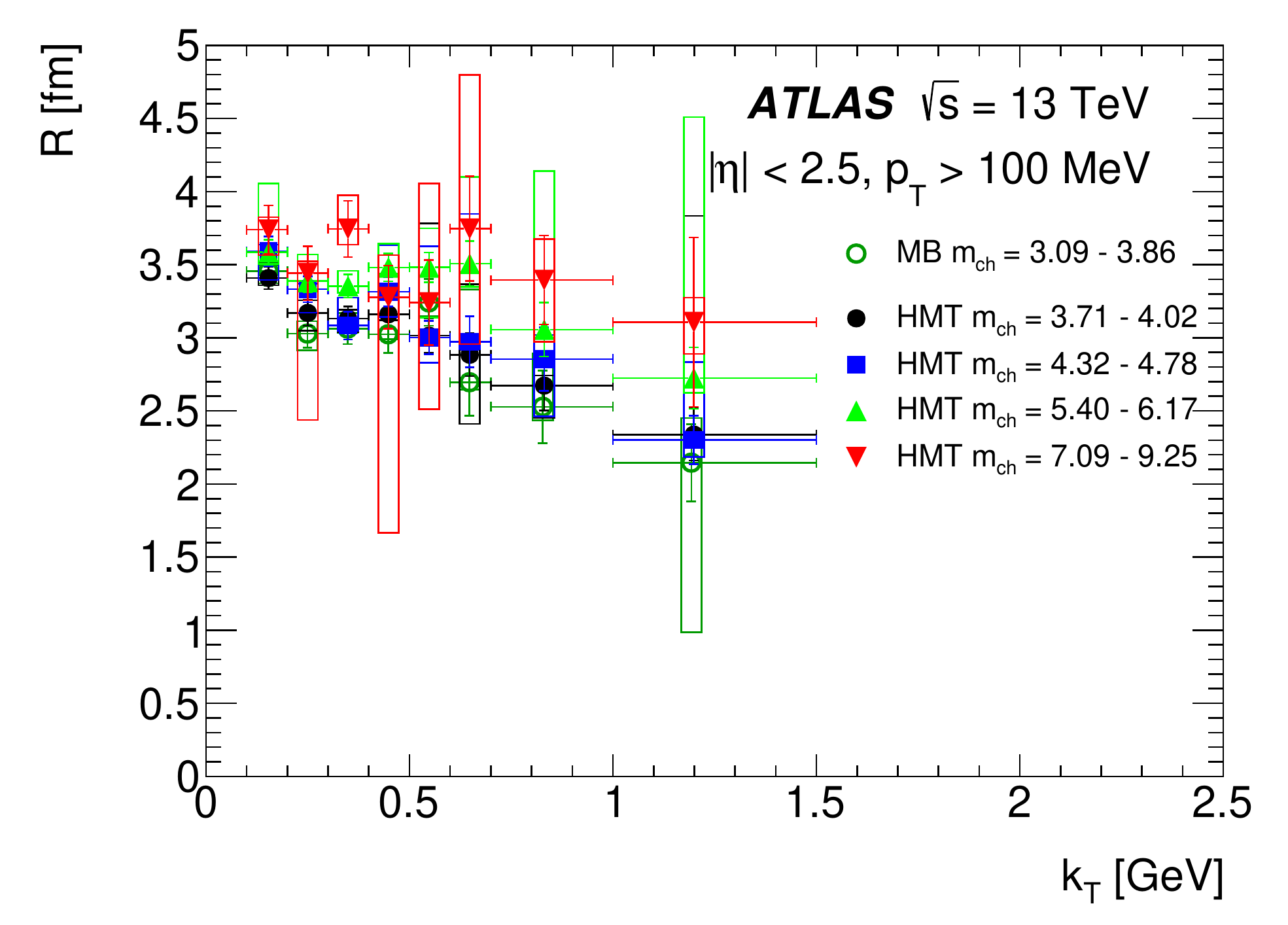}}
(b)
\\
\end{minipage}
\vfill
\begin{minipage}[h]{0.50\linewidth}
\center{\includegraphics[width=1\linewidth]{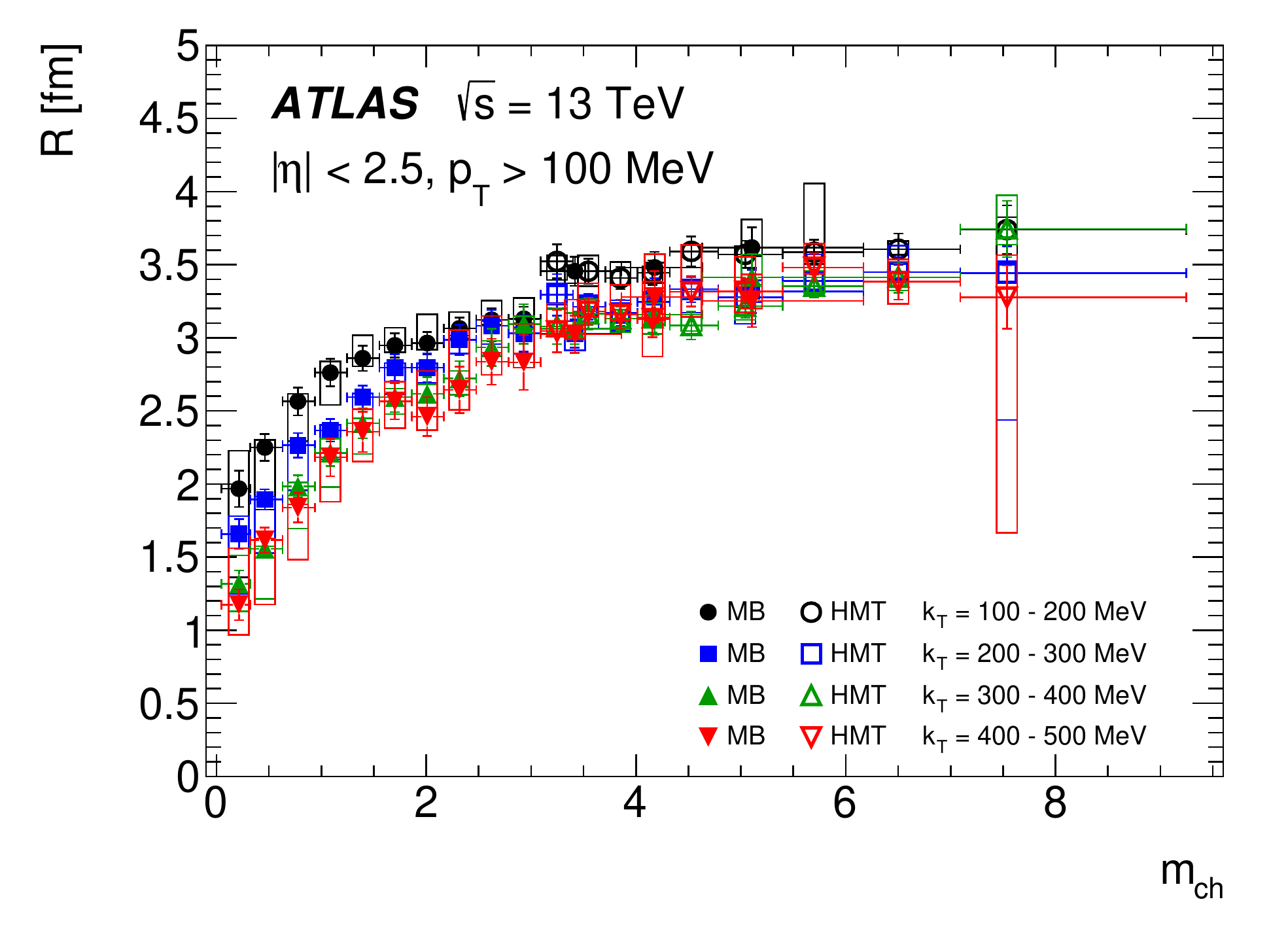}}
(c)
\\
\end{minipage}
\hfill
\begin{minipage}[h]{0.50\linewidth}
\center{\includegraphics[width=1\linewidth]{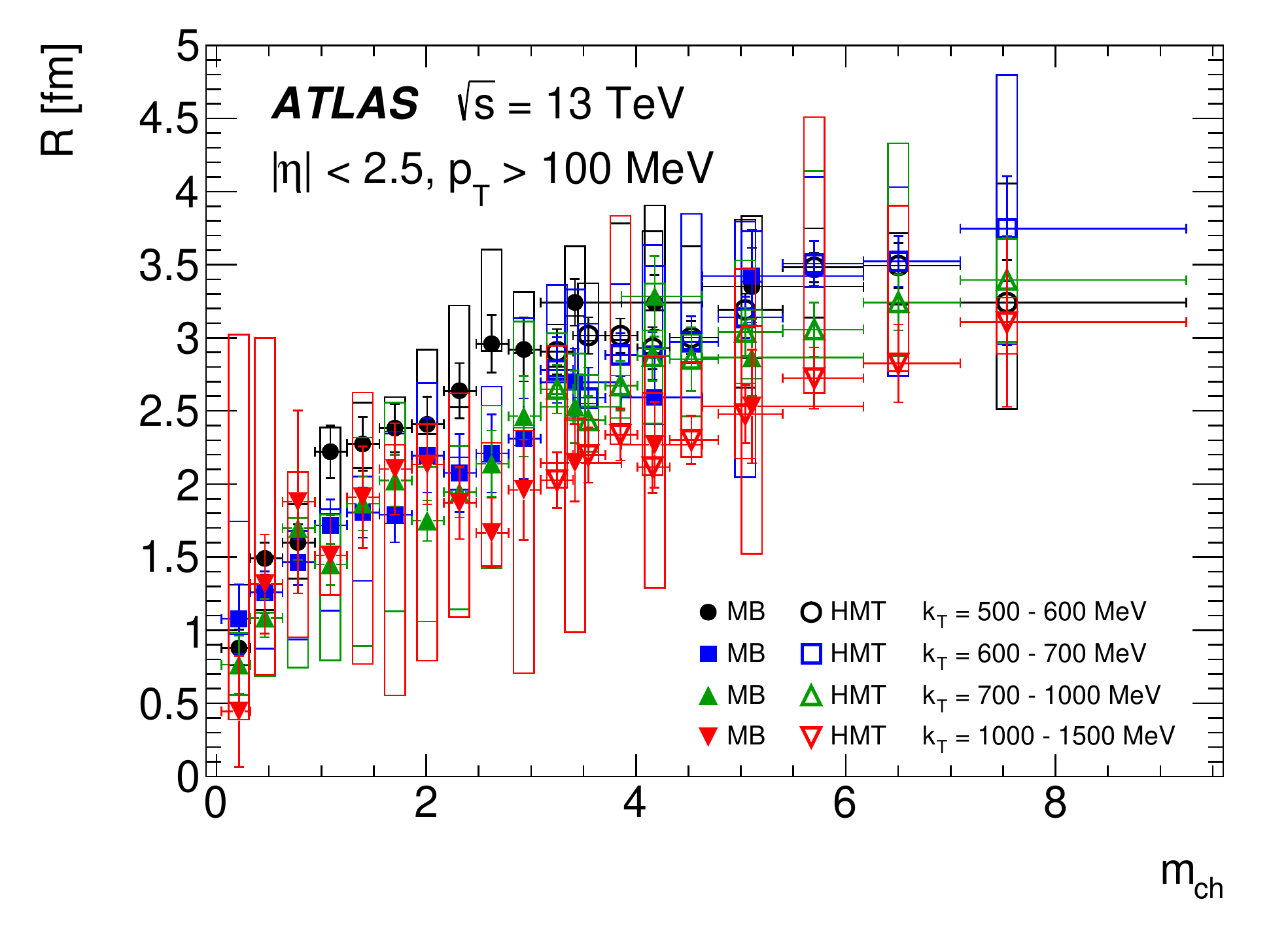}}
(d)
\\
\end{minipage}
\caption{
The parameter \(R\) for \(p_{\mathrm{T}} > 100\)~\MeV:
(a) as a function of \(k_{\mathrm{T}}\) in selected low \(m_{\mathrm{ch}}\) intervals,
(b) as a function of \(k_{\mathrm{T}}\) in selected high \(m_{\mathrm{ch}}\) intervals,
(c) as a function of \(m_{\mathrm{ch}}\) in \(k_{\mathrm{T}}\) intervals between 0.1 and 0.5~\GeV,
and (d) as a function of \(m_{\mathrm{ch}}\) in \(k_{\mathrm{T}}\) intervals between 0.5 and 1.5~\GeV.
The error bars and boxes represent the  statistical  and systematic  contributions, respectively.
}
\label{fig:R_1Dprojections_pt100}
\end{figure*}
 
 
\begin{figure*}[t!]
\begin{minipage}[h]{0.50\linewidth}
\center{\includegraphics[width=1\linewidth]{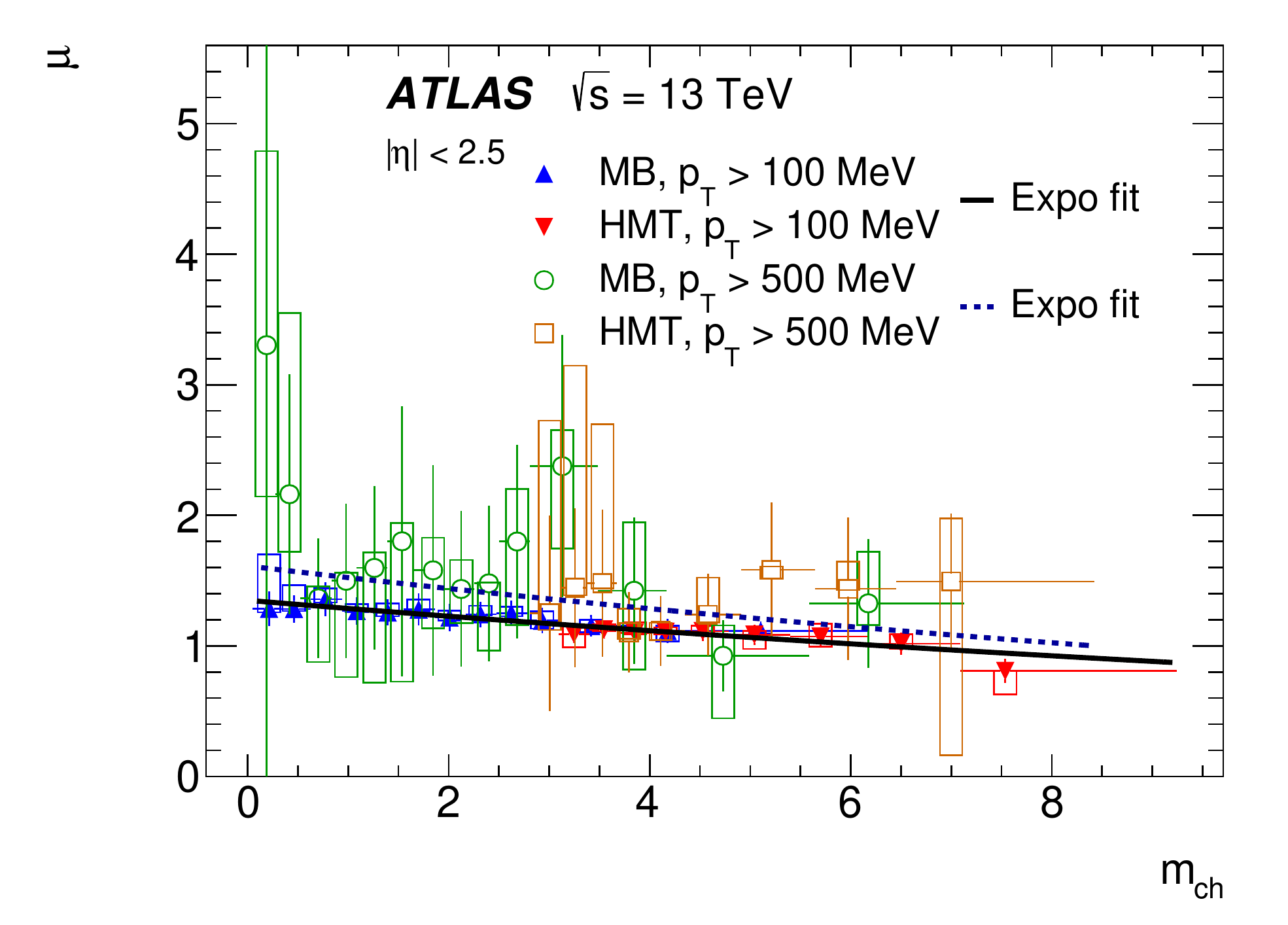}}
(a) \\
\end{minipage}
\hfill
\begin{minipage}[h]{0.50\linewidth}
\center{\includegraphics[width=1\linewidth]{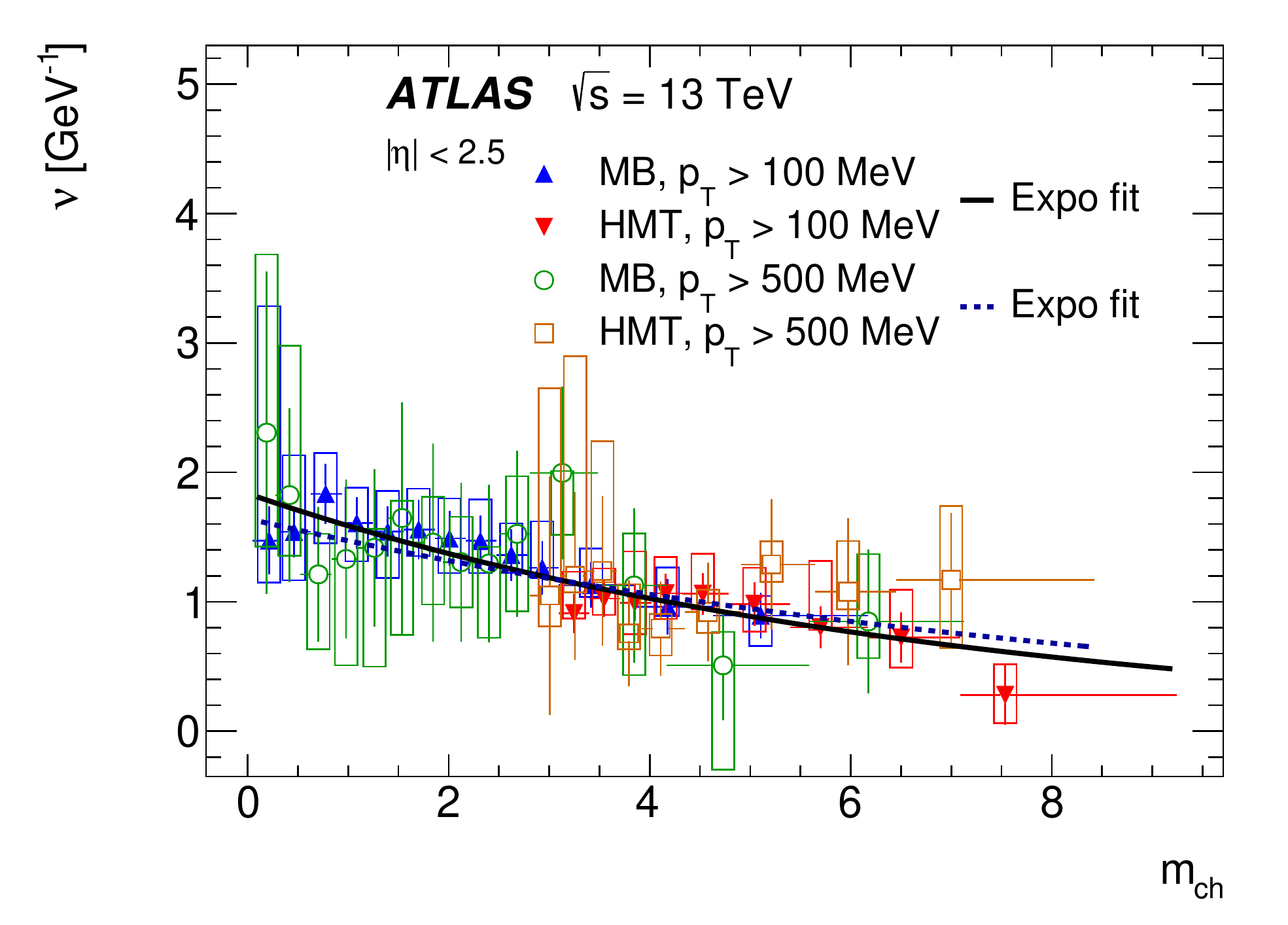}}
(b) \\
\end{minipage}
\caption{
The fit parameters (a) \(\mu\) and (b) \(\nu\) (Eq.~(\ref{eq_LkT})),
describing the dependence
of the correlation strength, \(\lambda\), on
charged-particle scaled multiplicity  (Eq.~(\ref{eq_Lnch})), for track \(p_{\mathrm{T}}>100\)~\MeV\
and track \(p_{\mathrm{T}}>500\)~\MeV\ in the minimum-bias   (MB) and high-multiplicity track (HMT)
samples at \(\sqrt{s}  = 13\)~\TeV.
The error bars and boxes represent the  statistical  and systematic  contributions, respectively.
The black solid (blue dashed) curve represents the exponential fit of the dependence of
parameter \(\mu\) (\(\nu\)) on  \(m_{\mathrm{ch}}\) for tracks
with \(p_{\mathrm{T}} >100\)~\MeV\  (\(p_{\mathrm{T}} >500\)~\MeV).
}
\label{Fig_LkT_nch}
\end{figure*}
\begin{figure*}[t!]
\begin{minipage}[h]{0.50\linewidth}
\center{\includegraphics[width=1\linewidth]{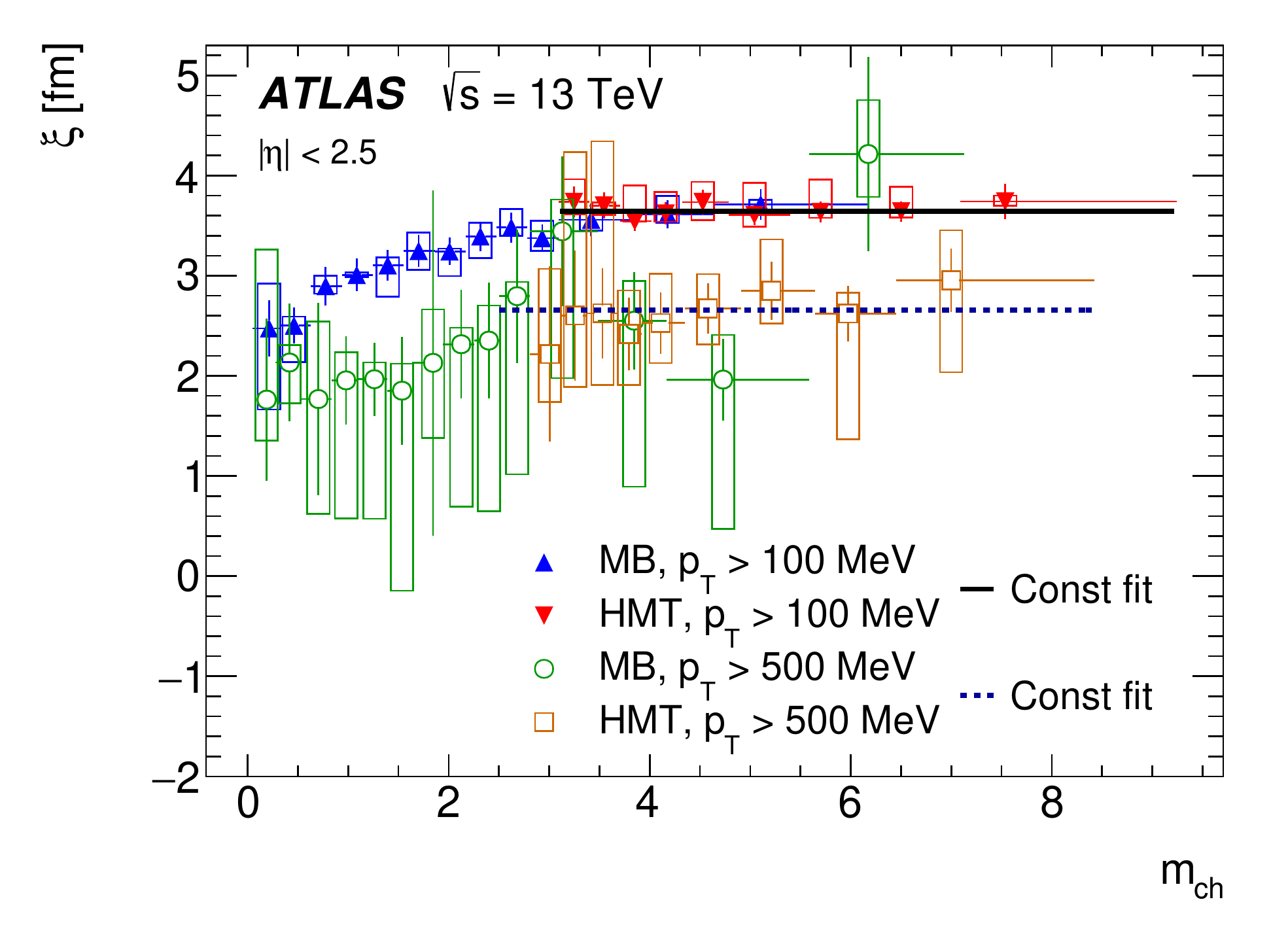}}
(a) \\
\end{minipage}
\hfill
\begin{minipage}[h]{0.50\linewidth}
\center{\includegraphics[width=1\linewidth]{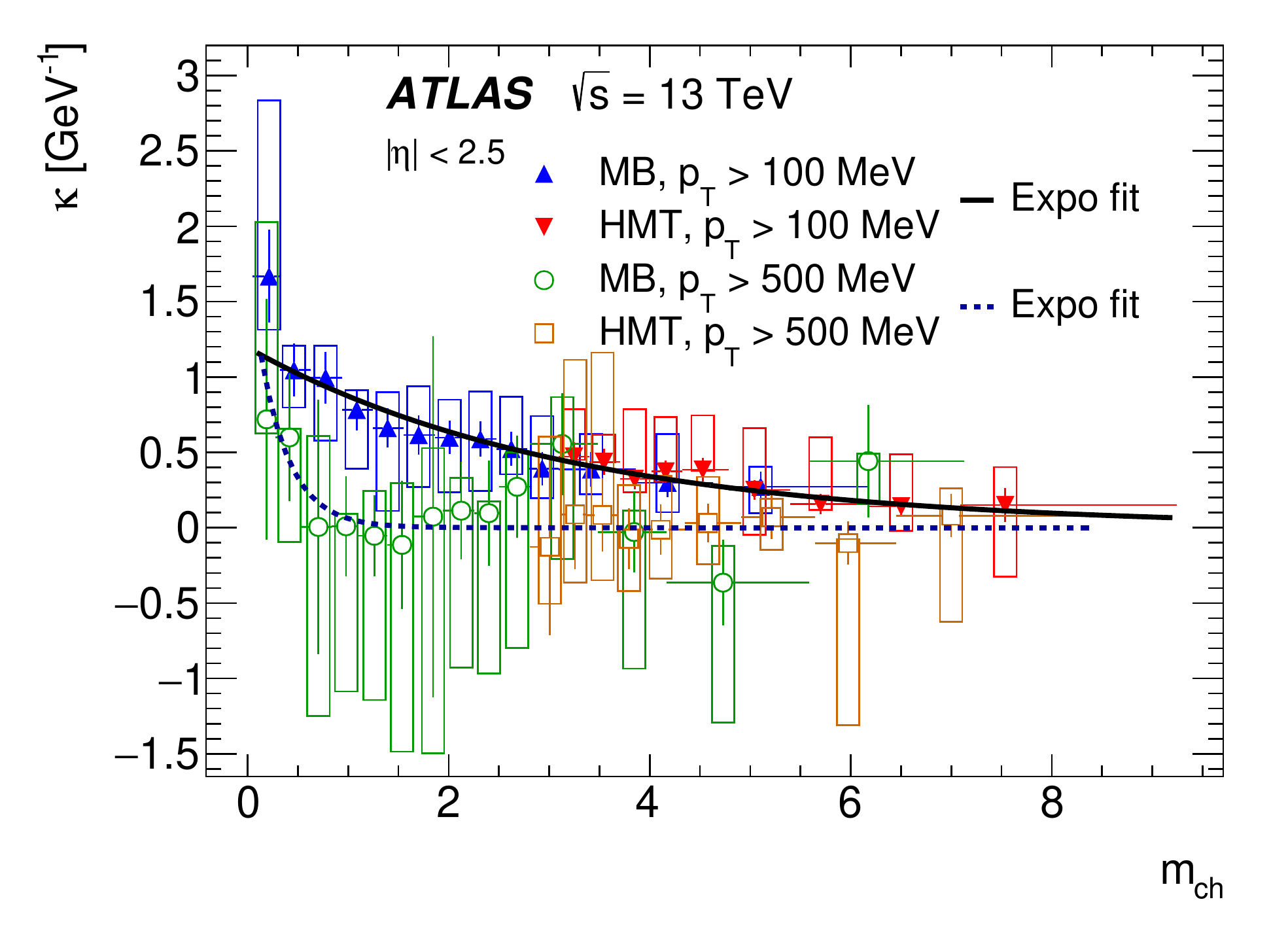}}
(b) \\
\end{minipage}
\caption{
The parameters (a) \(\xi\)  and (b) \(\kappa\) (Eq.~(\ref{eq_RkT})),
describing the dependence of the source radius,  \(R\),  on charged-particle   scaled multiplicity,
\(m_{\mathrm{ch}}\), for track \(p_{\mathrm{T}}>100\)~\MeV\ and track \(p_{\mathrm{T}}>500\)~\MeV\
in the minimum-bias (MB) and high-multiplicity track (HMT) samples at \(\sqrt{s} = 13\)~\TeV.
The error bars and boxes represent the  statistical  and systematic  contributions, respectively.
(a) The black solid and blue dashed curves represent the saturated value of the parameter
\(\xi\)  for \(m_{\mathrm{ch}} > 3.0\) for tracks with
\(p_{\mathrm{T}} >100\)~\MeV\ and for \(m_{\mathrm{ch}} > 2.8\)
for tracks with \(p_{\mathrm{T}} >500\)~\MeV, respectively.
(b)  The black solid and blue dashed curves represent the exponential fit to
the parameter \(\kappa\)  for tracks with \(p_{\mathrm{T}} >100\)~\MeV\ and for tracks with
\(p_{\mathrm{T}} >500\)~\MeV, respectively.
}
\label{Fig_RkT_nch}
\end{figure*}
 
These features can be seen more clearly in selected 1D projections in Figures~\ref{fig:Lambda_1Dprojections_pt100} and
\ref{fig:R_1Dprojections_pt100} for \(\lambda\) and \(R\), respectively.
MB and HMT data agree where they overlap.
 
Figures~\ref{fig:Lambda_1Dprojections_pt100}(c) and \ref{fig:Lambda_1Dprojections_pt100}(d) show the essentially flat behaviour of \(\lambda\)
as a function of \(m_{\mathrm{ch}}\) for low \(k_{\mathrm{T}}\),
and the slight rise at higher \(k_{\mathrm{T}}\),
which are described above.
As also noted above, Figure~\ref{Fig_pTcuts_com_nch}(a) shows
a slight but clear decrease of \(\lambda\) with \(m_{\mathrm{ch}}\) for \(p_{\mathrm{T}} >100\)~\MeV\
when integrated over all \(k_{\mathrm{T}}\).
This arises because particle \(p_{\mathrm{T}}\) and hence \(k_{\mathrm{T}}\) of the pair increase with \(m_{\mathrm{ch}}\),
but \(\lambda\) decreases as \(k_{\mathrm{T}}\) increases.
This leads to the overall decrease with \(m_{\mathrm{ch}}\) seen in
Figure~\ref{Fig_pTcuts_com_nch}(a) for \(p_{\mathrm{T}} >100 \)~\MeV.
 
The 2D behaviour can be parameterized  according to  Eqs.~(\ref{eq_LkT}) and (\ref{eq_RkT}):
\begin{equation}
\label{eq_LmchkT}
\lambda ( m_{\mathrm{ch}} , k_{\mathrm{T}} ) =
\mu(m_{\mathrm{ch}})\, {\mathrm{e}}^{- \nu( m_{\mathrm{ch}} ) k_{\mathrm{T}}}\ ,
\end{equation}
and
\begin{equation}
\label{eq_RmchkT}
R ( m_{\mathrm{ch}} , k_{\mathrm{T}}) =
\xi( m_{\mathrm{ch}})\, {\mathrm{e}}^{- \kappa( m_{\mathrm{ch}} ) k_{\mathrm{T}}}\ .
\end{equation}
 
The parameters \( \mu, \nu, \xi \) and \(\kappa\) are shown in Figures~\ref{Fig_LkT_nch}
and \ref{Fig_RkT_nch} for both the \(p_{\mathrm{T}} >100 \)~\MeV\ and
\(p_{\mathrm{T}} >500 \)~\MeV\ selections.
 
The parameters \( \mu, \nu, \xi \) and \(\kappa\) can themselves be fitted as functions of
\(m_{\mathrm{ch}}\).
For the \(\lambda\) parameters and \(p_{\mathrm{T}} >100 \)~\MeV:
\begin{equation}\nonumber
\mu (m_{\mathrm{ch}}) =
(1.35\pm 0.05 (\mathrm{stat.})~^{+ 0.14}_{-0.01} (\mathrm{syst.}))
\cdot
\exp{ \{ (-0.047\pm 0.009 (\mathrm{stat.})~^{+ 0.013}_{-0.013} (\mathrm{syst.}))\,
m_{\mathrm{ch}} \} },
\end{equation}
\begin{equation}\nonumber
\nu (m_{\mathrm{ch}}) =
(1.84\pm 0.12 (\mathrm{stat.})~^{+ 0.63}_{- 0.34} (\mathrm{syst.}))\,
{\mathrm{[GeV^{-1}]}}
\cdot
\exp{ \{  (-0.146\pm 0.020 (\mathrm{stat.})~^{+ 0.008}_{- 0.008} (\mathrm{syst.}))\,
m_{\mathrm{ch}} \} }.
\end{equation}
And for the \(R\) parameters and \(p_{\mathrm{T}} >100 \)~\MeV:
\begin{equation}\nonumber
\xi (m_{\mathrm{ch}}) = 3.64\pm0.04 (\mathrm{stat.})~^{+0.22}_{-0.08}
(\mathrm{syst.}) \,
{\mathrm{[fm]}} ,
\end{equation}
\begin{equation}\nonumber
\kappa (m_{\mathrm{ch}}) =
(1.20\pm 0.12 (\mathrm{stat.})~^{+ 0.09}_{- 0.45} (\mathrm{syst.}))\,
{\mathrm{[GeV^{-1}]}}
\cdot
\exp{ \{ (-0.31\pm 0.03 (\mathrm{stat.})~^{+ 0.16}_{- 0.03} (\mathrm{syst.}))\,
m_{\mathrm{ch}} \} },
\end{equation}
where the fit for \(\xi\) applies for \(m_{\mathrm{ch}} > 3.0\).
 
\begin{figure*}[t!]
\begin{minipage}[h]{0.50\linewidth}
\center{\includegraphics[width=1\linewidth]{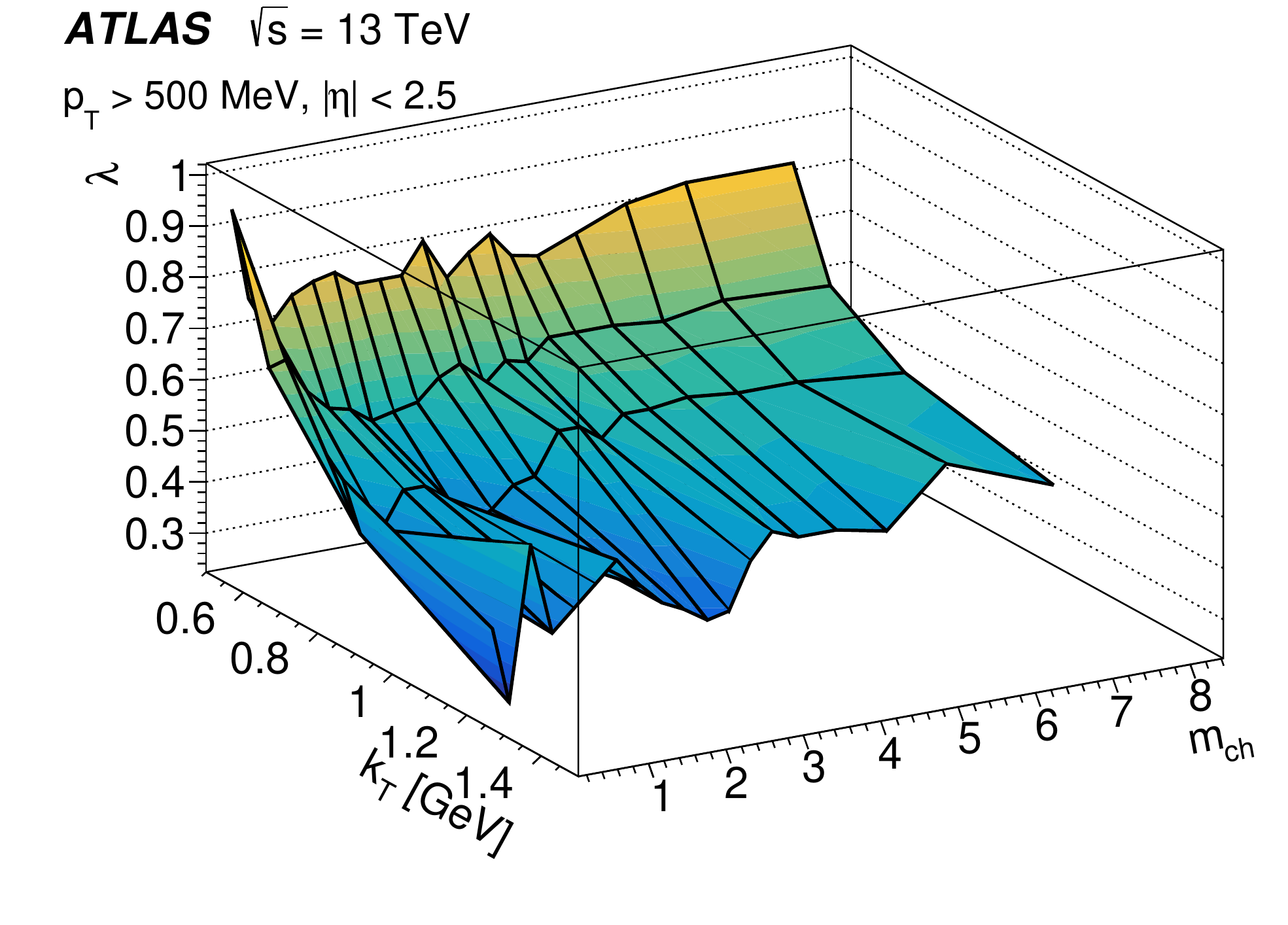}}
(a) \\
\end{minipage}
\hfill
\begin{minipage}[h]{0.50\linewidth}
\center{\includegraphics[width=1\linewidth]{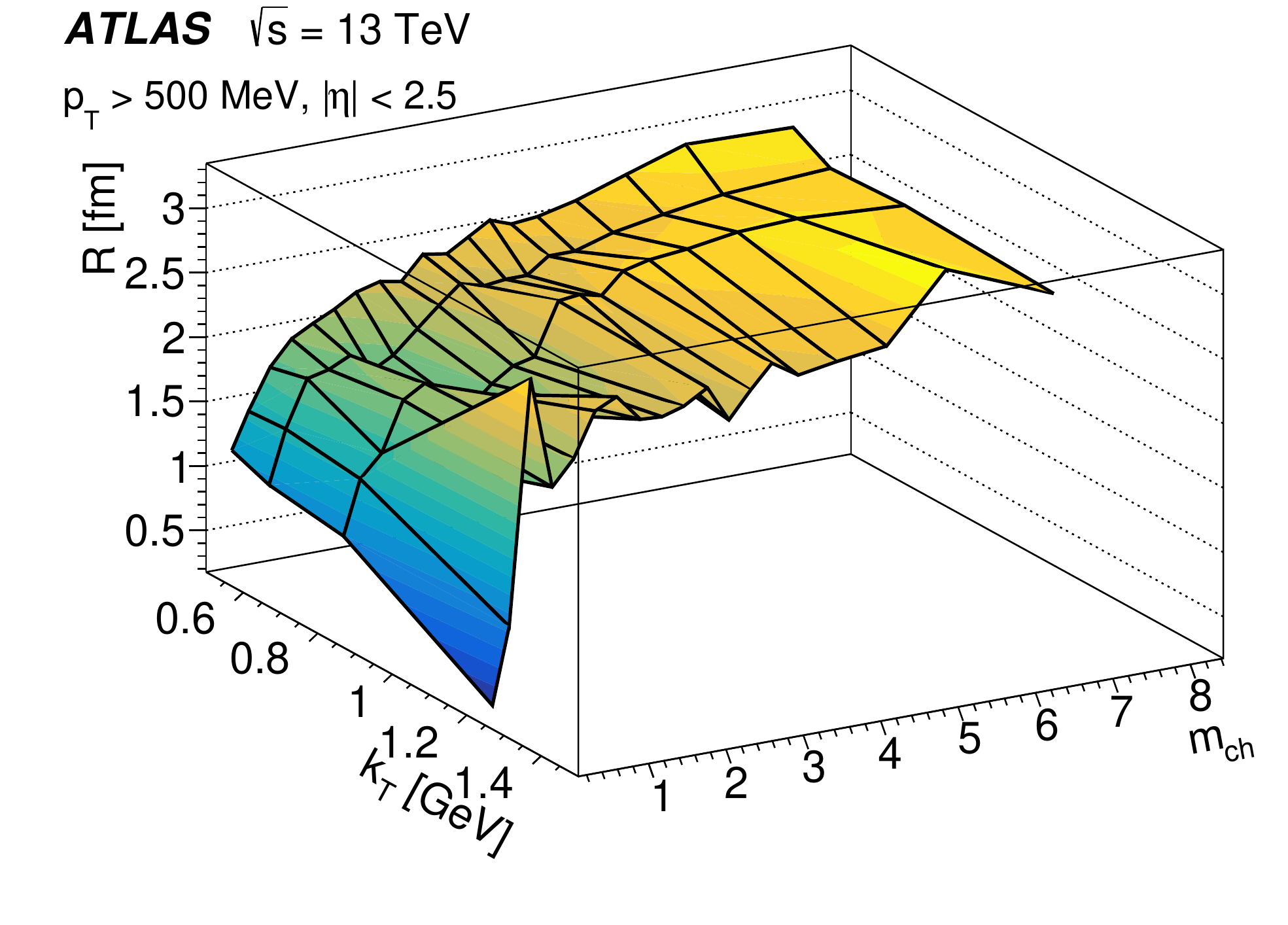}}
(b) \\
\end{minipage}
\caption{
The two-dimensional
dependence on \( m_{\mathrm{ch}}\) and \( k_{\mathrm{T}}\) for
\(p_{\mathrm{T}} > 500\)~\MeV\ for (a) the correlation strength,
\(\lambda \),
and (b) the source radius,
\(R\),
obtained from the exponential fit to the \(R_2(Q) \) correlation functions
using the MB sample for \( m_{\mathrm{ch}} \leq 3.08 \)
and the HMT sample for \( m_{\mathrm{ch}} > 3.08 \).
}
\label{fig:LambdaR_2Dsurfaces_pt500}
\end{figure*}
 
\begin{figure*}[t!]
\begin{minipage}[h]{0.50\linewidth}
\center{\includegraphics[width=1\linewidth]{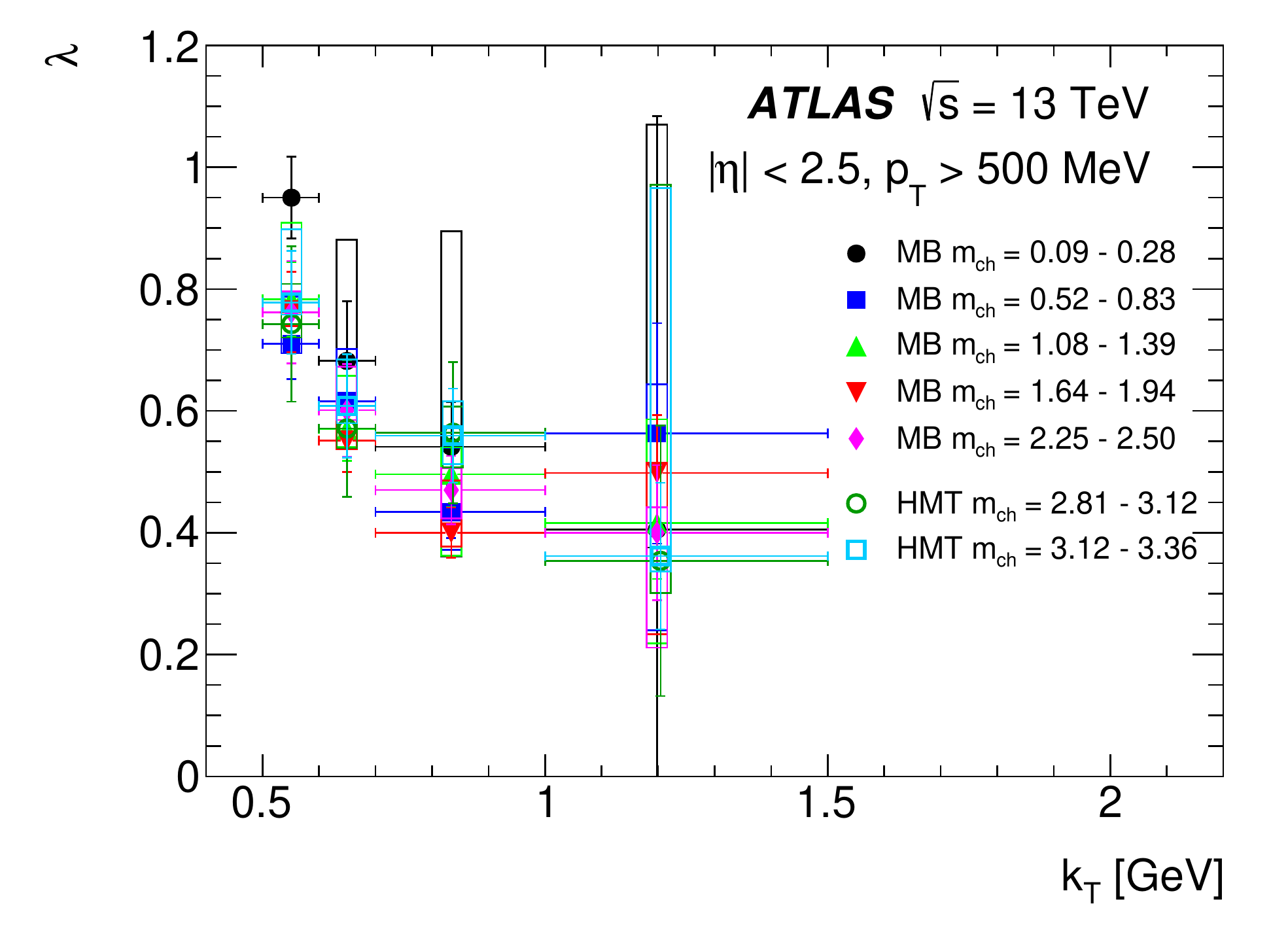}}
(a)
\\
\end{minipage}
\hfill
\begin{minipage}[h]{0.50\linewidth}
\center{\includegraphics[width=1\linewidth]{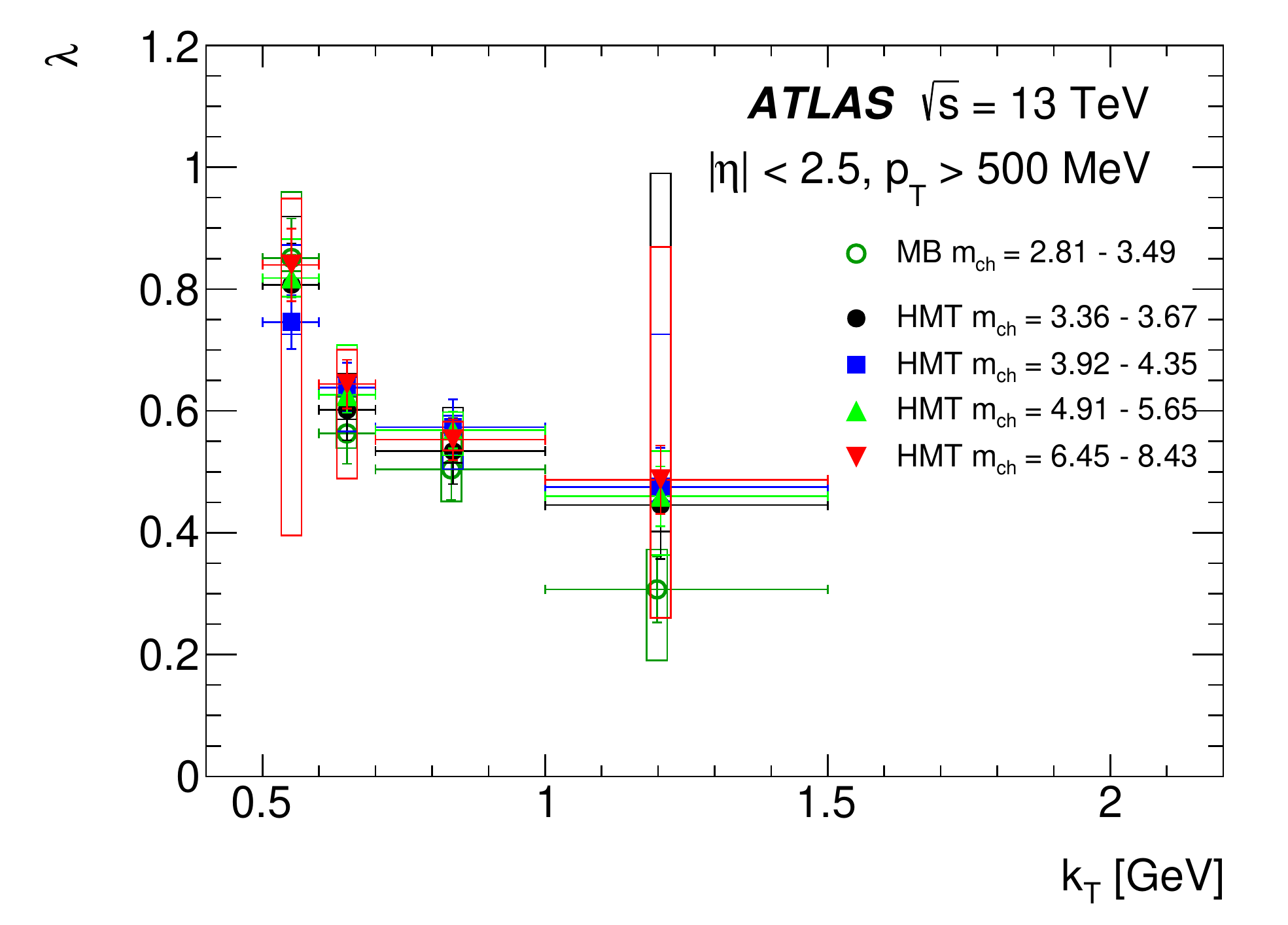}}
(b)
\\
\end{minipage}
\vfill
\begin{minipage}[h]{0.50\linewidth}
\center{\includegraphics[width=1\linewidth]{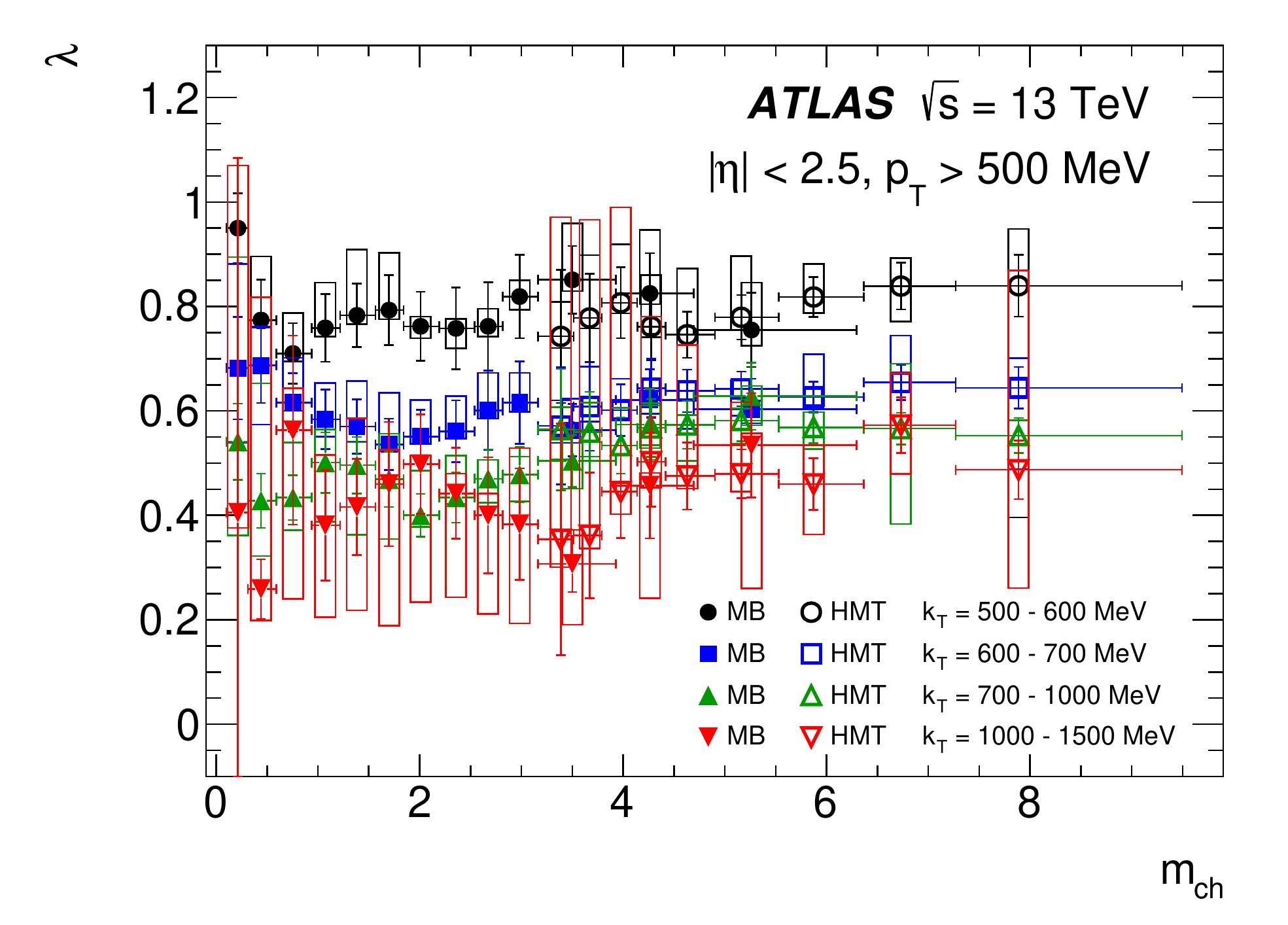}}
(c)
\\
\end{minipage}
\caption{
The parameter \(\lambda\) for \(p_{\mathrm{T}} > 500\)~\MeV:
(a) as a function of \(k_{\mathrm{T}}\) in selected low \(m_{\mathrm{ch}}\) intervals,
(b) as a function of \(k_{\mathrm{T}}\) in selected high \(m_{\mathrm{ch}}\) intervals,
and (c) as a function of \(m_{\mathrm{ch}}\) in \(k_{\mathrm{T}}\) intervals between 0.5 and 1.5~\GeV.
The error bars and boxes represent the  statistical  and systematic  contributions, respectively.
}
\label{fig:Lambda_1Dprojections_pt500}
\end{figure*}
 
\begin{figure*}[t!]
\begin{minipage}[h]{0.50\linewidth}
\center{\includegraphics[width=1\linewidth]{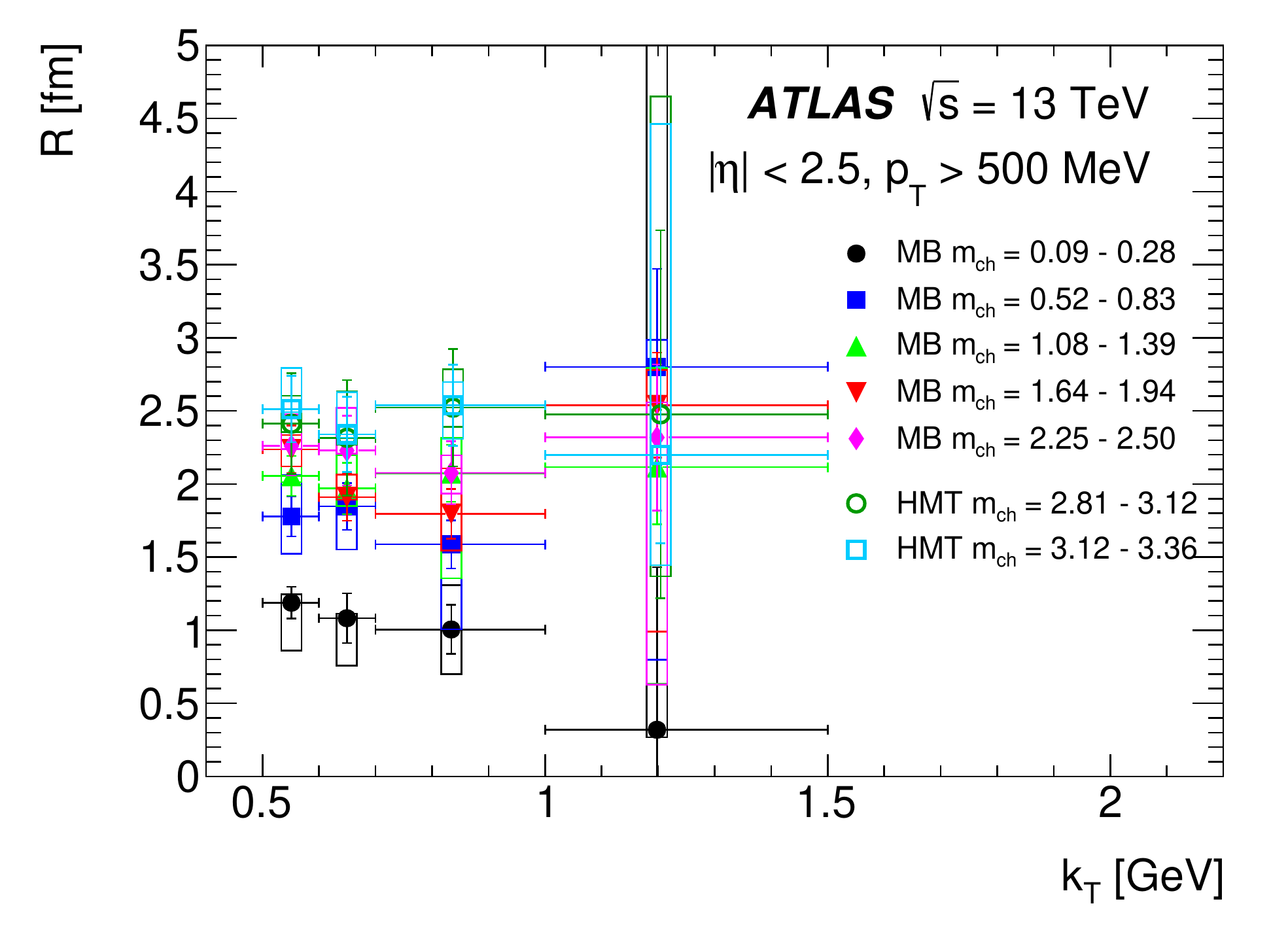}}
(a)
\\
\end{minipage}
\hfill
\begin{minipage}[h]{0.50\linewidth}
\center{\includegraphics[width=1\linewidth]{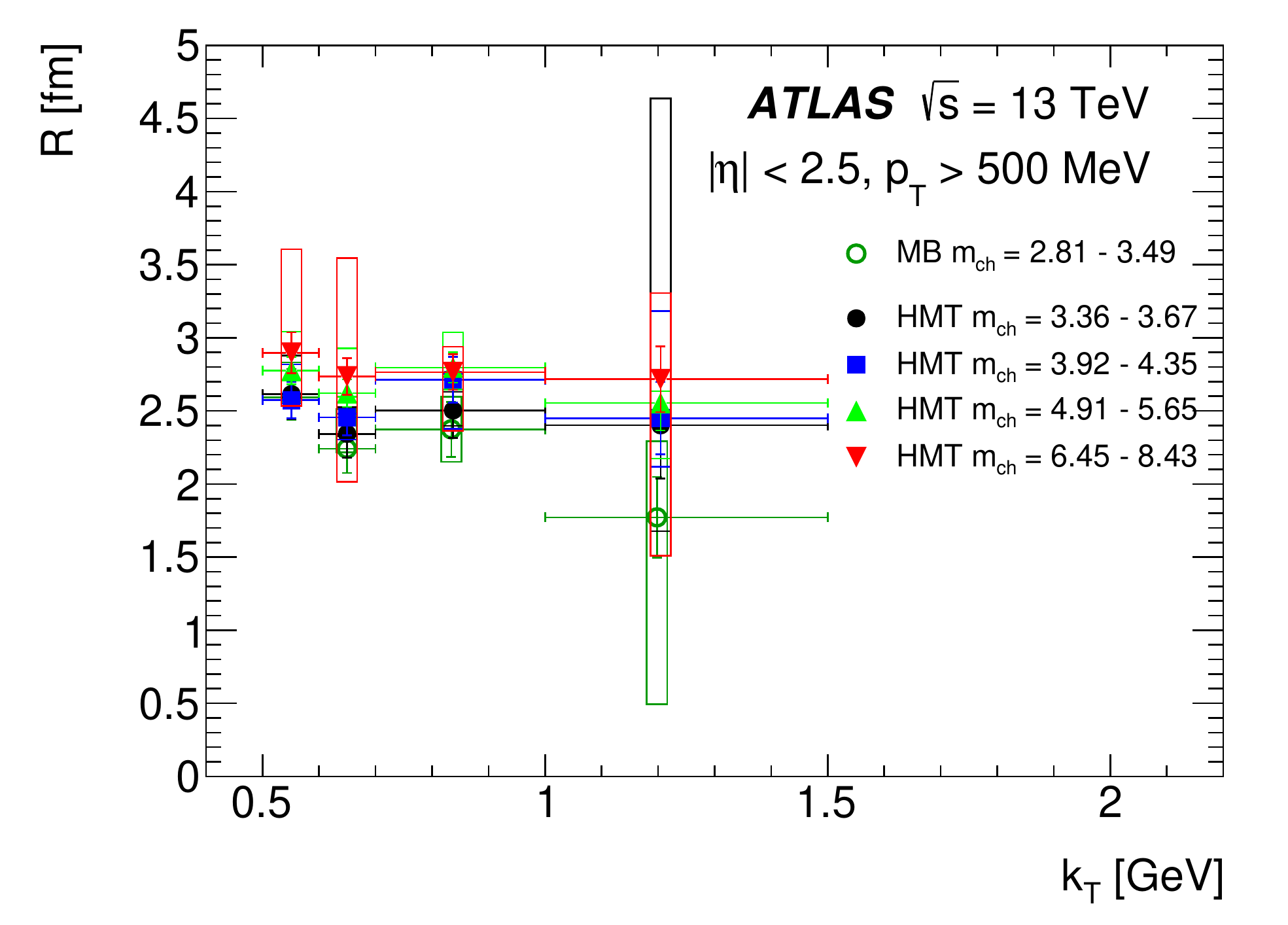}}
(b)
\\
\end{minipage}
\vfill
\begin{minipage}[h]{0.50\linewidth}
\center{\includegraphics[width=1\linewidth]{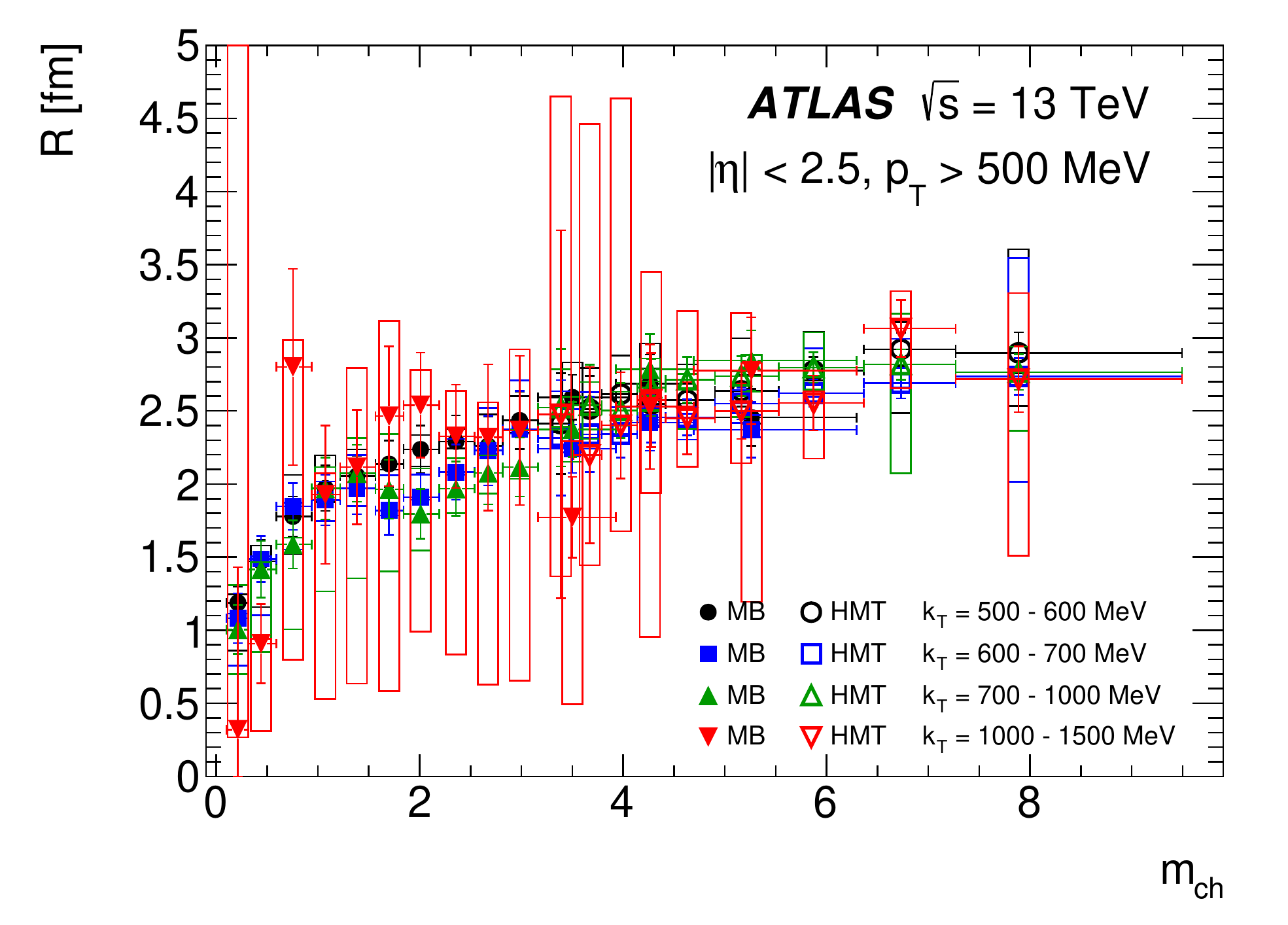}}
(c)
\\
\end{minipage}
\caption{
The parameter \( R \) for \( p_{\mathrm{T}} > 500 \)~\MeV:
(a) as a function of \( k_{\mathrm{T}} \) in selected low \( m_{\mathrm{ch}} \) intervals,
(b) as a function of \( k_{\mathrm{T}} \) in selected high \( m_{\mathrm{ch}} \) intervals,
and (c) as a function of \( m_{\mathrm{ch}} \) in \( k_{\mathrm{T}} \)
intervals between 0.5 and 1.5~\GeV.
The error bars and boxes represent the  statistical  and systematic  contributions, respectively.
}
\label{fig:R_1Dprojections_pt500}
\end{figure*}
 
The 2D dependence has also been studied for the  \(p_{\mathrm{T}} > 500\)~\MeV\  selection
and is shown in  Figure~\ref{fig:LambdaR_2Dsurfaces_pt500}.
As explained above, the \(p_{\mathrm{T}}\) selection constrains the BEC parameters to the region
\( k_{\mathrm{T}} > 500\)~\MeV.
 
The main features are similar to those observed in
Figure~\ref{fig:LambdaR_2Dsurfaces_pt100}.
In particular, the value of \(R\) tends to a plateau at large \(m_{\mathrm{ch}}\) in all \(k_{\mathrm{T}}\) intervals,
but the plateau level is systematically lower than for
\( p_{\mathrm{T}} > 100\)~\MeV,
in line with what is observed in Figures~\ref{Fig_pTcuts_com_nch}(b) and \ref{Fig_pTcuts_com_nch}(c).
Selected 1D projections are shown in
Figures~\ref{fig:Lambda_1Dprojections_pt500} and
\ref{fig:R_1Dprojections_pt500}
for \(\lambda\) and \(R\), respectively.
As with the lower \(p_{\mathrm{T}}\) selection,
MB and HMT data agree where they overlap.
 
The 2D behaviour is parameterized according to
Eqs.~(\ref{eq_LmchkT}) and (\ref{eq_RmchkT}).
The fit parameters \( \mu, \nu, \xi \) and \(\kappa\) for \(p_{\mathrm{T}} > 500\)~\MeV\
are shown in Figures~\ref{Fig_LkT_nch}
and \ref{Fig_RkT_nch}.
As before, the parameters \( \mu, \nu, \xi \) and \(\kappa\) are themselves  fitted as functions of
\(m_{\mathrm{ch}}\).
For the \(\lambda\) parameters and \(p_{\mathrm{T}} > 500\)~\MeV:
\begin{equation}\nonumber
\mu (m_{\mathrm{ch}}) =
(1.61\pm 0.32 (\mathrm{stat.})~^{+ 0.36}_{- 0.48} (\mathrm{syst.}))
\cdot
\exp{ \{ (-0.057\pm 0.051 (\mathrm{stat.})~^{+ 0.005}_{- 0.022}  (\mathrm{syst.}))\,
m_{\mathrm{ch}} \} },
\end{equation}
\begin{equation}\nonumber
\nu (m_{\mathrm{ch}}) =
(1.64\pm 0.34 (\mathrm{stat.})~^{+ 0.53}_{- 0.72}
(\mathrm{syst.}))\,
{\mathrm{[GeV^{-1}]}}
\cdot
\exp{ \{  (-0.110\pm 0.061 (\mathrm{stat.})~^{+ 0.009}_{- 0.039}
(\mathrm{syst.}))\,
m_{\mathrm{ch}} \} }.
\end{equation}
And for the \(R\) parameters and \(p_{\mathrm{T}} > 500\)~\MeV:
\begin{equation}\nonumber
\xi (m_{\mathrm{ch}}) =
2.65\pm 0.10 (\mathrm{stat.})~^{+0.51}_{-0.78} (\mathrm{syst.})\,
{\mathrm{[fm]}} .
\end{equation}
\begin{equation}\nonumber
\kappa (m_{\mathrm{ch}}) =
(1.78\pm 2.22 (\mathrm{stat.})~^{+1.11}_{-1.82} (\mathrm{syst.}))\, {\mathrm{[GeV^{-1}]}}
\cdot
\exp{ \{ (-3.37\pm 3.10 (\mathrm{stat.})~^{+3.20}_{-6.61} (\mathrm{syst.}))\,
m_{\mathrm{ch}} \} } ,
\end{equation}
where the fit for \(\xi\) applies for \(m_{\mathrm{ch}} > 2.8\).

 
\FloatBarrier
\subsection{Comparison with CMS}
\label{sect:CMScomparison}
 
\begin{figure*}[t!]
\begin{minipage}[h]{0.50\linewidth}
\center{\includegraphics[width=1\linewidth]{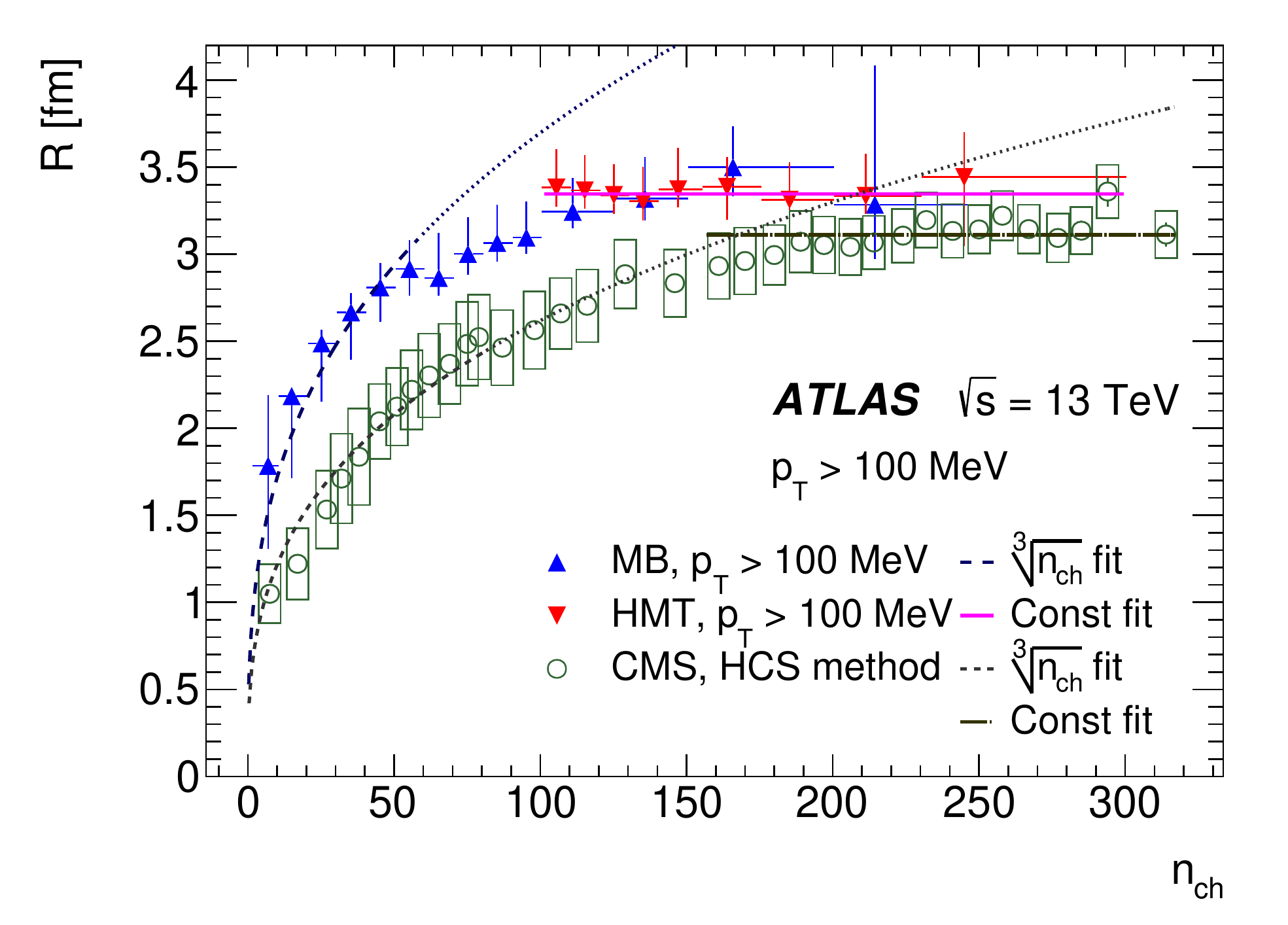}}
(a) \\
\end{minipage}
\hfill
\begin{minipage}[h]{0.50\linewidth}
\center{\includegraphics[width=1\linewidth]{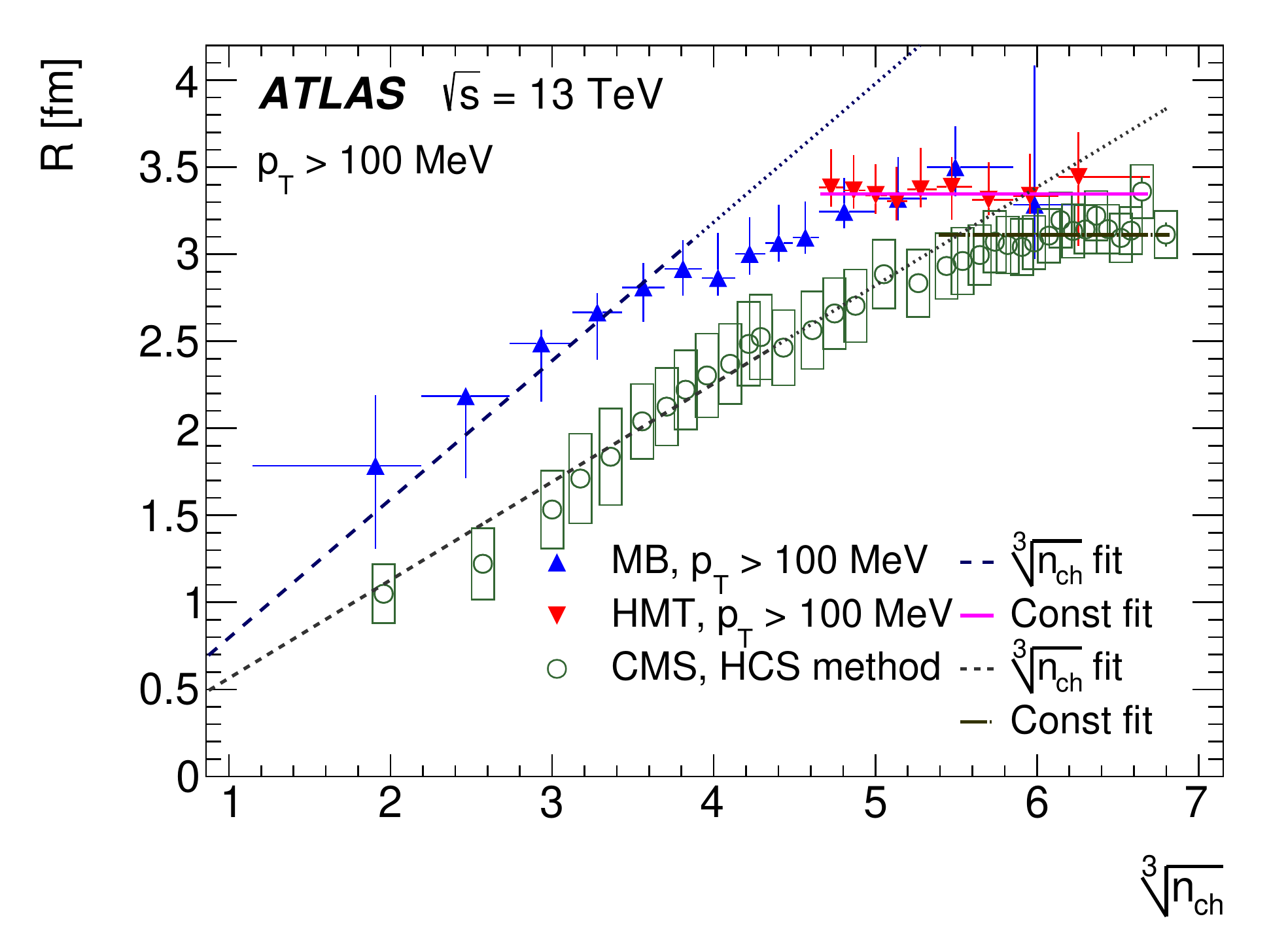}}
(b) \\
\end{minipage}
\caption{
ATLAS and CMS results for the source radius \(R\) as a function of (a) \(n_{\mathrm{ch}}\)
and (b) \(\sqrt[3]{n_{\mathrm{ch}}}\) in \(pp\) interactions at \(13\)~\TeV.
The CMS results (open circles) have been adjusted (by the CMS collaboration) to the ATLAS kinematic region:
\(p_{\mathrm{T}} > 100\)~\MeV\ and \(|\eta|<2.5\).
The ATLAS uncertainties are the sum in quadrature of the statistical and asymmetric systematic uncertainties.
For CMS, only the systematic uncertainties are shown since the statistical uncertainties
are smaller than the  marker size.
The dashed blue (ATLAS) and black (CMS) lines represent the fit to \(\sqrt[3]{n_{\mathrm{ch}}}\)
at low multiplicity, continued as dotted lines beyond the fit range.
The solid green (ATLAS) and broken black (CMS) lines indicate the plateau level at high multiplicity.
}
\label{Fig_R_ATLAS-CMS_comparison}
\end{figure*}
 
As noted in the Introduction, BEC have been studied extensively at the LHC.
In this section we compare our results on the BEC radius parameter with the only
other published study at \(13\)~\TeV.
The CMS Collaboration has studied BEC at \(13\)~\TeV\  in the
kinematic region \( p_{\mathrm{T}} > 200 \)~\MeV, \( |\eta|<2.4 \)
and \( k_{\mathrm{T}} < 1 \)~\GeV\  \cite{Sirunyan:2019umv}.
In the CMS publication comparison is made to the ATLAS \(7\)~\TeV\ results.
To do so, CMS has adjusted its results for \(R\) to match the ATLAS kinematic region,
\( p_{\mathrm{T}} > 100 \)~\MeV, \( |\eta| < 2.5 \),
by extrapolating the  \( n_{\mathrm{ch}} \) values to the lower
\( p_{\mathrm{T}} \) threshold and excluding the same \(Q\) regions as ATLAS in the fit to \(R_2 (Q)\).
These adjusted CMS values are shown together with the ATLAS results in
Fig.~\ref{Fig_R_ATLAS-CMS_comparison}.
The behaviour of \(R\) in the two experiments is qualitatively similar.
However, there is a clear difference in the results, particularly at low \( n_{\mathrm{ch}} \).
Results from this analysis (Fig.~\ref{Fig_pTcuts_com_nch}(b))
and earlier studies for the lower-energy ATLAS analysis \cite{Aad:2015sja}
indicate a direct dependence of \(R\) on the \( p_{\mathrm{T}}\) threshold.
So the effect of the difference in kinematic coverage as well as the differing approaches
to constructing the reference sample requires further investigation.

\FloatBarrier
\section{Summary  and conclusions}
\label{summary}
 
Results are presented of measurements  of two-particle  BEC  of like-charge hadrons
with track \(p_{\mathrm{T}}\)-thresholds of \(100\) and \(500\)~\MeV\
and \(|\eta|< 2.5\) produced  in \(pp\) collisions  at \(\sqrt{s}= 13\)~\TeV\
recorded by the ATLAS  detector at the CERN LHC.
The study is carried out using data collected with  the  MB and  HMT
triggers.
The integrated luminosities are
\(151\)~\(\mu\)b\(^{-1}\)
and
\(8.4\)~nb\(^{-1}\)
for the MB and HMT data samples, respectively.
 
The studies are performed using the  \(R_2 (Q)\) correlation function,
which is the ratio of the \(C_2^{\mathrm{data}} (Q)\) correlation function
obtained  from the data  to the  \(C_2^{\mathrm{MC}} (Q)\)
calculated using  MC events with no BEC effect.
The \(C_2\) function is itself a ratio of like-charge particle pairs to a reference sample, and
the reference sample is constructed using unlike-charge particle pairs.
A clear indication of  BEC is observed  in the region of small four-momentum difference,
\(Q \lesssim 0.2\)~\GeV.
The BEC parameters, \(R\) and  \(\lambda\),
are studied as  functions  of a scaled charged-particle multiplicity
\( m_{\mathrm{ch}} \),
the pair average transverse momentum
\( (k_{\mathrm{T}}) \)
and in
\( (m_{\mathrm{ch}}, k_{\mathrm{T}}) \)-intervals.
 
The  parameter \( R (m_{\mathrm{ch}}) \) is found to increase as
\(\alpha \sqrt[3]{ m_{\mathrm{ch}} }\) for low multiplicity for both \( p_{\mathrm{T}} \) thresholds.
For \(p_{\mathrm{T}} > 100\)~\MeV,
\(\alpha = 2.54^{+0.12}_{-0.22} (\mathrm{total})\)~fm,
for scaled multiplicity up to  \(m_{\mathrm{ch}} \approx 2\).
For \(m_{\mathrm{ch}} \ge 3\), the source radius \(R\) saturates
at a value
\(R = 3.35^{+0.20}_{-0.09} (\mathrm{total})\)~fm,
confirming the previous observation of a high-multiplicity plateau
by ATLAS at \(\sqrt{s} = 7\)~\TeV.
For \(p_{\mathrm{T}} > 500\)~\MeV, the behaviour of \(R\) is similar, but systematically lower.
The  parameter \(\lambda  (m_{\mathrm{ch}})\) decreases with multiplicity
for the lower \( p_{\mathrm{T}} \) threshold and is lower for the higher \( p_{\mathrm{T}} \)
threshold but increases slightly with multiplicity.
The parameters \(R (k_{\mathrm{T}})\) and  \(\lambda (k_{\mathrm{T}})\)
both decrease with increasing pair average transverse momentum \(k_{\mathrm{T}}\).
Considering the combined \((m_{\mathrm{ch}}, k_{\mathrm{T}})\) dependence,
\(\lambda\)  decreases with \(k_{\mathrm{T}}\) in all \( m_{\mathrm{ch}} \) intervals,
for both \( p_{\mathrm{T}} \) thresholds.
The decrease is more pronounced at lower multiplicity.
The  radius parameter \(R\) also decreases with
\(k_{\mathrm{T}}\) for \( p_{\mathrm{T}} >100\)~\MeV,
decreasing more strongly in the lower \( m_{\mathrm{ch}} \) intervals. For \(p_{\mathrm{T}} > 500 \)~\MeV\  the
behaviour of \(R\) is rather flat with \( k_{\mathrm{T}} \).
As a function of \( m_{\mathrm{ch}} \), \(R\) increases and reaches a plateau at large multiplicity
in all \( m_{\mathrm{ch}} \) intervals and for both \( p_{\mathrm{T}} \) thresholds.
Within the uncertainties of this analysis,
the turn-on of the plateau occurs at the same value of \( m_{\mathrm{ch}} \)
in all \( k_{\mathrm{T}} \) intervals
and for both \( p_{\mathrm{T}} \) thresholds.
A comparison of the \(R\) parameter as a function of \( n_{\mathrm{ch}} \) with CMS results at
\(13\)~\TeV\ shows qualitatively similar behaviour and the differences at low \( n_{\mathrm{ch}} \)
require further investigation.
 
The measurements presented here complement the ATLAS measurements at
\(\sqrt{s} = 0.9\) and \(7\)~\TeV\ by extending  the studies to higher multiplicities
and pair average transverse momenta, and provide a more detailed study of the BEC parameters
by  exploring the double-differential \((m_{\mathrm{ch}}, k_{\mathrm{T}})\)-dependence.
The present results confirm the saturation of the source radius at  higher  multiplicities
with  \( n_{\mathrm{ch}} \gtrsim 100\).
 
 
 
\clearpage
\section*{Acknowledgements}
 
We thank CERN for the very successful operation of the LHC, as well as the support staff from our institutions without whom ATLAS could not be operated efficiently.
 
We acknowledge the support of
ANPCyT, Argentina;
YerPhI, Armenia;
ARC, Australia;
BMWFW and FWF, Austria;
ANAS, Azerbaijan;
SSTC, Belarus;
CNPq and FAPESP, Brazil;
NSERC, NRC and CFI, Canada;
CERN;
ANID, Chile;
CAS, MOST and NSFC, China;
Minciencias, Colombia;
MEYS CR, Czech Republic;
DNRF and DNSRC, Denmark;
IN2P3-CNRS and CEA-DRF/IRFU, France;
SRNSFG, Georgia;
BMBF, HGF and MPG, Germany;
GSRI, Greece;
RGC and Hong Kong SAR, China;
ISF and Benoziyo Center, Israel;
INFN, Italy;
MEXT and JSPS, Japan;
CNRST, Morocco;
NWO, Netherlands;
RCN, Norway;
MEiN, Poland;
FCT, Portugal;
MNE/IFA, Romania;
JINR;
MES of Russia and NRC KI, Russian Federation;
MESTD, Serbia;
MSSR, Slovakia;
ARRS and MIZŠ, Slovenia;
DSI/NRF, South Africa;
MICINN, Spain;
SRC and Wallenberg Foundation, Sweden;
SERI, SNSF and Cantons of Bern and Geneva, Switzerland;
MOST, Taiwan;
TAEK, Turkey;
STFC, United Kingdom;
DOE and NSF, United States of America.
In addition, individual groups and members have received support from
BCKDF, CANARIE, Compute Canada and CRC, Canada;
COST, ERC, ERDF, Horizon 2020 and Marie Skłodowska-Curie Actions, European Union;
Investissements d'Avenir Labex, Investissements d'Avenir Idex and ANR, France;
DFG and AvH Foundation, Germany;
Herakleitos, Thales and Aristeia programmes co-financed by EU-ESF and the Greek NSRF, Greece;
BSF-NSF and GIF, Israel;
Norwegian Financial Mechanism 2014-2021, Norway;
NCN and NAWA, Poland;
La Caixa Banking Foundation, CERCA Programme Generalitat de Catalunya and PROMETEO and GenT Programmes Generalitat Valenciana, Spain;
Göran Gustafssons Stiftelse, Sweden;
The Royal Society and Leverhulme Trust, United Kingdom.
 
The crucial computing support from all WLCG partners is acknowledged gratefully, in particular from CERN, the ATLAS Tier-1 facilities at TRIUMF (Canada), NDGF (Denmark, Norway, Sweden), CC-IN2P3 (France), KIT/GridKA (Germany), INFN-CNAF (Italy), NL-T1 (Netherlands), PIC (Spain), ASGC (Taiwan), RAL (UK) and BNL (USA), the Tier-2 facilities worldwide and large non-WLCG resource providers. Major contributors of computing resources are listed in Ref. \cite{ATL-SOFT-PUB-2021-003}.

\clearpage
\appendix
\part*{Appendix}
\section{Effect of resonances on \(R_2(Q)\) correlation function}
\label{app:resonances}
 
\begin{figure*}[h!]
\centering
\begin{minipage}[h]{0.50\linewidth}
\center{\includegraphics[width=1\linewidth]{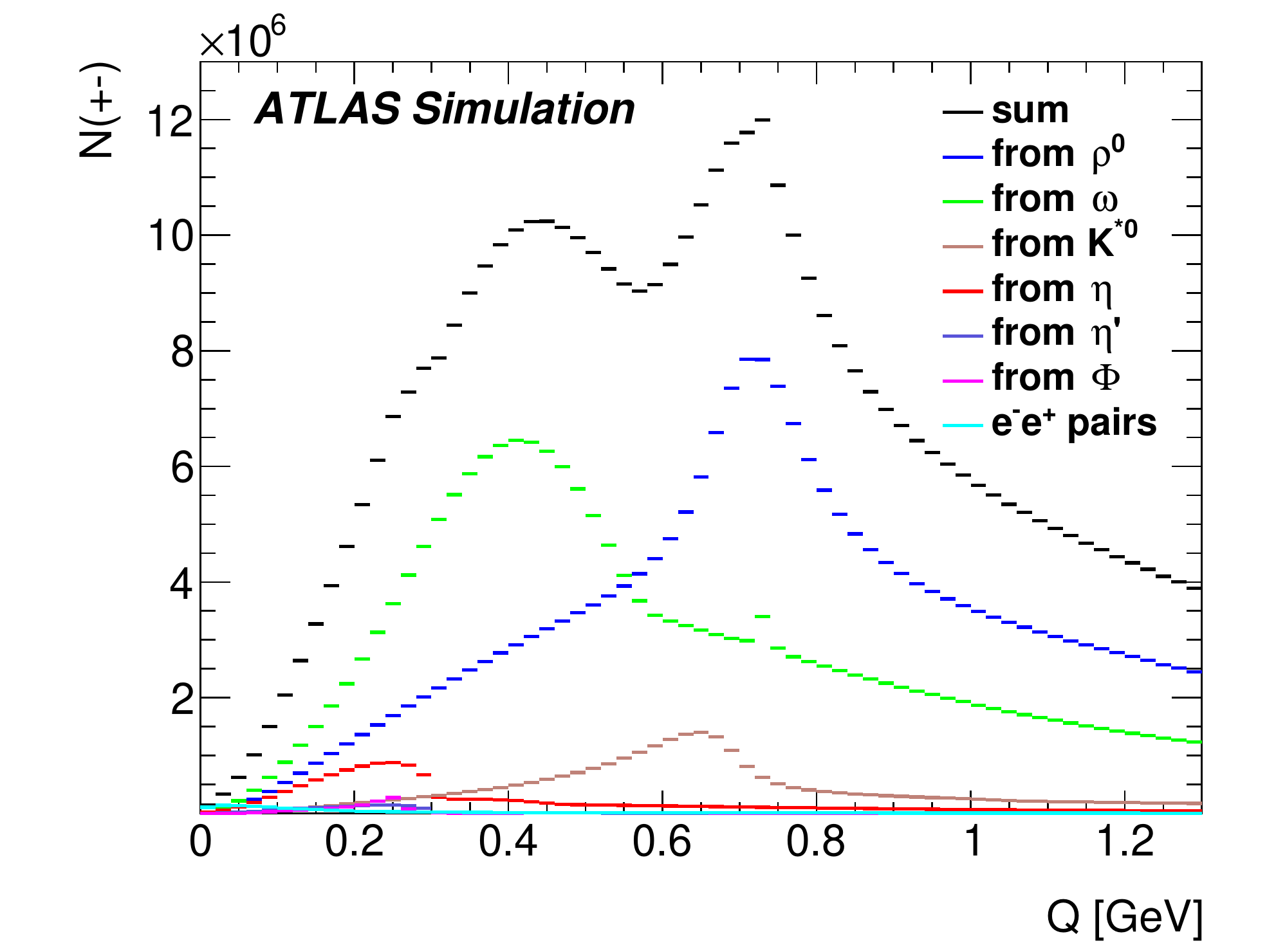}}
\\
\end{minipage}
\caption{The inclusive \(Q\) distribution of unlike-charge particle  pairs
for track pairs originating from the decays of various species of single resonances.
The full range of multiplicities contribute to each distribution.
}
\label{fig:resonances_1}
\end{figure*}

The potential influence of resonances on the  \(R_2 (Q)\)
is  evident by studying the \(Q\) distributions of
the  UCP pairs (\(h^\pm h^\mp\))  where both particles come from
the same species of resonance;
this is shown in    Figure~\ref{fig:resonances_1}.
The presented distributions are for track pairs  with \(p_{\mathrm{T}} > 100\)~\MeV.
The reflections of the resonance peaks  can be seen clearly, and arise  mainly
from  contributions from the   \(\eta\),    \(\omega\),   \(\rho\)   and  \(K^{*0}\)  resonances,
with much smaller   contributions from the  \(\eta^{\prime}\) and \(\phi\) mesons.
The contributions  of \(f_0 (980) \rightarrow \pi^+ \pi^-\) and
\(f_2 (1270) \rightarrow \pi^+ \pi^-\) are  insignificant.
 
\newpage

 
\clearpage
\newpage
 
\begin{flushleft}
\hypersetup{urlcolor=black}
{\Large The ATLAS Collaboration}

\bigskip

\AtlasOrcid[0000-0002-6665-4934]{G.~Aad}$^\textrm{\scriptsize 102}$,    
\AtlasOrcid[0000-0002-5888-2734]{B.~Abbott}$^\textrm{\scriptsize 129}$,    
\AtlasOrcid[0000-0002-7248-3203]{D.C.~Abbott}$^\textrm{\scriptsize 103}$,    
\AtlasOrcid[0000-0002-2788-3822]{A.~Abed~Abud}$^\textrm{\scriptsize 71a,71b}$,    
\AtlasOrcid[0000-0002-1002-1652]{K.~Abeling}$^\textrm{\scriptsize 53}$,    
\AtlasOrcid[0000-0002-2987-4006]{D.K.~Abhayasinghe}$^\textrm{\scriptsize 94}$,    
\AtlasOrcid[0000-0002-8496-9294]{S.H.~Abidi}$^\textrm{\scriptsize 168}$,    
\AtlasOrcid[0000-0002-8279-9324]{O.S.~AbouZeid}$^\textrm{\scriptsize 40}$,    
\AtlasOrcid{N.L.~Abraham}$^\textrm{\scriptsize 157}$,    
\AtlasOrcid[0000-0001-5329-6640]{H.~Abramowicz}$^\textrm{\scriptsize 162}$
% ...
\bigskip
\\

% ...
$^{40}$Niels Bohr Institute, University of Copenhagen, Copenhagen; Denmark.\\
% ...
$^{53}$II. Physikalisches Institut, Georg-August-Universit\"{a}t G\"ottingen, G\"ottingen; Germany.\\
% ...
$^{71}$$^{(a)}$INFN Sezione di Pavia;$^{(b)}$Dipartimento di Fisica, Universit\`a di Pavia, Pavia; Italy.\\
% ...
$^{94}$Department of Physics, Royal Holloway University of London, Egham; United Kingdom.\\
% ...
$^{102}$CPPM, Aix-Marseille Universit\'e, CNRS/IN2P3, Marseille; France.\\
$^{103}$Department of Physics, University of Massachusetts, Amherst MA; United States of America.\\
% ...
$^{129}$Homer L. Dodge Department of Physics and Astronomy, University of Oklahoma, Norman OK; United States of America.\\
% ...
$^{157}$Department of Physics and Astronomy, University of Sussex, Brighton; United Kingdom.\\
% ...
$^{162}$Raymond and Beverly Sackler School of Physics and Astronomy, Tel Aviv University, Tel Aviv; Israel.\\
% ...
$^{168}$Department of Physics, University of Toronto, Toronto ON; Canada.\\
% ...
$^{a}$ Also at Borough of Manhattan Community College, City University of New York, New York NY; United States of America.\\
$^{b}$ Also at CERN, Geneva; Switzerland.\\
% ...
$^{*}$ Deceased

\end{flushleft}


\end{document}